\documentclass[a4paper,11pt]{article}
\pdfoutput=1
\usepackage{jheppub} 
\usepackage{lineno}

\usepackage{amsmath,amsfonts,amssymb,amsthm}

\usepackage{graphicx}
\usepackage{booktabs}
\usepackage{hyperref}
\usepackage{verbatim}
\usepackage[normalem]{ulem}
\usepackage[scr=boondoxo]{mathalpha}
\usepackage{slashed}

\usepackage{tikz}
\usetikzlibrary{arrows.meta}
\usepackage{pgfplots}
\pgfplotsset{compat=1.18}
\usepgfplotslibrary{groupplots}
\definecolor{wedgefill}{RGB}{216,216,255}
\definecolor{quadfill}{RGB}{253,236,219}

\newcommand\fft[2]{\frac{#1}{#2}}
\newcommand\ft[2]{{\textstyle\frac{#1}{#2}}}
\newcommand\nn{\nonumber}

\renewcommand\Im{\mbox{Im}}
\DeclareMathOperator{\Tr}{Tr}

\newcommand{\fsl}[1]{{\ooalign{\(#1\)\cr\hidewidth\(/\)\hidewidth\cr}}}

\preprint{LITP-26-19}

\title{Giant Gravitons in AdS$_5\times$SE$_5$ and the Superconformal Index of Toric Quiver Gauge Theories}

\author{Evan Deddo,}
\emailAdd{evdedd@umich.edu}
\author{Sabarenath Jayaprakash,}
\emailAdd{sabare@umich.edu}
\author{James T. Liu}
\emailAdd{jimliu@umich.edu}
\author{and Leopoldo A. Pando Zayas}
\emailAdd{lpandoz@umich.edu}

\affiliation[]{Leinweber Institute for Theoretical Physics, University of Michigan, Ann Arbor, MI 48109-1040, USA}

\abstract{We derive giant-graviton contributions to superconformal indices by quantizing probe D3-branes in type IIB backgrounds
$\mathrm{AdS}_{5}\times SE_{5}$. For $SE_{5}=T^{1,1}$ and $Y^{p,q}$,
we determine the scalar, vector, and fermionic fluctuations of maximal
giant gravitons and extract their
BPS spectra. The resulting single-letter indices agree with those predicted by the giant-graviton expansions of the dual toric quiver gauge theories. These results show that the single-giant terms in finite-$N$ supersymmetric counting functions can be recovered directly from quantized D3-brane degrees of
freedom and point toward a universal worldvolume origin of giant-graviton expansions directly on the supergravity side. We view the results as a step in an intrinsically gravitational framework, much like the open-closed-open triality,  for addressing supersymmetric microstate-counting problems that are conventionally outsourced to the dual field theory.}

\begin{document}

\maketitle

\section{Introduction and Summary}
\label{sec:introduction}

Supersymmetric indices provide some of the most precise probes of the microscopic content of holographic dualities. Because they are invariant
under continuous supersymmetry-preserving deformations, they can often be
computed exactly in the boundary theory and compared with semiclassical or quantum calculations in the bulk. At infinite $N$, the superconformal index of the boundary theory is reproduced by the
Kaluza-Klein spectrum of the dual supergravity background \cite{Kinney:2005ej}. The finite-$N$
structure of the superconformal index naturally probes bulk quantum aspects due to finite Newton constant effects and seems to require genuinely extended objects which might be viewed as emblems of non-perturbative bulk physics. 

A particularly useful organization of these finite-$N$ effects is provided
by the giant-graviton expansion~\cite{Gaiotto:2021xce,Imamura:2021ytr,Murthy:2022ien} (see relevant  precedent in \cite{Bourdier:2015wda, Bourdier:2015sga}). Schematically, for toric
quiver gauge theories this expansion takes the form
\begin{equation}
  \frac{Z_N}{Z_\infty}
  =
  \sum_{\{n_I\}}
  \left(\prod_I v_I^{N n_I}\right)
  \hat I_{(\{n_I\})}^{\mathrm{GG}} .
  \label{eq:intro-GGE}
\end{equation}
Here $Z_N$ is the finite-$N$ superconformal index, while $Z_\infty$
captures the perturbative Kaluza-Klein spectrum in the corresponding
supergravity background. The factors $v_I^{N n_I}$ denote contributions of fugacities $v_I$ with $\{n_I\}$ a set of integers. It is remarkable that the giant graviton index $\hat I_{(\{n_I\})}^{\mathrm{GG}}$ is independent of $N$. On the bulk side, the factors $v_I^{N n_I}$ encode the classical
charges of D3-branes wrapped on supersymmetric three-cycles, and
$\hat I_{(\{n_I\})}^{\mathrm{GG}}$ counts their BPS
excitations. Equation~\eqref{eq:intro-GGE} thus suggests a direct
gravitational interpretation of finite-$N$ corrections: they arise from
giant gravitons, namely D3-branes wrapping nontrivial cycles in the
internal geometry and carrying angular momentum along transverse
directions.

This interpretation has been investigated most extensively in the maximally supersymmetric $\mathrm{AdS}_{5}\times S^{5}$ background. Direct
probe-brane calculations have reproduced the single-giant and finite-$N$  contributions to the half-BPS and more general superconformal indices~\cite{Eleftheriou:2023jxr,Lee:2024hef,Eleftheriou:2025lac}.
A complementary derivation based on  the fully backreacted Lin-Lunin-Maldacena
geometries demonstrated how finite-$N$ discreteness emerges from the
quantization of the gravitational moduli space~\cite{Deddo:2024liu}. More recently,
the quantization of probe D3-, M2-, and M5-brane fluctuations has revealed
a common organization of the bosonic and fermionic BPS modes contributing
to giant-graviton indices of other prototypical examples of AdS/CFT pairs~\cite{Deddo:2025lfm}. Together, these developments provide substantial evidence that giant gravitons are
not merely a useful field-theoretic reorganization of finite-$N$ corrections, but are the corresponding microscopic degrees of freedom in
the gravitational description.

It remains important to determine which aspects of this picture rely on the high degree of symmetry of sphere compactifications (AdS$_5\times S^5$, AdS$_{4}\times S^{7}$ and AdS$_7\times S^4$)  and which persist
in more general holographic backgrounds. Theories with only 
$\mathcal{N}=1$ supersymmetry furnish a natural setting in which to address
this question. Type IIB compactifications on
$\mathrm{AdS}_{5}\times SE_{5}$, where $SE_{5}$ is a five-dimensional
Sasaki-Einstein manifold, give rise to a broad class of four-dimensional
$\mathcal{N}=1$ superconformal field theories \cite{Martelli:2004wu,Benvenuti:2004dy}. In contrast with the
$\mathrm{AdS}_{5}\times S^{5}$ example, these backgrounds generally lack
integrability~\cite{Basu:2011di,Basu:2011fw}, and their wrapped branes probe cycles with
nontrivial topology, squashing, and nontrivial normal-bundle structure. A discussion of the  finite-$N$ half-BPS analogue in these theories was recently addressed in \cite{GonzalezLezcano:2026wje}. More generally, a successful
derivation of giant-graviton indices in this setting would therefore test whether the worldvolume mechanism underlying the expansion is robust
under a substantial reduction of symmetry.

In this work, we perform such a derivation for
$\mathrm{AdS}_{5}\times T^{1,1}$ and for the infinite family of $\mathrm{AdS}_{5}\times Y^{p,q}$ backgrounds. The first example is dual to
the Klebanov-Witten theory~\cite{Klebanov:1998hh} and provides the simplest
non-spherical Sasaki-Einstein compactification. The $Y^{p,q}$ spaces form
an infinite family of toric Sasaki-Einstein manifolds~\cite{Martelli:2004wu}, whose
dual descriptions are $\mathcal{N}=1$ quiver superconformal field theories \cite{Benvenuti:2004dy}. In both cases, the toric divisors determine distinguished
supersymmetric three-cycles on which maximal giant gravitons may wrap. The expected contributions of these branes to the field-theory index are
known from the toric formulation of the giant-graviton expansion~\cite{Arai:2019aou,Fujiwara:2023azx,Kim:2024ucf,Imamura:2021ytr}. Our objective is to reproduce these contributions
directly from the quantized degrees of freedom of the wrapped D3-branes.

We work in the probe approximation and expand the D3-brane Dirac-Born-Infeld and Wess-Zumino actions to quadratic order around each
maximal giant-graviton embedding. We quantize the resulting scalar, worldvolume gauge-field, and fermionic fluctuations, determine their
charges under the Cartan generators appearing in the superconformal index, and isolate the modes that saturate the relevant BPS bound. The
single-letter giant-graviton index is then obtained by summing the bosonic and fermionic BPS excitations with the appropriate signs and
fugacities. This procedure treats the single-brane sector of \eqref{eq:intro-GGE}. Contributions involving coincident giants would
require the corresponding non-Abelian worldvolume theory and lie beyond the scope of the present analysis.

For $\mathrm{AdS}_{5}\times T^{1,1}$, there are four types of maximal
giants associated with the four external vertices of the toric diagram. Their worldvolumes are squashed three-spheres, and the reduced symmetry of
the geometry leads to a nontrivial reorganization of the fluctuation spectrum relative to the $S^{5}$ case. Nevertheless, after imposing the
BPS condition, the contributions of the scalar, vector, and fermionic
modes combine into precisely the single-letter indices predicted by the giant-graviton expansion. This agreement includes the
dependence on the mesonic and baryonic fugacities and provides a direct
bulk derivation of all four single-giant sectors. Unlike the giants in $S^5$ which wrap topologically trivial $S^3$'s the giants in this case wrap non trivial cycles. Such non-trivial cycles are the geometric origin of baryon number in the dual field theory which tracks it in the index with a particular fugacity \cite{Gubser:1998fp, Berenstein:2002ke}. Even though this symmetry is not directly visible from the $T^{1,1}$ or $Y^{p,q}$ geometries, the giants definitely probe it. The separation of  the configuration of giants and their fluctuations into baryon number sectors is a genuinely new feature of our analysis which is not present in the simpler $S^5$ giants.

The $Y^{p,q}$ backgrounds further provide a considerably subtler structure. The
supersymmetric cycles wrapped by the maximal giants are lens spaces
$L(k_I,1)$, with $k_I$ determined by the toric data. Correspondingly, the
single-letter contribution associated with a cycle $S_I$ is expected to
take the form of an orbifold projection,
\begin{equation}
  f_I(\mathfrak p, \mathfrak q;v_i)
  =
  \frac{1}{k_I}
  \sum_{s=0}^{k_I-1}
  f\left( \mathfrak
    p, \mathfrak q;
    \omega^s (w_{I-\frac12})^{1/k_I},
    \widetilde\omega^s (w_{I+\frac12})^{1/k_I}
  \right),
  \label{eq:intro-orbifold-index}
\end{equation}
where $\omega$ and $\widetilde\omega$ are appropriately related
$k_I$-th roots of unity, and
\begin{equation}
  f( \mathfrak p, \mathfrak q;w,w')
  =
  1-
  \frac{(1-ww'/\mathfrak{pq})(1-\mathfrak{p})(1-\mathfrak{q})}
       {(1-w)(1-w')}
  \label{eq:intro-universal-letter}
\end{equation}
is the unprojected single-letter expression; all other arguments $\mathfrak{p,q}, v_i, w, w'$  and $w_{I\pm\fft12}$ are fugacities associated to the superconformal and flavor symmetries. We show that both the
universal function~\eqref{eq:intro-universal-letter} and the projection
in~\eqref{eq:intro-orbifold-index} arise directly from the D3-brane
fluctuation problem. In particular, the topology of the wrapped cycle
determines the allowed lattice of worldvolume quantum numbers and hence
implements the required orbifold projection without additional
field-theoretic input.

The four $Y^{p,q}$ giants separate naturally into two geometrically
distinct classes. Class one giants are localized at the endpoints of the
$y$ interval (in the conventional parametrization of $Y^{p,q}$) and wraps a lens-space orbit generated by the remaining
angular coordinates. Its spectrum can be obtained through harmonic
analysis on a squashed lens space and yields the expected
$\mathbb{Z}_{p+q}$ and $\mathbb{Z}_{p-q}$ projections. Class two giants are
localized at poles of the round two-sphere and wraps a cycle parametrized in
part by the nontrivial $y$ direction. In this case, the scalar and
fermionic fluctuation equations reduce to Heun-type problems, and the
full spectrum is not available in closed form. The BPS spectrum can
nevertheless be determined exactly.  It follows that the regular zero modes exhaust the bosonic BPS spectrum. An analogous analysis identifies the fermionic BPS
modes. Their charge lattice produces the required $\mathbb{Z}_p$
projection, and the complete scalar, vector, and fermionic contributions
again combine into~\eqref{eq:intro-universal-letter}.

Our results establish that the single-giant contributions predicted by
the field-theoretic expansion are encoded in the perturbative
worldvolume dynamics of wrapped D3-branes, even in backgrounds with
reduced supersymmetry and nontrivial cycle topology. They also reveal a
useful separation between local and global information. The unprojected
single-letter function has the same form for all giants considered here,
whereas the topology and embedding of the wrapped cycle determine the
allowed charge lattice and its orbifold projection. This structure
suggests a geometric origin for the apparent universality of
giant-graviton indices. It would be desirable to formulate this
universality more conceptually, perhaps by relating the worldvolume BPS
zero modes to holomorphic functions or sections on the Calabi-Yau cone following the original formulation in \cite{Mikhailov:2000ya, Beasley:2002xv} for the half-BPS sector.
The present results provide evidence for such a general framework across
toric Sasaki-Einstein compactifications, although an analysis of
arbitrary $SE_{5}$ geometries and of the interacting multi-giant sectors
remains an open problem. Our results can be viewed as an important step in the open-closed-open triality  \cite{Gopakumar:2022djw,Gopakumar:2024jfq} that allows one to, in principle, tackle the microstate counting directly in the giant graviton framework and expect a finite-$N$ reconstruction of the original open string formulation. 

The remainder of the paper is organized as follows. In
section~\ref{sec:KWreview}, we review the superconformal index of the
Klebanov-Witten theory and its giant-graviton expansion. In
section~\ref{sec:T11-giants}, we quantize the fluctuations of maximal
D3-branes in $\mathrm{AdS}_{5}\times T^{1,1}$ and reconstruct the
corresponding single-letter indices. Section~\ref{sec:Ypq-review} reviews
the $Y^{p,q}$ geometries, their dual quiver theories, and the
field-theory prediction for the giant-graviton expansion. In
section~\ref{sec:Ypq-giants}, we analyze the two classes of maximal
$Y^{p,q}$ giants and derive their orbifold single-letter indices from the
quantized BPS spectra. We conclude in section~\ref{sec:conclusions}.
The appendices collect the harmonic analysis on squashed lens spaces in appendix~\ref{app:lens},
details of the internal fluctuation equations in appendix~\ref{app: internal modes} and fermions in appendix~\ref{App:FermionsClassII}.

\section{Review of Klebanov-Witten Theory}\label{sec:KWreview}

Type IIB string theory on AdS$_5\times T^{1,1}$ gives rise to the Klebanov-Witten theory \cite{Klebanov:1998hh}.  This is an $\mathcal N=1$ super-Yang-Mills quiver gauge theory with gauge group $SU(N)\times SU(N)$. We write the ten dimensional metric as
\begin{equation}
    ds_{10}^2 = L^2 (ds_{\mathrm{AdS}_5}^2 + ds_{T^{1,1}}^2),
\end{equation}
where $L$ is the curvature scale.
In the standard presentation, the $T^{1,1}$ metric can be written as a circle bundle over $S^2\times S^2$
\begin{equation}
    ds^2 =\dfrac{1}{6} (d\theta_1^2 + \sin^2\theta_1 d\phi_1^2) + \dfrac{1}{6} (d\theta_2^2 + \sin^2\theta_2 d\phi_2^2)+ \dfrac{1}{9} (d\psi + \cos\theta_1 \, d\phi_1 + \cos\theta_2 \, d\phi_2)^2,
\end{equation}
with the identifications $\psi \to \psi+4\pi$ and $(\psi,\phi_i)\to (\psi+2\pi, \phi_i + 2\pi)$.  This  metric admits the isometry group $SU(2)_A\times SU(2)_B\times U(1)_r$, realized as isometries of the two $S^2$ factors and shifts in $\psi$. The isometry of the $U(1)_r$ corresponds to the $R$-symmetry of $\mathcal N=1$ SYM.  From the gauge theory point of view, the two $SU(2)$'s are flavor symmetries, often denoted as mesonic symmetries.  In addition, the Klebanov-Witten theory also has a $U(1)_B$ baryonic symmetry, which is not realized geometrically on $T^{1,1}$. Instead its bulk origin is topological; it arises from the reduction of the Ramond-Ramond four form on the nontrivial three cycles of $T^{1,1}$ \cite{Gubser:1998fp}.  The Abelian part of these global symmetries is then
\begin{equation*}
    U(1)_r \times U(1)_{F_A} \times U(1)_{F_B} \times U(1)_B,
\end{equation*}
corresponding to one $R$-symmetry and three flavors where $U(1)_{F_A}$ and $U(1)_{F_B}$ are the Cartan subgroups of $SU(2)_A$ and $SU(2)_B$ respectively.

The background is supported by a self dual five form, whose internal components are given by
\begin{equation}
    F_{[5]} = \dfrac{L^4}{27} \sin\theta_1\, \sin\theta_2 \, d\theta_1 \wedge d\theta_2 \wedge d\phi_1 \wedge d\phi_2 \wedge d\psi. 
\end{equation}
Flux quantization fixes the relation between the curvature scale $L$, D$3$ brane charge $N$ and the brane tension,
\begin{equation}
    T_3 L^4 =  \dfrac{27N}{32 \pi^2} \label{eq: tension}.
\end{equation}
We will use this normalization while analyzing the probe brane action in later sections.

\subsection{The field theory index}
We now briefly review the superconformal index of the KW theory. Fairly explicit discussions, including single-letter indices, exist in the literature  \cite{Nakayama:2005mf,Nakayama:2006ur,Gadde:2010en,Eager:2012hx,ArabiArdehali:2013vyp}. In particular, \cite{Gadde:2010en,Eager:2012hx,ArabiArdehali:2013vyp} include a supergravity derivation of the infinite-$N$ index.

On the field theory side, in radial quantization, the bosonic subgroup of the $\mathcal N=1$ superconformal group, $SU(2,2|1)$, decomposes as
\begin{align}
    SU(2,2|1)\supset SO(2,4)\times U(1)_r\supset& SO(2)\times SO(4)\times U(1)_r,\nn\\
    &\quad\Delta\qquad j_1,j_2\qquad r
\end{align}
where we use $(\Delta,j_1,j_2,r)$ to denote the corresponding charges.  The $\mathcal N=1$ superconformal index is a refined Witten index \cite{Romelsberger:2005eg,Kinney:2005ej}
\begin{equation}
    Z(\mathfrak p,\mathfrak  q)=\Tr[(-1)^Fe^{-\beta\mathcal H}\mathfrak p^{j_1+\fft12r}\mathfrak q^{j_2+\fft12r}],
\label{eq:N=1sci}
\end{equation}
where
\begin{equation}
    \mathcal{H} = \Delta - j_1 - j_2 - \dfrac{3}{2}r.
\end{equation}
Note that the only states that contribute to the index are the BPS ones that saturate the BPS bound $\mathcal{H} \geq 0$.

The Klebanov-Witten theory has three additional flavor symmetries, which can be used to further refine the index. Along with the $R$-symmetry, we denote these by $\rho, u, v,\zeta$ corresponding to the $r,F_A,F_B,B$ charges. The refined index may be written as,
\begin{align}
    Z(\mathfrak p,\mathfrak q;v_i) = \Tr [(-1)^F e^{-\beta \mathcal{H}}\mathfrak p^{j_1} \mathfrak q^{j_2} \rho^{3r/2}u^{F_A} v^{F_B} \zeta^B],
\end{align}
with the constraint $\rho^3 = \mathfrak p \mathfrak q$. It is convenient, however, to introduce four vertex fugacities, $v_i$, $i=1,\ldots,4$, satisfying the constraint $v_1v_2v_3v_4=\mathfrak p \mathfrak q$. This will make the comparison with the giant graviton expansions in AdS$_5 \times S^5$ and  existing literature straightforward. The index then takes the form
\begin{equation}
    Z(\mathfrak p,\mathfrak q,v_i)=\Tr[(-1)^Fe^{-\beta\mathcal H}\mathfrak p^{j_1} \mathfrak q^{j_2}v_1^{R_1}v_2^{R_2}v_3^{R_3}v_4^{R_4}],
\end{equation}
where the $R$-charge is given by $r=\fft12(R_1+R_2+R_3+R_4)$.  Note that the flavor charges are orthogonal to $r$.  Using the notation of \cite{Arai:2019aou}, we have
\begin{align}
    r&=\fft12(R_1+R_2+R_3+R_4),& F_A&=R_1-R_3,\nn\\
    B&=\fft1N(R_1-R_2+R_3-R_4),& F_B&=R_2-R_4,
\end{align}
and
\begin{align}
    u = \sqrt{\dfrac{v_1}{v_3}}, \quad v = \sqrt{\dfrac{v_2}{v_4}}, \quad \rho = (v_1v_2v_3v_4)^{1/3}, \quad \zeta = \left(\dfrac{v_1 v_3}{v_2v_4}\right)^{N/4}. \label{eq: uv to vi fugacity t11}
\end{align}
The normalization of $B$ is chosen so that a determinant operator constructed from an $A$ field has baryonic charge $+1$, while one constructed from a $B$ field has charge $-1$.

\begin{figure}[t]
\centering
\begin{tikzpicture}[
  node/.style={circle,draw,thick,minimum size=18mm,inner sep=0pt},
  arr/.style={-Latex,thick},
  lab/.style={midway,fill=white,inner sep=1pt}
]
  \node[node] (G1) at (0,0) {$SU(N)_1$};
  \node[node] (G2) at (4,0) {$SU(N)_2$};

  \draw[arr] (G1) to[bend left=40] node[lab,above] {$A_1$} (G2);
  \draw[arr] (G1) to[bend left=15] node[lab,above] {$A_2$} (G2);

  \draw[arr] (G2) to[bend left=40] node[lab,below] {$B_1$} (G1);
  \draw[arr] (G2) to[bend left=15] node[lab,below] {$B_2$} (G1);

\end{tikzpicture}
\caption{Klebanov-Witten quiver: $SU(N)_1\times SU(N)_2$ with bifundamentals $A_{1,2}:(N,\overline N)$ and $B_{1,2}:(\overline N,N)$.}
\label{fig:KW-quiver}
\end{figure}
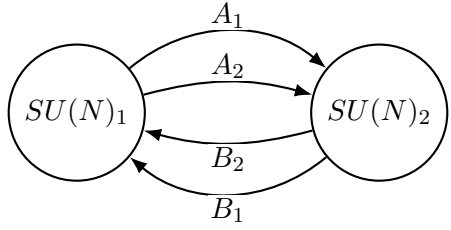

The quiver diagram for Klebanov-Witten theory is shown in Figure~\ref{fig:KW-quiver}.  The bifundamentals $A_1,A_2$ transform as an $SU(2)_A$ doublet, while $B_1,B_2$ transform as an $SU(2)_B$ doublet.  The quantum numbers of these fields are given in Table~\ref{tbl:KW-charges}.  The $T^{1,1}$ index is then built out of single letter indices for the adjoint $SU(N)\times SU(N)$ gauge bosons and the chiral bifundamentals.  The vector multiplet index is
\begin{equation}
    i_v(\mathfrak p,\mathfrak q)=\left(1-\fft{1-\mathfrak p\mathfrak q}{(1-\mathfrak p)(1-\mathfrak q)}\right)\chi_{\mathrm{adj}}(G), \label{eq: vecmult index}
\end{equation}
where $\chi_{\mathrm{adj}}(G)$ is the adjoint character of $SU(N)$.  For a chiral multiplet, we have
\begin{equation}
    i_c(\mathfrak p,\mathfrak q,G,F)=\fft{(\mathfrak p\mathfrak q)^{\fft12r}\chi_{\mathcal R}(G,F)-(\mathfrak p\mathfrak q)^{1-\fft12r}\chi_{\bar{\mathcal R}}(G,F)}{(1-\mathfrak p)(1-\mathfrak q)}, \label{eq:chiral mult index}
\end{equation}
where $\chi_{\mathcal R}(G,F)$ is the character in representation $\mathcal R$ of the combined gauge and flavor groups.  Combining everything together and plethystically exponentiating to get the full index gives
\begin{align}
    Z_N(\mathfrak p,\mathfrak q,v_i)&=\fft1{(N!)^2}(\mathfrak p;\mathfrak p)_\infty^{2(N-1)}(\mathfrak q;\mathfrak q)_\infty^{2(N-1)}\nn\\
    &\qquad\times\int\prod_i^{N-1}\fft{dz_i^{(1)}}{2\pi iz_i^{(1)}}\fft{dz_i^{(2)}}{2\pi iz_i^{(2)}}\fft{\prod_{i,j}^N\Gamma\Bigl(\fft{z_i^{(1)}}{z_j^{(2)}}v_1\Bigr)\Gamma\Bigl(\fft{z_i^{(2)}}{z_j^{(1)}}v_2\Bigr)\Gamma\Bigl(\fft{z_i^{(1)}}{z_j^{(2)}}v_3\Bigr)\Gamma\Bigl(\fft{z_i^{(2)}}{z_j^{(1)}}v_4\Bigr)}{\prod_{i\ne j}^N\Gamma\Bigl(\fft{z_i^{(1)}}{z_j^{(1)}}\Bigr)\Gamma\Bigl(\fft{z_i^{(2)}}{z_j^{(2)}}\Bigr)}.
\label{eq:T11sci}
\end{align}
The $\mathfrak q$-Pochhammer symbol is given by
\begin{equation}
    (z;\mathfrak q)_n=\prod_{k=0}^{n-1}(1-z\mathfrak q^k),
\end{equation}
and the elliptic Gamma function is defined by
\begin{equation}
    \Gamma(z;\mathfrak p,\mathfrak q)=\prod_{m,n\ge0}\fft{1-\mathfrak p^{m+1}\mathfrak q^{n+1}/z}{1-\mathfrak p^m\mathfrak q^nz}.
\end{equation}
Note that, for brevity, we have used the shorthand notation $\Gamma(z)\equiv\Gamma(z;\mathfrak p,\mathfrak q)$.

There is one important comment to make about the integrals over the gauge groups.  The gravity picture of a stack of D3-branes at the tip of the conifold yields a $U(N)\times U(N)$ theory.  However, this can be written as
\begin{equation}
    U(N)\times U(N)=SU(N)\times SU(N)\times U(1)_B\times U(1)_{\mathrm{diagonal}}.
\end{equation}
The diagonal $U(1)$ decouples, while the other $U(1)$ is the baryon symmetry.  If we were to integrate over $U(N)\times U(N)$ in (\ref{eq:T11sci}), we would then project into the $B=0$ sector.  In order to keep the baryon number free we must only integrate over $SU(N)\times SU(N)$, which corresponds to taking $\prod_i z_i^{(1)}=1$ and $\prod_i z_i^{(2)}=1$.

\begin{table}[t]
\centering
\begin{tabular}{l|rrrr|rrrr}
Field&$R_1$&$R_2$&$R_3$&$R_4$&$r$&$F_A$&$F_B$&$NB$\\
\hline
$A_1$&1&&&&$\fft12$&1&0&1\\
$B_1$&&1&&&$\fft12$&0&1&$-1$\\
$A_2$&&&1&&$\fft12$&$-1$&0&1\\
$B_2$&&&&1&$\fft12$&0&$-1$&$-1$
\end{tabular}
\caption{Charge assignments for the bifundamental fields.}
\label{tbl:KW-charges}
\end{table}

We now turn to the organization of the finite $N$ corrections. The leading contribution at large $N$ is given by
\begin{equation}
    Z_\infty(\mathfrak p,\mathfrak q;v_i)=\fft1{(v_1v_2;v_1v_2)_\infty(v_2v_3;v_2v_3)_\infty(v_3v_4;v_3v_4)_\infty(v_4v_1;v_4v_1)_\infty}.
\end{equation}
Holographically, this corresponds to the contribution of Kaluza-Klein gravitons to the index.  In addition to these KK gravitons, we must also include giant gravitons, which in this case would be D3-branes wrapped along supersymmetric three cycles inside $T^{1,1}$.  It was argued in Refs.~\cite{Arai:2019aou,Fujiwara:2023azx,Kim:2024ucf,Imamura:2021ytr}, based on toric geometry, that the giant graviton expansion for $T^{1,1}$ should take the form
\begin{equation}
    \fft{Z_N(\mathfrak p,\mathfrak q;v_i)}{Z_\infty(\mathfrak p,\mathfrak q;v_i)}=\sum_{n_i\ge0}v_1^{n_1N}v_2^{n_2N}v_3^{n_3N}v_4^{n_4N}\hat I_{(n_1,n_2,n_3,n_4)}^{GG}(\mathfrak p,\mathfrak q;v_i),
\end{equation}
where the giant graviton index, $\hat I_{(n_1,n_2,n_3,n_4)}^{GG}(\mathfrak p,\mathfrak q;v_i)$,  is independent of $N$ and can be interpreted as contributions from stacks of $\{n_I\}$  D3-branes wrapping a basis of four distinct 3-cycles in $T^{1,1}$.  Note that each term in this expansion carries baryon charge $B=n_1+n_3-n_2-n_4$.  The single giant graviton indices, $\hat I_{(1,0,0,0)}^{GG}(\mathfrak p,\mathfrak q;v_i)$ and $\hat I_{(0,0,1,0)}^{GG}(\mathfrak p,\mathfrak q;v_i)$ have $B=1$ while $\hat I_{(0,1,0,0)}^{GG}(\mathfrak p,\mathfrak q;v_i)$ and $\hat I_{(0,0,0,1)}^{GG}(\mathfrak p,\mathfrak q;v_i)$ have $B=-1$. As an example $\hat I_{(0,0,1,0)}^{GG}(\mathfrak p,\mathfrak q;v_i)$ can be obtained as the plethystic exponentiation of the single-letter index 

\begin{equation}
    f_{(0,0,1,0)} = 1 - \dfrac{(1- v_1/v_3)(1 - \mathfrak p)(1 - \mathfrak q)}{(1 - v_1v_2)(1 - v_1v_4)}.
\end{equation}
Our goal in the next section is to derive all single-letter giant graviton indices directly from the probe D3-brane action. 

\section{The giant graviton expansion on \texorpdfstring{$\mathrm{AdS}_5\times T^{1,1}$}{AdS5xT11}}\label{sec:T11-giants}

In this section, we obtain the single giant contribution to the superconformal index by quantizing the fluctuations of the probe D3 branes.  The interest in such branes in AdS$_5\times T^{1,1}$ dates back to the original works using the language of holomorphic sections \cite{Mikhailov:2000ya,Beasley:2002xv}. A more explicit analysis of their embeddings and scalar fluctuations was  carried out in \cite{Hamilton:2010sv, Canoura:2005uz} while a recent study considered a protected subsector and its non-Abelian generalization  \cite{GonzalezLezcano:2026wje}.

As in the case of AdS$_5\times S^5$ giants, we will focus on maximal giants. In $T^{1,1}$, however, the relevant branes wrap nontrivial three cycles rather than contractible $S^3$'s. To identify these three cycles, we follow the construction of \cite{Berenstein:2002ke,Hamilton:2010sv}. The conifold can be described by complex coordinates $z^i$ satisfying $z^1z^2=z^3z^4$.  For $T^{1,1}$ on the base, we can take
\begin{align}
    z^1&=\sin\ft{\theta_1}2\sin\ft{\theta_2}2e^{\fft12i(\psi-\phi_1-\phi_2)},&z^2&=\cos\ft{\theta_1}2\cos\ft{\theta_2}2e^{\fft12i(\psi+\phi_1+\phi_2)},\nn\\
    z^3&=\cos\ft{\theta_1}2\sin\ft{\theta_2}2e^{\fft12i(\psi+\phi_1-\phi_2)},&z^4&=\sin\ft{\theta_1}2\cos\ft{\theta_2}2e^{\fft12i(\psi-\phi_1+\phi_2)}.
\end{align}
Maximal giants correspond to $z^i=0$ for fixed $i$.  For example, $z^1=0$ is solved by either $\theta_1=0$ or $\theta_2=0$.  As argued in \cite{Hamilton:2010sv}, this is the union of two D3-branes wrapped on different two-spheres and the $U(1)$ fiber.  Pulling this apart, we find four distinct maximal single giants, corresponding to
\begin{align}
    \theta_1=0:&\quad\mbox{wrapping }(t,\theta_2,\phi_2,\tilde \psi\equiv\psi+\phi_1)\mbox{ and moving around }\phi_1,\nn\\
    \theta_1=\pi:&\quad\mbox{wrapping }(t,\theta_2,\phi_2,\tilde \psi\equiv\psi-\phi_1)\mbox{ and moving around }-\phi_1,\nn\\
    \theta_2=0:&\quad\mbox{wrapping }(t,\theta_1,\phi_1,\tilde \psi\equiv\psi+\phi_2)\mbox{ and moving around }\phi_2,\nn\\
    \theta_2=\pi:&\quad\mbox{wrapping }(t,\theta_1,\phi_1,\tilde \psi\equiv\psi-\phi_2)\mbox{ and moving around }-\phi_2.
\end{align}
Here ``motion'' along $\phi_i$ should be understood as rotation of the local normal plane at the corresponding pole. The combinations $\psi \pm \phi_i$ remain regular on the brane and provide the appropriate fiber coordinates.

In the coordinate conventions used here, the two mesonic charges and the R-charge act as
\begin{align}
    F_A = -2i\partial_{\phi_1}, \quad F_B = -2i\partial_{\phi_2}, \quad r = -2i\partial_{\psi}. \label{eq: T11 charges}
\end{align}
These normalizations agree with the charge assignments of the elementary bifundamental fields summarized in Table \ref{tbl:KW-charges}.

As each one of these four types of giants can be treated similarly, we will focus on the $\theta_1=0$ giant and its contribution to $\hat I^{\mathrm{GG}}_{(0,0,1,0)}$.  The spatial world volume can be parametrized by
\begin{equation}
    ds^2_3 = \dfrac{2}{3} \left[\dfrac{1}{4} \left(d\theta_2^2 + \sin^2\theta_2\, d\phi_2^2 + \dfrac{2}{3}(d\tilde \psi + \cos\theta_2\, d\phi_2)^2\right)\right]. \label{eq: t11 squahed s3}
\end{equation}
Two features distinguish this from the maximal giants in $S^5$. One, its worldvolume is a squashed sphere instead of a round $S^3$. Two, the three cycle that it wraps has a non trivially fibered normal bundle. The squashing modifies the harmonic spectra, while the normal-bundle connection produces additional couplings for the internal scalar and fermionic fluctuations.

\subsection{Scalar fluctuations}\label{sec:scalars}
The bosonic dynamics of a probe D3-brane is governed by the DBI and Wess–Zumino actions. In the absence of a worldvolume gauge field, the action is
\begin{equation}
    S_{B} = - T_3 \int d^4 \xi \sqrt{-\det P[g]} \,+ T_3 \int P[A_{4}], \label{eq: DBI Action}
\end{equation}
where $P$ is the pullback to the worldvolume of the giant graviton. We choose static gauge, with worldvolume coordinates $(t,\theta_2,\phi_2, \tilde\psi)$, and expand about the maximal giant localized at the center of $\mathrm{AdS}_5$ and at $\theta_1=0$ in $T^{1,1}$.

We will consider small fluctuations of the brane and expand the action to quadratic order in these fluctuations, $S_{DBI} = S_{(0)} + S_{(2)}$. The classical contributions to the index $v_i^N$ come from the leading classical piece,
\begin{equation}
    S_{(0)} = - \dfrac{T_3 L^4}{18} \int d^4\xi\, \sin\theta_2 \left( 1 + \dfrac{4}{3} \dot \phi_1\right),
\end{equation}
where $T_3$ is the tension given by Eq.~\eqref{eq: tension}.

The D3-brane has six transverse bosonic fluctuations. Four, denoted $y_k$ with $k=1,\ldots,4$, describe motion away from the center of $\mathrm{AdS}_5$. The remaining two describe motion in the plane normal to the wrapped cycle inside $T^{1,1}$. Because the spherical coordinates $(\theta_1,\phi_1)$ are singular at $\theta_1=0$, it is convenient to introduce local Cartesian coordinates on the plane
\begin{equation}
    y_5 = \delta\theta_1 \cos\phi_1, \qquad y_6 =\delta\theta_1 \sin\phi_1,
\label{eq:y5y6}
\end{equation}
which can be assembled into the complex coordinates
\begin{equation}
    y^{(\pm)} = y_5 \mp i y_6.
\end{equation}
Each $y_i$, $i=1,\ldots,6$, is a function of the worldvolume coordinates $\xi^\alpha=(t,\theta_2,\phi_2,\widetilde\psi)$. Expanding in these fluctuations, the quadratic part of the action takes the form,
\begin{align}
        S_{(2)} = - \dfrac{3N}{128\pi^2} \int_{\Sigma} d^4\xi \sin\theta_2\Bigl[ &\sum_k \left(y_k^2 - \dot y_k^2 -y_k\nabla^2 y_k \right) + \dfrac{1}{6} \sum_{i=5,6} \left(- \dot y_i^2 - y_i\nabla^2 y_i\right)\nn\\
    &+ \dfrac{2}{3} \left(y_6 \dot y_5 - y_5 \dot y_6\right) + \left(y_6 \partial_{\tilde \psi} y_5 - y_5\partial_{\tilde \psi} y_6\right) \Bigr],
\label{eq:S2scalar}
\end{align}
where $\nabla^2$ is the scalar Laplacian on the squashed sphere with metric \eqref{eq: t11 squahed s3}. Notice that the internal ($y_5$ and $y_6$) fluctuations are coupled to each other through two first order terms. The term involving the time derivative comes from the WZ action and is similar to the case of $S^5$ giants. The terms involving $\partial_{\tilde\psi}$, however, are  absent for the $S^5$ case, and arise because the fiber is wrapped nontrivially.

To quantize the scalar fluctuations, we expand them in eigenfunctions of the scalar Laplacian on the squashed three sphere wrapped by the giant. The relevant harmonic analysis is reviewed in Appendix~\ref{app:lens}. Comparing \eqref{eq: t11 squahed s3} with \eqref{eq:squashed}, we identify the squashing parameter to be $e^{2\lambda} = \frac{2}{3}$ and $(\phi,\psi) = (\phi_2,\tilde \psi)$ with an overall volume scaling of $\frac{2}{3}$. Let $\Phi_{lmn}$ denote the scalar harmonics satisfying $\nabla^2 \Phi_{lmn} = - \Lambda_{lmn} \Phi_{lmn}$. Then in terms of $z = \sin^2(\frac{\theta_2}{2})$
\begin{equation}
    \Phi(z,\phi_2,\tilde \psi)_{l m n} = z^{\frac{1}{2}|m-n|}(1-z)^{\frac{1}{2}|m+n|} {}_2F_1(a,b;c;z)\; e^{in\tilde \psi} e^{im\phi_2},
\end{equation}
where $a,b,c$ are the parameters of the hypergeometric function given by \eqref{eq: hypergeom params} and $m,n$ satisfy the quantization conditions,
\begin{equation}
    m + n \in \mathbb Z \quad \text{and} \quad n \in \frac{1}{2} \mathbb Z.
\end{equation}
Taking into account the volume scaling, the spectrum is given by
\begin{equation}
    \dfrac{2}{3}\Lambda_{l mn} = 4l (l + 1) + 2 n^2, \quad l = j + \max(|m|,|n|), \quad j = 0,1,2,\ldots. \label{eq: t11 scalar eigenvalue}
\end{equation}
We expand the four AdS$_5$ scalar fluctuations and the two complex internal fluctuations in this basis
\begin{align}
    y^{(k)} &= \sum_{l m n} y^{(k)}_{l m n} \Phi_{l m n}, \quad k = 1,\ldots,4,\nn\\
    y^{(\pm)} &= \sum_{l m n} y^{(\pm)}_{l m n} \Phi_{l m n}, \quad y^{(\pm)} = y_5 \mp i y_6.
\end{align}
Substituting the harmonic expansions into the quadratic action and using orthonormality reduces the scalar sector to a collection of decoupled quantum-mechanical modes
\begin{align}
    S_{2} = \dfrac{3N}{64\pi^2} \int_{\Sigma} dt &\sum_k \sum_{l m n}\left[ \dfrac{1}{2} \dot y^{(k)}_{l m n} \dot y^{(k)}_{l -m -n} -\dfrac{1}{2} (\Lambda_{l mn}+ 1)\;y^{(k)}_{l m n} y^{(k)}_{l -m -n}\right]\nn\\
    &+ \dfrac{1}{12} \sum_{lmn} \Bigl[ \dot y^{(+)}_{l m n} \dot y^{(-)}_{l -m -n} -2i (y^{(+)}_{l m n} \dot y^{(-)}_{l -m -n} - y^{(-)}_{l m n} \dot y^{(+)}_{l -m -n}) \nn\\
    &\kern4em-(\Lambda_{l mn} + 6n)\; y^{(+)}_{l m n} y^{(-)}_{l -m -n}\Bigr].
\label{eq:T11sact2}
\end{align}
To simplify the subsequent canonical quantization, we first rescale the fields, $y^{(k)} \to 8\pi y^{(k)}/\sqrt{3N}$ and $y^{(\pm)} \to 16\pi y^{(\pm)}/\sqrt{N}$. Then the conjugate momenta are,
\begin{align}
    p^{(k)}_{l m n} &= \dot y^{(k)}_{l -m -n},\nn\\
    p^{(\pm)}_{l m n} &= \dot y^{(\mp)}_{l - m -n} \pm 2i\, y^{(\mp)}_{l - m -n}.
\end{align}
This allows one to write the Hamiltonian for the fluctuations as
\begin{align}
    H =& \sum_{l mn} \sum_k\left[ \dfrac{1}{2} p^{(k)}_{l m n}  p^{(k)}_{l -m -n} +\dfrac{1}{2} (\Lambda_{l mn}+1) y^{(k)}_{l mn}y^{(k)}_{l-m-n} \right]\nn\\
    &+\sum_{l mn} \biggl[ p^{(+)}_{l mn} p^{(-)}_{l -m-n} + 2i\, (y^{(+)}_{l m n} p^{(+)}_{l m n} - y^{(-)}_{l m n} p^{(-)}_{l m n}) \nn\\
    &\kern3em+ \left(\Lambda_{l m n} + 6n + 4\right)\, y^{(+)}_{l m n} y^{(-)}_{l-m-n}\biggr].
\end{align}
We now define the raising and lowering operators
\begin{align}
    y^{k}_{l m n} &= \dfrac{1}{\sqrt{2\sqrt{\Lambda_{l mn} + 1}}} \,(a^{(k)}_{l m n} + a^{(k)}{}_{l -m -n}^\dagger),\nn \\
    p^{k}_{l m n} &= -i\sqrt{\dfrac{\sqrt{\Lambda_{l mn} + 1}}{2}} \,(a^{(k)}_{l -m -n} - a^{(k)}{}_{l m n}^\dagger),\nn \\
    y^{(+)}_{l m n} &= \dfrac{1}{\sqrt{2\sqrt{(\Lambda_{l mn} + 6n + 4)}}}\, (b_{l m n} + c^\dagger_{l -m -n}) = y^{(-)}{}^\dagger_{l -m -n },\nn\\
    p^{(+)}_{l m n} &= i \sqrt{\dfrac{\sqrt{\Lambda_{l mn} + 6n + 4}}{2}}\, (b^\dagger_{l m n} - c_{l -m -n}) = p^{(-)}{}^\dagger_{l -m -n }.
\end{align}
For convenience, one could go to the helicity basis and define,
\begin{align}
    a^{(1\pm)}_{l m n} = \dfrac{1}{\sqrt{2}} (a^{(1)}_{l m n} \pm i a^{(2)}_{l mn}),\quad
    a^{(2\pm)}_{l m n} = \dfrac{1}{\sqrt{2}} (a^{(3)}_{l m n} \pm i a^{(4)}_{l mn}). \label{eq: helicity basis}
\end{align}
Then the Hamiltonian takes a more suggestive form,
\begin{align}
        H &= \sum_{k} \sum_{l m n} \sqrt{\Lambda_{l mn} + 1}\; (N^{(k+)}_{l m n} + N^{(k-)}_{l m n})\nn\\
    &+\sum_{l m n} \Bigl[\left(\sqrt{\Lambda_{l mn}+6n + 4}-2\right) N^+_{l m n}+ \left(\sqrt{\Lambda_{l mn}-6n + 4}+2\right)N^-_{l m n}\Bigr],
\end{align}
where $N^x_{lmn}$ are the corresponding number operators and now $k=1,2$.

\subsubsection*{BPS modes and their contribution to the index}
In terms of the gravitational charges, the supersymmetric Hamiltonian for $\mathcal{N} = 1$ theories takes the form,
\begin{equation}
    \mathcal{H} = H - J_1 - J_2 - \dfrac{3}{2}r \label{eq: susyH}.
\end{equation}
These charges can be written in terms of the number operators in the helicity basis. For the AdS angular momenta we have
\begin{equation}
    J_i = \sum_{l m n} N^{(i+)}_{l m n} - N^{(i-)}_{l m n}.
\end{equation}
From \eqref{eq: T11 charges}, the $r$ charge and the flavor charges are respectively,
\begin{align}
    r &= 2 \sum_{l m n} n \Bigl[\sum_i N^{(i+)}_{l m n} + N^{(i-)}_{l m n} + N^+_{l m n} + N^-_{l m n}\Bigr],\nn\\
        F^A &= 2 \sum_{l m n}  \Bigl[(n+1)N^{+}_{l m n}+ (n-1) N^{-}_{l m n} + n \sum_i N^{(i+)}_{l m n} + N^{(i-)}_{l m n}\Bigr],\nn\\
    F^B &= 2 \sum_{l m n} m \Bigl[\sum_i N^{(i+)}_{l m n} + N^{(i-)}_{l m n}+ N^+_{l m n} + N^-_{l m n}\Bigr].
\end{align}
Furthermore, these excitations do not change the topology of the giants and hence do not carry any additional Baryon charge.

The supersymmetric Hamiltonian can then be written as
\begin{align}
    \mathcal{H}& = \sum_{lmn} \Bigl[\left(\sqrt{\Lambda_{lmn} + 1} \,-(3n+1)\right)(N^{(1+)}_{lmn} + N^{(2+)}_{lmn}) + \left(\sqrt{\Lambda_{lmn} + 6n +4} - (3n+2)\right)N^+_{lmn}\nn\\
    & \quad+\left(\sqrt{\Lambda_{lmn} + 1} \,-(3n-1)\right)(N^{(1-)}_{lmn} + N^{(2-)}_{lmn}) + \left(\sqrt{\Lambda_{lmn} - 6n +4} -(3n-2)\right)N^-_{lmn} \Bigr],
\label{eq: susy H t11 scalar}
\end{align}
where we have dropped the zero point energy.
Using  \eqref{eq: t11 scalar eigenvalue}, it is useful to rewrite
\begin{align}
    \Lambda_{lmn} + 1 & = \fft3{2} \left[\left(2l  +1\right)^2 - \left(2|n| + 1\right)^2\right] + (3|n|+1)^2 ,\nn\\
    \Lambda_{lmn} + 6n + 4 &= \fft3{2} \left[\left(2l  +1\right)^2 - \left(2n + 1\right)^2\right] + (3n+2)^2.
\label{eq: t11 scalar lambda useful}
\end{align}
BPS states saturate the bound $\mathcal{H} \geq 0$. Since $l \geq |n|$, the bracket in the first line of \eqref{eq: t11 scalar lambda useful} is non-negative and vanishes only when $l = |n|$, while the bracket in the second line vanishes only when $l = n \geq 0$ (for $n<0$ it is strictly positive). It follows that among the AdS scalars the positive-helicity modes $y_{(1+)}$ and $y_{(2+)}$ saturate the BPS bound for $n \geq 0$ and $l = n$. Among the internal modes, $y_{(+)}$ saturates the bound under the same conditions. The remaining modes have strictly positive $\mathcal{H}$. The quantum numbers of the states with $l = |n|$ are summarized in Table~\ref{tab: D3 scalars T11}.

Notice that the condition $l = |n|$ forces the radial excitation number to vanish and restricts $|m| \leq |n|$. Accordingly, the residual hypergeometric function is constant. These are the lowest radial harmonics at fixed angular charges, although they are not generally constant functions on the wrapped three-sphere. This agrees with the intuition that BPS states are the ground states of the system. This is a principle that we will see again and again in other cases and we expect it to be quite universal in the whole class of AdS$_5\times SE_5$.

\begin{table}[t]
    \centering
    \begin{tabular}{c c c c c c c c c }
        \toprule
        {}  & $H$ & $J_1$ & $J_2$ & $r$& $F_A$& $F_B$ & BPS?  \\
        \midrule
        $y_{(1+)\ell m n}$ & $3|n|+1$ & $1$ & $0$ & $2n$ & $2n$ &$2m$ & \checkmark\\
        $y_{(2+)\ell m n}$ & $3|n|+1$  & $0$ & $1$ & $2n$ & $2n$ &$2m$ & \checkmark\\
        $y_{(1-)\ell m n}$ & $3|n|+1$  & $-1$ & $0$ & $2n$ & $2n$ &$2m$ & {}\\
        $y_{(2-)\ell m n}$ & $3|n|+1$ & $0$ & $-1$ & $2n$ & $2n$ &$2m$ & {}\\
        \midrule
        $y_{(+)\ell m n}$ & $3n$ &   $0$ & $0$ & $2n$ & $2n+2$ &$2m$ & \checkmark\\
        $y_{(-)\ell m n}$ & $\sqrt{9n^2+4}+2$  & $0$ & $0$ & $2n$ & $2n-2$ &$2m$ & {}\\
        \bottomrule
    \end{tabular}
    \caption{Quantum numbers of the transverse scalar fluctuations of giant graviton on $T^{1,1}$. We highlight the BPS states by restricting to the sector with $l = |n|$. Check marks indicate that the modes satisfy $\mathcal{H} = 0$.}
    \label{tab: D3 scalars T11}
\end{table}

Having identified the BPS excitations in the scalar sector, we can now construct the single-letter index,
\begin{equation}
    f = \Tr[(-1)^F e^{-\beta \mathcal{H}} \mathfrak p^{J_1} \mathfrak q^{J_2} \rho^{3r/2}u^{F_A}v^{F_B} \zeta^{B}].
\end{equation}
From Table \ref{tab: D3 scalars T11}, it follows that,
\begin{align}
    f_{\mathrm{scalar}} = (\mathfrak p + \mathfrak q + u^{2}) \sum_{n\ge0}\sum_{|m|\leq n} \rho^{3n} u^{2n} v^{2m}.
\end{align}
We can rewrite this in terms of the $v_i$ fugacities using \eqref{eq: uv to vi fugacity t11}. It is useful to introduce the integer quantum numbers
\begin{equation}
    x = m + n, \quad y = n - m, \label{eq: T11xy}
\end{equation}
so that the summation is over $x,y \geq 0$. Then the single letter index takes the form,
 \begin{align}
     f_{\mathrm{scalar}}& = (\mathfrak p + \mathfrak q + v_1/v_3) \sum_{x,y \geq 0} (v_1v_2)^x (v_1 v_4)^y = \dfrac{\mathfrak p + \mathfrak q + v_1/v_3}{(1 - v_1 v_2) (1 - v_1 v_4)}. \label{eq: scalar sl t11}
 \end{align}
The scalar BPS spectrum consists of two positive-helicity AdS$_5$ modes and one internal mode. All lie in the lowest radial sector $l = n$, with $n\ge0$ and $|m| \le n$.

\subsection{Vector fluctuations} \label{sec:vectors}

We next consider fluctuations of the Abelian gauge field on the D3-brane worldvolume. Expanding the DBI action to quadratic order in the field strength gives Maxwell theory on the induced worldvolume geometry,
\begin{equation}
    S = -\dfrac{1}{4} \int d^4\xi\;\sqrt{-g}\, F_{\mu\nu}^2.
\end{equation}
We follow a treatment similar to \cite{Deddo:2025lfm} by quantizing the Abelian gauge field in the Coulomb gauge,
\begin{equation}
    A^t = 0, \quad \nabla\cdot \vec A = 0.
\end{equation}
The gauge fixed action then reduces to 
\begin{equation}
    S = \int d^4\xi \, \sqrt{-g}\, \left(\dfrac{1}{2} |\dot {\vec{ A}}|^2 + \dfrac{1}{2} A_i( \gamma^{ij} \nabla^2 - R^{ij})A_j\right).\label{eq: t11 vect action}
\end{equation}
The harmonic analysis of transverse one forms on a squashed sphere is discussed in detail in Appendix~\ref{app:lens}.  Recall that the squashing factor is $e^{2\lambda} = \frac{2}{3}$ and there is an overall volume factor of $\frac{2}{3}$. 
The two independent polarizations transform in the $l$ of $SU(2)_l$ with appropriate $U(1)_r$ quantum numbers. We expand
\begin{equation}
    A^{(\pm)} = \sum_{l m n} A^{(\pm)}_{l m n} X^{(\pm)}_{l mn}.
\end{equation}
The spectrum of these excitations, after some manipulations, can be written as
\begin{align}
    \Lambda^{(\pm)} & = \fft32 \left[ \left(2l + 1\right)^2 -  \left(2|n| - 1\right)^2 \right] + (3|n| - 1)^2+ 1 \nn\\
    &\quad\mp2 \sqrt{\fft32 \left[ \left(2l + 1\right)^2 -  \left(2|n| - 1\right)^2 \right] + (3|n| - 1)^2}, \nn\\
    \text{where}& \quad |n| \leq l \mp 1, \quad |m| \leq l, \quad l = j  + |m|, \quad j = 0,1,2,\ldots.
    \label{eq: t11 vect eigen useful}
\end{align}
Here the  `$(\pm)$' label corresponds to the two polarizations. Notice that the bounds of the quantum number $n$ depend on the type of polarization. However, for both cases, $m$ and $n$ follow the same quantization conditions dictated by the regularity of the wave functions on the squashed sphere,
\begin{equation}
    m + n \in \mathbb Z, \quad n \in \frac{1}{2}\mathbb Z.
\end{equation}
Using orthonormality of the vector harmonics, the Maxwell action reduces to a collection of harmonic oscillators,
\begin{equation}
    S = \dfrac{1}{2} \int dt \sum_{l m n}' \sum_{k = \pm} (\dot A^{(k)}_{l mn} \dot A^{(k)}_{l -m -n} - \Lambda^{(k)}_{l m n}A^{(k)}_{l mn}A^{(k)}_{l -m-n}),
\end{equation}
where $\displaystyle \sum'$ denotes that the summation range for $n$ is different for $(\pm)$ states, and where we have integrated over the spatial squashed sphere. One can follow the standard quantization procedure and write the Hamiltonian as
\begin{equation}
    H = \sum_{l mn}'\left(\sqrt{\Lambda^{(+)}_{l mn}} \, n^{(+)}_{l mn}+ \sqrt{\Lambda^{(-)}_{l mn}} \, n^{(-)}_{l mn}\right),
\end{equation}
where $n^{(\pm)}_{lmn}$ are the number operators corresponding to the respective polarizations.

\subsubsection*{BPS modes and their contribution to the index}
We now proceed to write other charges in terms of these number operators to identify the modes that saturate the BPS bound $\mathcal{H} \geq 0$ with $\mathcal H$ given in \eqref{eq: susyH}. 
These longitudinal modes do not carry any AdS angular momentum so $J_1 = J_2 = 0$, but carry associated flavor and $R$ charges,
\begin{align}
    r = F_A = \sum_{l mn}' 2n (n^{(+)}_{l mn} + n^{(-)}_{l mn}), \quad F_B = \sum_{l mn}' 2m (n^{(+)}_{l mn} + n^{(-)}_{l mn}).
\end{align}
Notice that for a given state to be BPS, $\mathcal H = 0$, we want  $H = 3n$. It therefore requires $n \geq 0$. The $+$ tower obeys $|n| \leq l - 1$ and cannot reach the extremal state needed to saturate the bound. For the $-$ tower, there are two distinguished end points, $|n| = l$ or $|n| = l+1$.  From \eqref{eq: t11 vect eigen useful}, it follows that
\begin{equation}
    \sqrt{\Lambda^{(-)}_{lm(|n| = l)}} = 3|n| + 2, \quad \text{ or } \quad\sqrt{\Lambda^{(-)}_{lm(|n| = l+1)}}=3|n|.
\end{equation}
Thus only the $-$ polarization modes with $n = l + 1 > 0$ are BPS. The quantum numbers of these states are summarized in Table \ref{tab: D3 vectors T11}.
It is then straightforward to write the contribution to the single letter index from these vector modes
\begin{equation}
    f_{\mathrm{vector}} = \sum_{n\ge1} \sum_{|m| \leq n - 1} \rho^{3n} u^{2n}v^{2m}.
\end{equation}
Once again, this can be re-expressed in terms of the $v_i$ fugacities and integer quantum numbers $x$, $y$ given in \eqref{eq: T11xy} as
\begin{equation}
    f_{\mathrm{vector}} = \sum_{x,y\geq 1}(v_1v_2)^x(v_1v_4)^y=\fft{(v_1v_2)(v_1v_4)}{(1-v_1v_2)(1-v_1v_4)}
    =\fft{\mathfrak p \mathfrak q \,(v_1/v_3)}{(1-v_1v_2)(1-v_1v_4)}. \label{eq: vector sl t11}
\end{equation}

\begin{table}[t]
    \centering
    \begin{tabular}{c c c c c c c c c }
        \toprule
        {}  & $H $ & $J_1$ & $J_2$ & $r$& $F_A$& $F_B$ & BPS?  \\
        \midrule
        $A^{(-)}_{\ell m (|n| = l)}$ & $3|n|+2$ & $0$ & $0$ & $2n$ & $2n$ &$2m$ & \\
        $A^{(-)}_{\ell m (|n| = l+1)}$ & $3|n|$  & $0$ & $0$ & $2n$ & $2n$ &$2m$ & \checkmark\\
        \bottomrule
    \end{tabular}
    \caption{Quantum numbers of the longitudinal fluctuations (vector modes) of the giant graviton on $T^{1,1}$. We highlight the BPS states by restricting to the sector with $l = |n|$ and $l+1 = |n|$. As discussed in the main text, the $(+)$ states do not exist in this sector.}
    \label{tab: D3 vectors T11}
\end{table}


\subsection{Fermion fluctuations}
\label{sec:fermions}

The final contribution to the index comes from fermionic fluctuations on the D3-brane worldvolume.  These fluctuations were analyzed in Ref.~\cite{Deddo:2025lfm} in the context of AdS$_5\times S^5$, and we follow the same procedure here.  The starting point is the fermionic part of the D3-brane action of \cite{Martucci:2003gc,Marolf:2003ye,Marolf:2003vf,Marolf:2004jb},
\begin{equation}
    S=\fft{iT}2\int d^4\xi\, e^{-\phi}\sqrt{-(g+F)}\;\bar\theta\,(1-\tilde\Gamma_{D3})\left(\Gamma^i\hat D_i-\hat\Delta+L_3\right)\theta,
\label{eq:D3fermionact}
\end{equation}
where $\theta=(\theta_1,\theta_2)$ is a pair of IIB spinors.  Here, $L_3$ is built from the worldvolume gauge field $F$, and $\hat\Delta$ is built from the axidilaton and complex three-form of IIB theory \cite{Marolf:2004jb}.  Following the conventions of \cite{Deddo:2025lfm}, where the D3-brane worldvolume directions are labeled by $0,1,2,3$, the covariant derivative and $\kappa$-symmetry projector are given explicitly by
\begin{align}
    \hat D_m&=\nabla_m+\fft18\fft1{2\cdot5!}F_5\cdot\Gamma\,\Gamma_m(i\sigma^2),\nn\\
    \tilde\Gamma_{D3}&=\Gamma_{0123}(-i\sigma^2),
\end{align}
where the Pauli matrices act on the pair of IIB spinors.

Since the AdS$_5\times T^{1,1}$ background only incorporates the metric and self-dual five-form, we set the dilaton, $\phi$, and $\hat\Delta$ to zero in the action, (\ref{eq:D3fermionact}).  Furthermore, since we are only interested in the spectrum of linearized fluctuations, we set the worldvolume gauge field $F=0$.  Noting that the self-dual five-form has the universal structure
\begin{equation}
    F_5=\fft4L(1+*_{10})\epsilon_{\mathrm{AdS}_5},
\end{equation}
the fermionic action reduces to
\begin{equation}
    S=iT\int d^4\xi\sqrt{-g}\;\bar\theta\fft{1-\tilde\Gamma_{D3}}2\Gamma^i\hat D_i\theta,
\end{equation}
where the Dirac operator takes the form
\begin{equation}
    \Gamma^i\hat D_i=\Gamma^i\nabla_i-\fft1L\Gamma^{12389}(i\sigma^2).
\end{equation}
Here we have used the IIB Weyl condition, $\Gamma^{11}\theta=-\theta$ in the conventions of Ref.~\cite{Deddo:2025lfm}.  Note that the AdS$_5$ directions are labeled $0,4,5,6,7$ and the internal directions are $1,2,3,8,9$.  Finally, incorporating the $\kappa$-symmetry projection $\tilde\Gamma_{D3}\theta=\theta$ brings the fermionic action to the form
\begin{equation}
    S=iT\int dt\,d^3\xi\,L^3\sqrt h\;\bar\theta_1\left(\Gamma^0\partial_t+\slashed\nabla+\fft1L\Gamma^{089}\right)\theta_1.
\end{equation}
Here $0,1,2,3$ are the world-volume tangent-space indices, and $8,9$ are the two directions normal to the brane inside $T^{1,1}$, namely the $(y_5,y_6)$ plane, (\ref{eq:y5y6}).  Decomposing the spinor $\theta_1$ into four-dimensional Weyl spinors, following the conventions of \cite{Deddo:2025lfm}, gives
\begin{equation}
    S=\fft{iT}L\sum_p\int dt\,d^3\xi\,L^3\sqrt h\;\psi^\dagger_p\left(-\partial_t+\fsl\nabla+2iQ\right)\psi_p+\mathrm{c.c.},
    \qquad Q=\fft i2\Gamma^{89}, \label{eq: fermion action}
\end{equation}
so that $Q=\pm\ft12$ measures the charge under rotations of the normal plane.

So far, the analysis is identical to that of Ref.~\cite{Deddo:2025lfm}.  However, two features distinguish this from the round $S^5$ case.  First, $\slashed\nabla$ is now the Dirac operator on the \emph{squashed} three-sphere, whose spectrum is obtained in Appendix~\ref{app:spinor}.  Second, and more importantly, the normal bundle of the giant is non-trivially fibered over its world-volume.  To see this, take the bosonic \textit{coordinates} $\theta_1$ and $\phi_1$ and combine them into the complex normal coordinate $W=(\theta_1/\sqrt6)e^{i\phi_1}$ near the North pole, $\theta_1=0$.  Since $\cos\theta_1\,d\phi_1=d\phi_1-\ft12\theta_1^2d\phi_1+\cdots$, the $T^{1,1}$ metric becomes, to quadratic order in the fluctuation,
\begin{equation}
    ds^2=\fft16\left(d\theta_2^2+\sin^2\theta_2\,d\phi_2^2\right)+|dW|^2+\fft19\Bigl(\hat e^3-3\,\mathrm{Im}(\bar W dW)\Bigr)^2,
    \qquad \hat e^3\equiv d\tilde \psi+\cos\theta_2\,d\phi_2.
\end{equation}
The cross term $-\ft23\,\hat e^3\,\mathrm{Im}(\bar WdW)$ is precisely the origin of the coupling $\left(y_6\partial_{\tilde \psi} y_5-y_5\partial_{\tilde\psi} y_6\right)$ in the scalar action \eqref{eq:S2scalar}: transporting the normal frame along the fiber rotates it.  Equivalently, the normal bundle carries the connection $\mathcal A=\ft13\hat e^3$, which is the unit-normalized fiber one-form of $T^{1,1}$.  For the internal scalar fluctuations $y^{(\pm)}$, whose normal-plane charge is $\pm1$, this is what produced the shift $\Lambda\to\Lambda+6n$ in (\ref{eq:T11sact2}).  For the spinors, whose charge is $Q=\pm\ft12$, the same connection enters the world-volume Dirac operator through the normal-bundle piece $\omega^{89}\propto\hat e^3$ of the spin connection, and is captured by the spectrum (\ref{eq:diractwist}) with charge
\begin{equation}
    q=\ft13Q.
\end{equation}

Putting this together, and using $e^{2\lambda}=\ft23$ together with the overall volume factor of $\ft23$ in \eqref{eq: t11 squahed s3}, the world-volume Hamiltonian for the fermionic fluctuations is
\begin{equation}
    H=\ft12\pm\sqrt{6j(j+1)+3n^2+6n Q+\ft74}\;-2Q,
\label{eq:fermH}
\end{equation}
where the two signs correspond to the two towers of spinor harmonics labeled under $SU(2)_l\times SU(2)_r$ with $SU(2)_r$ broken to $U(1)_r$ by the squashing as
\begin{align}
    \mbox{`$+$':}&\quad(j,j+\ft12),&\mbox{`$-$':}&\quad(j,j-\ft12),\nn\\
    &\quad|m|\le j,~|n|\le j+\ft12&&\quad|m|\le j,~|n|\le j-\ft12
\end{align}
where $j$ is half-integer quantized.
The associated charges follow from \eqref{eq: T11 charges}.  Since $\partial_\psi$ does not rotate the normal plane while $\partial_{\phi_1}$ does, we have
\begin{equation}
    r=2n,\qquad F_A=2(n+Q),\qquad F_B=2m,
\end{equation}
together with $J_1,J_2=\pm\ft12$ from the spinor representation in AdS$_5$.

\subsubsection*{BPS modes and their contribution to the index}

Exactly as in the scalar and vector sectors, a supersymmetric state satisfying (\ref{eq: susyH}) requires the energy to be half-integer quantized.  This can only happen when we have the square-root of a perfect square in \eqref{eq:fermH}.  The quantum numbers for which this generically happens are shown in Table~\ref{tbl:perfectsq}.  To identify the BPS states, we first note that, since the fermions originated as ten-dimensional IIB fermions, the spinor weights $J_1,J_2,Q=\pm\fft12$ are correlated in that there are an odd number of minus signs on the `$+$' branch.  In addition, the `$-$' branch modes have negative $H$; these are the filled Dirac-sea states, and the corresponding physical excitations are the antiparticles, obtained by reversing all charges.  In this case, the spins $J_1,J_2,Q$ have an even number of minus signs.  Comparing the half-integer quantized tower of states in Table~\ref{tbl:perfectsq} with the supersymmetric Hamiltonian, (\ref{eq: susyH}), we find four BPS towers of fermionic excitations as listed in Table~\ref{tab: D3 fermions T11}.

\begin{table}[t]
\centering
\begin{tabular}{r|r|c|c|c}
$n\kern1.2em$&$Q$&$\sqrt{6j(j+1)+3n^2+6nQ+\fft74}$&$H_+$&$H_-$\\
\hline
$j+\fft12\phantom{)}$&$\fft12$&$3j+2$&$3j+\fft32$&\\
$-(j+\fft12)$&$-\fft12$&$3j+2$&$3j+\fft72$&\\
\hline
$j+\fft12\phantom{)}$&$-\fft12$&$3j+1$&$3j+\fft52$\\
$j-\fft12\phantom{)}$&$\fft12$&$3j+1$&$3j+\fft12$&$-(3j+\fft32)$\\
$-(j-\fft12)$&$-\fft12$&$3j+1$&$3j+\fft52$&$-(3j-\fft12)$\\
$-(j+\fft12)$&$\fft12$&$3j+1$&$3j+\fft12$
\end{tabular}
\caption{Quantum numbers $n$ and $Q$ for which the Hamiltonian, (\ref{eq:fermH}), reduces to a rational expression.  Note that the $|n|=j+\fft12$ states are excluded from the `$-$' branch.}
\label{tbl:perfectsq}
\end{table}

\begin{table}[t]
    \centering
    \begin{tabular}{c c c c c c c c c}
        \toprule
        {} & branch & $H$ & $J_1$ & $J_2$ &$Q$& $r$ & $F_A$ & $F_B$\\
        \midrule
        $\psi_{(1)j mn}$ & $+$ & $3n+1$ & $\ft12$ & $\ft12$ &$-\fft12$& $2n$ & $2n-1$ & $2m$\\
        $\psi_{(2)j mn}$ & $+$ & $3n$ & $\ft12$ & $-\ft12$ &$\fft12$& $2n$ & $2n+1$ & $2m$\\
        $\psi_{(3)j mn}$ & $+$ & $3n$ & $-\ft12$ & $\ft12$&$\fft12$ & $2n$ & $2n+1$ & $2m$\\
        $\psi_{(4)j mn}$ & $-$ & $3n+1$ & $\ft12$ & $\ft12$ &$\fft12$& $2n$ & $2n+1$ & $2m$\\
        \bottomrule
    \end{tabular}
    \caption{The four BPS towers of fermionic fluctuations.  The first three saturate the extremal weight $n=j+\ft12$ of the `$+$' branch, so that $|m|\le j= n-\ft12$.  The fourth is the antiparticle of a `$-$' branch mode, for which $n=j-\ft12$ and hence $|m|\le j= n+\ft12$, subject to $j\ge\ft12$.}
    \label{tab: D3 fermions T11}
\end{table}

As before, we introduce
\begin{equation}
    x=\ft12r+\ft12F_B=n+m,\qquad y=\ft12r-\ft12F_B=n-m,
\end{equation}
in terms of which a state contributes $(v_1v_2)^x(v_1v_4)^y$ to the index, up to the extra factors carried by $\mathfrak p^{J_1}\mathfrak q^{J_2}$ and by the normal-plane charge.  For the first three towers $|m|\le n-\ft12$ gives $x,y\ge\ft12$, whereas for the fourth $|m|\le n+\ft12$ gives $x,y\ge-\ft12$, with the single value $x=y=-\ft12$ excluded because it corresponds to $j=0$.  Using $u^2=v_1/v_3$, $\rho^{3/2}uv=v_1v_2$ and $\rho^{3/2}u/v=v_1v_4$, the four towers contribute 
\begin{align}
    \psi_{(1)}:&\quad (\mathfrak p \mathfrak q)^{\fft12}u^{-1}\!\!\sum_{x,y\ge\fft12}\!\!(v_1v_2)^{x}(v_1v_4)^{y}=\fft{\mathfrak p \mathfrak q}{(1-v_1v_2)(1-v_1v_4)},\nn\\
    \psi_{(2)}:&\quad \left(\fft{\mathfrak p}{\mathfrak q}\right)^{\fft12}\!\!u\!\!\sum_{x,y\ge\fft12}\!\!(v_1v_2)^{x}(v_1v_4)^{y}=\fft{(v_1/v_3)\,\mathfrak p}{(1-v_1v_2)(1-v_1v_4)},\nn\\
    \psi_{(3)}:&\quad \left(\fft{\mathfrak q}{\mathfrak p}\right)^{\fft12}\!\!u\!\!\sum_{x,y\ge\fft12}\!\!(v_1v_2)^{x}(v_1v_4)^{y}=\fft{(v_1/v_3)\,\mathfrak q}{(1-v_1v_2)(1-v_1v_4)},\nn\\
    \psi_{(4)}:&\quad (\mathfrak p \mathfrak q)^{\fft12}u\!\!\!\!\sum_{\substack{x,y\ge-\fft12\\(x,y)\ne(-\fft12,-\fft12)}}\!\!\!\!(v_1v_2)^{x}(v_1v_4)^{y}=\fft1{(1-v_1v_2)(1-v_1v_4)}-1,
\end{align}
where we used $\mathfrak p \mathfrak q=v_1v_2v_3v_4$ throughout.  Adding them up, the single-letter fermionic index is
\begin{equation}
    f_{\mathrm{fermion}}=\fft{v_1v_2+v_1v_4+\mathfrak p \mathfrak q+\fft{v_1}{v_3}(\mathfrak p+\mathfrak q)-\fft{v_1}{v_3}\mathfrak p \mathfrak q}{(1-v_1v_2)(1-v_1v_4)}.
\label{eq: fermion sl t11}
\end{equation}
The final term in the numerator, $-\ft{v_1}{v_3}\mathfrak p \mathfrak q=-(v_1v_2)(v_1v_4)$, is precisely the subtraction of the $j=0$ state from the fourth tower; it plays the same role as the analogous term in the AdS$_5\times S^5$ and AdS$_4\times S^7$ giants of \cite{Deddo:2025lfm}, where it prevents an overcounting of BPS states.

\subsection{The SCI from giant graviton excitations}\label{sec:SCI}

Having examined the scalar, vector and fermion fluctuations on the D3-brane worldvolume, we are now in a position to assemble them into the single letter index.  Summing the individual contributions from the scalars \eqref{eq: scalar sl t11}, the vectors \eqref{eq: vector sl t11} and the fermions \eqref{eq: fermion sl t11} gives
\begin{align}
    f_{(0,0,1,0)} & = f_{\mathrm{scalar}}+f_{\mathrm{vector}}-f_{\mathrm{fermion}}\nn\\
    &=\fft{(\mathfrak p + \mathfrak q + (v_1/v_3)) + \mathfrak p \mathfrak q (v_1/v_3)-\left(v_1v_2 + v_1v_4 + \mathfrak p \mathfrak q + (v_1/v_3)(\mathfrak p + \mathfrak q) - (v_1/v_3) \mathfrak p \mathfrak q \right)}{(1-v_1v_2)(1-v_1v_4)}\nn\\
    &=1 - \fft{(1 - v_1/v_3)(1 - \mathfrak p)(1 - \mathfrak q)}{(1-v_1v_2)(1-v_1v_4)},
\end{align}
where in the last step we used the constraint $v_1v_2v_3v_4 = \mathfrak p \mathfrak q$.  This is precisely the expected result, and reproduces the structure of the single-letter index of the D3 giants in $S^5$
\begin{equation}
    f_{(1,0,0)}^{S^5}=1-\fft{(1-y_1^{-1})(1 - \mathfrak p)(1 - \mathfrak q)}{(1-y_2)(1-y_3)},
\end{equation}
under the identification
\begin{equation}
    y_1^{-1}\ \leftrightarrow\ \fft{v_1}{v_3},\qquad y_2\ \leftrightarrow\ v_1v_2,\qquad y_3\ \leftrightarrow\ v_1v_4,
\end{equation}
which maps $y_1y_2y_3 = \mathfrak p \mathfrak q$ onto $v_1v_2v_3v_4 = \mathfrak p \mathfrak q$.  The remaining three giants are obtained by the obvious permutations, giving $f_{(1,0,0,0)}$, $f_{(0,1,0,0)}$ and $f_{(0,0,0,1)}$, and the corresponding indices follow by plethystic exponentiation.

There are two limitations to our construction for recovering the full finite-$N$ superconformal index. The first is that the index that is constructed from $f_{(1,0,0,0)}$ and its permutation above is for a single giant. If we have a stack of  $m$ giants, we need a non Abelian generalization of the DBI action and its quantization to construct $\hat I^{GG}_{(m,0,0,0)}$. Nevertheless, one can effectively promote these fluctuations as strings extending between the coincident branes by appending $U(m)$ adjoint characters to the single letter index and projecting onto $U(m)$ singlets. This allows one to construct $\hat I^{GG}_{(m,0,0,0)}$ by plethystic exponentiation \cite{Imamura:2021ytr,Gaiotto:2021xce,Eleftheriou:2023jxr,Deddo:2025lfm}. The second is about intersecting brane configurations which contribute to, e.g., $\hat I^{GG}_{(m_1,m_2,0,0)}$. These are bi-fundamental string modes extending between the intersecting giants. Although \cite{Hamilton:2010sv} gives a description of scalar modes that extend between adjoining giants in $T^{1,1}$, we found that treatment and its generalization to be insufficient to recover the full index.  This is similar to the limitations noted in \cite{Deddo:2025lfm}. However, unlike $S^5$, $T^{1,1}$ has non-intersecting giants that do not have strings extending between them (the $\theta_i = 0,\pi$ giants). So one can construct indices of the form $\hat I_{(m_1, 0, m_3,0)}^{GG}$ and its permutations from the single letters above.

\section{Review of the SCFT dual to \texorpdfstring{$Y^{p,q}$}{Y(p,q)}}\label{sec:Ypq-review}

Having established how the giant graviton expansion emerges from quantized D3-brane fluctuations in $AdS_5 \times T^{1,1}$, we now turn to the $Y^{p,q}$ family. These backgrounds preserve the exact same amount of supersymmetry, but provide a substantially broader test of the worldvolume construction. Their supersymmetric three cycles turn out to be general lens spaces rather than squashed three spheres and their non trivial fibration leads to more intricate normal bundles and fluctuation equations. The $Y^{p,q}$ family therefore allows us to separate the local features of the D3-brane BPS spectrum from its global topological features.

Type IIB string theory on AdS$_5\times Y^{p,q}$ gives rise to a dual $\mathcal N=1$ superconformal quiver gauge theory. The spaces $Y^{p,q}$, labeled by positive coprimes $p,q$, $(0 < q < p)$, are five dimensional Sasaki-Einstein manifolds with an $S^2 \times S^3$ topology.
The unit radius $Y^{p,q}$ metric was written originally as \cite{Gauntlett:2004yd}
\begin{align}
    ds^2&=\fft{1-y}6(d\theta^2+\sin^2\theta d\phi^2)+\fft1{w(y)q(y)}dy^2+\fft{q(y)}9(d\psi-\cos\theta d\phi)^2\nn\\
    &\quad+w(y)\left(d\alpha+f(y)(d\psi-\cos\theta d\phi)\right)^2,
\label{eq:Ypq}
\end{align}
where
\begin{equation}
    w(y)=\fft{2(a-y^2)}{1-y},\qquad q(y)=\fft{a-3y^2+2y^3}{a-y^2},\qquad f(y)=\fft{a-2y+y^2}{6(a-y^2)}.
\end{equation}
Locally the $U(1)$ coordinates $\phi, \psi$ has period $2\pi$ while $\alpha$ have period $2\pi\ell$. the remaining coordinates obey $0 \leq \theta \leq \pi$ and $y_1 \leq y \leq y_2$ where $y_1 \leq y_2 \leq y_3$ are the roots of the cubic polynomial, $a-3y^2+2y^3$. 

The $(\theta,\phi)$ parametrize a round two sphere. The pair $(y,\psi)$ describes a second topological $S^2$ because the $\psi$ circle degenerates at $y = y_1$ and $y = y_2$. This sphere is non trivially fibered over the round $S^2$ as can be seen in the last term of the first line of \eqref{eq:Ypq}. Although the resulting four dimensional base is topologically  $S^2 \times S^2$ \cite{Gauntlett:2004yd},  its fibration structure plays an essential role in the fluctuation analysis.

The $U(1)$ fibration is well defined if we choose integer Chern numbers $p$ and $q$ for the $U(1)$ bundle over the $S^2\times S^2$ base.  This gives rise to a discrete set of manifolds, $Y^{p,q}$, with

\begin{align}
 y_1 = \dfrac{1}{2} \left(1 - \dfrac{3q}{2p} - \sqrt{1 - \dfrac{3q^2}{4p^2}}\right), \quad y_2 = \dfrac{1}{2} \left(1 + \dfrac{3q}{2p} - \sqrt{1 - \dfrac{3q^2}{4p^2}}\right), \quad y_3 = \dfrac{1}{2} + \sqrt{1 - \dfrac{3q^2}{4p^2}},
\end{align}
and
\begin{equation}
    \ell = \dfrac{q}{3q^2-2p^2 + p \sqrt{4p^2-3q^2}}. \label{eq: ell}
\end{equation}
Note that $y_1$ is negative while $y_2$ is positive.  In addition, the constant $a$ is given implicitly by $a=3y_i^2-2y_i^3$ for either $y_1$, $y_2$ or $y_3$.

The background is supported by a self-dual five-form whose internal component is
\begin{equation}
    F_{[5]} = \dfrac{2}{9} L^4 (1-y) \sin\theta\, d\theta \wedge d\phi \wedge d\psi \wedge d\alpha \wedge dy.
\end{equation}
The metric, (\ref{eq:Ypq}), has isometries $SU(2)\times U(1)\times U(1)$ acting on the round $S^2$ given by $(\theta,\phi)$ and shifts along $\psi$ and $\alpha$.  On the gauge theory side, the dual $\mathcal N=1$ SCFT has a global $SU(2)_{F_A}\times U(1)_{F_B}\times U(1)_{r_*}\times U(1)_B$ symmetry where $F_A$ and $F_B$ are mesonic symmetries, $r_*$ denotes the R-symmetry and $B$ is the baryonic symmetry.  In particular, note that we denote the true R-charge by $r_*$.   Since the baryonic symmetry is not realized geometrically on $Y^{p,q}$ we find a natural match between the $SU(2)\times U(1)\times U(1)$ isometry and $SU(2)_{F_A}\times U(1)_{F_B}\times U(1)_r$.  Note, however, that the $U(1)$ factors can mix, although $a$-maximization can be used to obtain the correct $U(1)_{r_*}$ \cite{Intriligator:2003jj, Martelli:2005tp}.

The $\mathcal N=1$ index, (\ref{eq:N=1sci}), can be further refined in the presence of additional flavor symmetries.  Following \cite{Arai:2019aou}, we may introduce three additional fugacities corresponding to $U(1)_{F_A}\times U(1)_{F_B}\times U(1)_B$ where $U(1)_{F_A}$ is the Cartan subgroup of $SU(2)_{F_A}$.  As in the $T^{1,1}$ case, it is convenient to introduce four fugacities, $v_i$, $i=1,\ldots,4$, satisfying the constraint $v_1v_2v_3v_4=\mathfrak p \mathfrak q$.  The index then takes the form
\begin{equation}
    Z(\mathfrak  p, \mathfrak q,v_i)=\Tr[(-1)^Fe^{-\beta\mathcal H}\mathfrak p^{j_1} \mathfrak q^{j_2}v_1^{R_1}v_2^{R_2}v_3^{R_3}v_4^{R_4}],
\end{equation}
where the $R_I$ charges correspond to the vertices of the toric diagram and
\begin{equation}
    \mathcal{H} = \Delta - J_1 - J_2 - \dfrac{3}{2}r_*. \label{eq: SUSY H ypq}
\end{equation}
Note that the constraint $v_1v_2v_3v_4 = \mathfrak  p \mathfrak q$ is related to the normalization of the true R-charge
\begin{equation}
    r_*=\sum_{I=1}^4c_IR_I,\qquad \sum_{I=1}^4c_I=2.
\end{equation}
However, for convenience, we will mostly use the fake R-charge
\begin{equation}
    r=\fft12(R_1+R_2+R_3+R_4),
\end{equation}
which also satisfies the relation $\sum_I c_I=2$.

\begin{figure}[t]
\centering
\begin{tikzpicture}[scale=0.8]

\def\p{5}
\def\q{2}

\pgfmathtruncatemacro{\a}{\p-\q-1}
\pgfmathtruncatemacro{\b}{\p-\q}

\draw[step=1, gray!25, very thin] (-0.5,-0.5) grid (\p+0.5,\p+0.5);

\draw[->] (-0.5,0) -- (\p+0.8,0) node[right] {$x$};
\draw[->] (0,-0.5) -- (0,\p+0.8) node[above] {$y$};

\coordinate (V1) at (0,0);
\coordinate (V2) at (1,0);
\coordinate (V3) at (\p,\p);
\coordinate (V4) at (\a,\b);

\draw[thick, fill=blue!10]
  (V1) -- (V2) -- (V3) -- (V4) -- cycle;

\fill (V1) circle (2pt) node[below left] {$1$};
\fill (V2) circle (2pt) node[below right] {$2$};
\fill (V3) circle (2pt) node[above right] {$3$};
\fill (V4) circle (2pt) node[above left] {$4$};

\node[below=10pt] at (V1) {\scriptsize $(0,0)$};
\node[below=10pt] at (V2) {\scriptsize $(1,0)$};
\node[above=10pt] at (V3) {\scriptsize $(p,p)$};
\node[above=20pt,left=-24pt] at (V4) {\scriptsize $(p-q-1,p-q)$};

\end{tikzpicture}
\caption{The toric diagram for $Y^{p,q}$ (given here for $p=5$ and $q=2$).}
\label{fig:Ypq}
\end{figure}
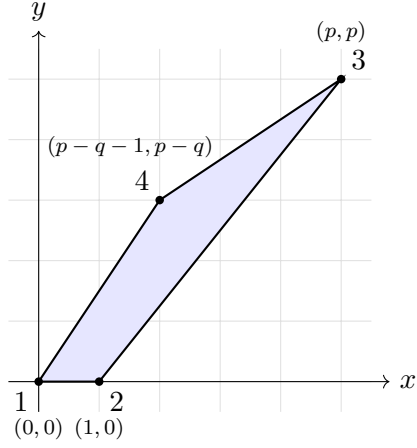

The toric diagram for $Y^{p,q}$ is shown in Figure~\ref{fig:Ypq}.    Following \cite{Arai:2019aou,Kim:2024ucf}, we label the vertices by $I=1,\ldots,4$.  The coordinates of the vertices are
\begin{equation}
    V_1=(0,0,1),\qquad V_2=(1,0,1),\qquad V_3=(p,p,1),\qquad V_4=(p-q-1,p-q,1),
\end{equation}
and each vertex is associated with a charge $R_I$ and the corresponding fugacity $v_I$.  Each edge between vertices $I$ and $I+1$ can be labeled by $r=I+\fft12$, and is associated with the fugacity combination
\begin{equation}
    w_r=\prod_{I=1}^4v_i^{V_I\cdot g_r}.
\end{equation}
The vectors $g_r$ are defined by
\begin{equation}
    g_r\cdot V_{r-\fft12}=0=g_r\cdot V_{r+\fft12},\qquad g_r\cdot V_I\ge0\mbox{ for }I=1,\ldots,4.
\end{equation}
For $Y^{p,q}$ we find explicitly
\begin{equation}
    g_{\fft12}=(p-q,1+q-p,0),\quad g_{1+\fft12}=(0,1,0),\quad g_{2+\fft12}=(-p,p-1,p),\quad g_{3+\fft12}=(q,-q-1,p),
\end{equation}
which gives
\begin{equation}
    w_{\fft12}=v_2^{p-q}v_3^p,\qquad w_{1+\fft12}=v_3^pv_4^{p-q},\qquad w_{2+\fft12}=v_4^{p+q}v_1^p,\qquad w_{3+\fft12}=v_1^pv_2^{p+q}.
\label{eq:wfug}
\end{equation}

Note that each node $I$ is associated with a three-cycle $S_I$ which is topologically a lens space $S^3/\mathbb Z_{k_I}$ where
\begin{equation}
    k_1=p-q,\qquad k_2=p,\qquad k_3=p+q,\qquad k_4=p.
\end{equation}
These are the three-cycles that the giant graviton D3-branes will wrap.

Following Eq.~(4.9) in Ref.~\cite{Martelli:2004wu}, we identify the charge $R_I$ generators by
\begin{align}
    R_1&=-i(\partial_\psi+\ft12(p-q)\ell\partial_\alpha),\qquad R_2=-i(-\partial_\phi+\partial_\psi),\nn\\
    R_3&=-i(\partial_\psi-\ft12(p+q)\ell\partial_\alpha),\qquad R_4=-i(\partial_\phi+\partial_\psi).
\end{align}
(Note that vertices 1 and 4 in the toric diagram, Figure~\ref{fig:Ypq}, are swapped compared to Ref.~\cite{Martelli:2004wu}.)  The coordinates $\theta$, $\phi$ and $\alpha$ are those of the metric, (\ref{eq:Ypq}).  The two flavor charges, fake R-charge and baryon charge are given by
\begin{align}
    F_A&=\ft12(R_4-R_2)&&=-i\partial_\phi,\nn\\
    F_B&=\ft12(R_3-R_1)&&=-i(-\ft12p\ell\partial_\alpha),\nn\\
    r&=\ft12(R_1+R_2+R_3+R_4)&&=-i(2\partial_\psi-\ft12q\ell\partial_\alpha),\nn\\
    NB&=-(p+q)R_1+pR_2-(p-q)R_3+pR_4&&=0.
\label{eq:charges}
\end{align}
For BPS considerations, we note that the true R-charge, which is related to the Reeb vector \cite{Martelli:2005tp}, is given by
\begin{equation}
    \ft32r_*=-i(3\partial_\psi-\ft12\partial_\alpha).
\label{eq:rstar}
\end{equation}
Moreover, we can introduce fugacities $u$, $v$, $\rho$ and $\zeta$ associated with $F_A$, $F_B$ and $r$ and $B$, respectively, according to
\begin{equation}
    v_1^{R_1} v_2^{R_2} v_3^{R_3} v_4^{R_4} = \rho^{\fft32r} u^{2F_A} v^{2F_B} \zeta^B.
\label{eq:fugmap}
\end{equation}
In particular, we have
\begin{equation}
    v_1=\rho^{\fft34}v^{-1}\zeta^{-(p+q)/N},\qquad v_2=\rho^{\fft34}u^{-1}\zeta^{p/N},\qquad v_3=\rho^{\fft34}v\zeta^{-(p-q)/N},\qquad v_4=\rho^{\fft34}u\zeta^{p/N}.
\end{equation}
The edge fugacities, (\ref{eq:wfug}), are then
\begin{align}
    w_{\fft12}&=\rho^{\fft34(2p-q)}u^{-(p-q)}v^p,\qquad w_{1+\fft12}=\rho^{\fft34(2p-q)}u^{p-q}v^p,\nn\\
    w_{2+\fft12}&=\rho^{\fft34(2p+q)}u^{p+q}v^{-p},\kern.6em\qquad w_{3+\fft12}=\rho^{\fft34(2p+q)}u^{-(p+q)}v^{-p}.
\label{eq:wtoquv}
\end{align}
The dual gauge theory is an $SU(N)^{2p}$ quiver gauge theory that can be constructed from the toric data  of $Y^{p,q}$.  The corresponding index is then built out of single letter indices for the adjoint gauge bosons \eqref{eq: vecmult index} and the chiral bifundamentals \eqref{eq:chiral mult index}. 

\section{The giant graviton expansion on \texorpdfstring{$\mathrm{AdS}_5\times Y^{p,q}$}{AdS5xY(p,q)}}\label{sec:Ypq-giants}

The giant graviton expansion starts with the infinite $N$ index, $Z_\infty(\mathfrak p, \mathfrak q;v_i)$.
Holographically, this corresponds to the contribution of Kaluza-Klein gravitons to the index.  In addition to these KK gravitons, we must also include giant gravitons, which in this case would be D3-branes wrapped along the three-cycles $S_I$. These giants sit where the Killing vectors $R_I$ degenerate, which is $y = y_1, y_2$ and $\theta = 0,\pi$, the vertices of the toric diagram.  The giant graviton expansion takes the form \cite{Arai:2019aou}
\begin{equation}
    \fft{Z_N(\mathfrak p,\mathfrak q;v_i)}{Z_\infty(\mathfrak p,\mathfrak q;v_i)}=\sum_{n_i\ge0}v_1^{n_1N}v_2^{n_2N}v_3^{n_3N}v_4^{n_4N}\hat I_{(n_1,n_2,n_3,n_4)}^{GG}(\mathfrak p,\mathfrak q;v_i).
\end{equation}
We consider single giants wrapping $S_I$.  Since $S_I$ is a lens space, $L(k_I,1)$, the single letter index can be written as an orbifold index
\begin{equation}
    f_I(\mathfrak p,\mathfrak q;v_i)=\fft1{k_I}\sum_{s=0}^{k_I-1}f(\mathfrak p,\mathfrak q;\omega^s(w_{I-\fft12})^{1/k_I},\tilde\omega^s(w_{I+\fft12})^{1/k_I}),
\label{eq:orbindex}
\end{equation}
where $\omega^{k_I}=1$ and $\tilde\omega^{k_I}=1$ are primitive $k_I$-th roots of unity with some appropriate relation between $\omega$ and $\tilde\omega$ (see Ref. \cite{Arai:2019aou} for further details).  Here, $f(\mathfrak p,\mathfrak q;w,w')$ is the 1/16-BPS single letter index specialized to the giant graviton
\begin{equation}\label{Eq:Yp-single-letter-index}
    f(\mathfrak p,\mathfrak q;w,w')=1-\fft{(1-ww'/\mathfrak p\mathfrak q)(1-\mathfrak p)(1-\mathfrak q)}{(1-w)(1-w')}.
\end{equation}
Our goal is to reproduce the above expressions, \eqref{Eq:Yp-single-letter-index}  and \eqref{eq:orbindex}, from the quantized action.

As we will see below, the four types of giants fall into two classes:
\begin{itemize}
    \item Class one: localized at $y = y_1$ (the $S_3$ giant) or $y = y_2$ (the $S_1$ giant) and with world volume wrapping $t,\theta,\phi,\alpha + f(y_i)\psi$.
    \item  Class two: localized at $\theta=0$ (the $S_4$ giant) or $\theta = \pi$ (the $S_2$ giant) and with world volume wrapping $t,y,\psi\pm\phi,\alpha$.
\end{itemize}
We first consider the class one giants as their worldvolume description is more direct.

\subsection{Class one giants}

We begin with the giants $S_1$ and $S_3$, localized at $y = y_2$ and $y = y_1$ respectively. To treat these embeddings uniformly, let $\tilde y$ denote $y_1$ or $y_2$. For these giant gravitons, the worldvolume is given by time and the three-dimensional space
\begin{equation}
    ds_3^2=\fft{1-\tilde y}6(d\theta^2+\sin^2\theta d\phi^2)+\fft{2(a-\tilde y^2)}{1-\tilde y}\left(d\alpha+\fft{a-2\tilde y+\tilde y^2}{6(a-\tilde y^2)}(d\psi-\cos\theta d\phi)\right)^2.
\end{equation}
Noting that
\begin{equation}
    a-y_i^2=2y_i^2(1-y_i),\qquad a-2y_i+y_i^2=-2y_i(1-y_i)^2,
\end{equation}
and furthermore that
\begin{equation}
    1-y_i^{-1}=-3c_i\ell,\quad\mbox{where}\quad c_1=-(p+q),\quad c_2=(p-q),
\end{equation}
we can rewrite the metric as
\begin{equation}
    ds_3^2=2\tilde y\tilde c\ell\left[\fft14\left(d\theta^2+\sin^2\theta d\phi^2+\fft{2\tilde y\ell}{\tilde c}(d\tilde\alpha+\tilde c\cos\theta d\phi)^2\right)\right],
\label{eq:lsm}
\end{equation}
where $\tilde c$ denotes $c_1$ or $c_2$ as appropriate, and we have defined the shifted fiber coordinate
\begin{equation}
    \tilde\alpha=\fft{2\alpha}\ell-\tilde c\psi.
\end{equation}
We can now identify this as the metric on the lens space $L(|\tilde c|,1)$.  We now look in turn at the scalar, vector and fermion fluctuations of the D3-brane giant wrapped on this lens space.

\subsubsection{Scalar fluctuations}

Starting with the scalar fluctuations, we expand the D3-brane action, \eqref{eq: DBI Action}, for the six scalar fluctuations about the $y=\tilde y$ giant.  These fluctuations are the four AdS$_5$ scalars $v_k$ and the two internal $y$ and $\psi$ modes. Because the $(y,\psi)$ two sphere is wrapped non trivially on the base, the internal fluctuations require some care. Unlike the AdS modes which are scalar fields, these internal modes are sections of the normal bundle carrying a charge $\sigma = \pm  1$ with its connection lying along $\tilde \alpha$. A careful treatment of these sections is detailed in Appendix \ref{app: internal modes}. 

We introduce rectangular coordinates $(A,B)$ centered at $y =\tilde y$ and define the complex combination $W = A + iB$. Consider small fluctuations $\epsilon A, \epsilon B$ on $Y^{p,q}$ and $\epsilon v_k$ on AdS$_5$. Then, up to quadratic order, the components of the DBI action take the form,
\begin{align}
    \sqrt{-\det P[g]} &= \dfrac{L^3}{3} |\tilde y| (1-\tilde y)\sin\theta \nn\\
    &\quad - \dfrac{L^3}{6} |\tilde y| (1-\tilde y) \sin\theta \biggl[L^2 |\dot W|^2 - |\mathcal{D}W|^2 - M_0^2 |W|^2\biggr]\nn\\
    &\quad- \dfrac{L}{6} |\tilde y| (1-\tilde y) \sin\theta \sum_k \Bigl[ L^2 \dot v_k^2 + v_k \nabla^2 v_k - v_k^2\Bigr],
\label{eq:dbiact}
\end{align}
while the Wess-Zumino term in an appropriate gauge choice becomes
\begin{equation}
    P[C_4] = \dfrac{2}{3}L^4 |\tilde y| (1 - \tilde y)\sin\theta\, \mathrm{Im}(\bar W \dot W).
\label{eq:wzact}
\end{equation}
Here we have chosen the orientation of each brane such that it is a BPS giant rather than an anti-giant. For a fixed orientation of the worldvolume the Wess-Zumino term is proportional to $\tilde y(\tilde y-1)$ and therefore changes sign between $\tilde y=y_1<0$ and $\tilde y=y_2>0$; the $S_1$ giant at $y_2$ thus carries the opposite orientation to the $S_3$ giant at $y_1$.

Also note that, $\nabla^2$ is the Laplacian on the lens space $L(|\tilde c|,1)$ with metric (\ref{eq:lsm}).
For the internal mode $W$, we have introduced a covariant derivative (see Appendix \ref{app: internal modes} for a detailed discussion),
\begin{equation}
    \mathcal{D}_i = \partial_i - i \sigma \mathcal{A}_i, \quad \mathcal{A} = (1 - 2\tilde c \tilde y \ell)\cos\theta \, d\phi - 2\tilde y \ell d \tilde \alpha, \quad \sigma = \pm 1, \label{eq: ypq cov der}
\end{equation}
and an effective mass term,
\begin{equation}
    M_0^2 = \dfrac{7 \tilde y -1}{1 - \tilde y} = -7 + \dfrac{2}{\tilde y \tilde c \ell}. \label{eq: ypq eff mass}
\end{equation}
Because of these differences, we will study the AdS modes and internal modes separately.

\subsubsection*{Scalar fluctuations into AdS}

The fluctuations into  AdS are similar in structure to those of $T^{1,1}$, so we start with those. They are described by the quadratic action
\begin{equation}
    S_2^{\mathrm{AdS}} = \dfrac{L T_3}{6} |\tilde y| (1-\tilde y) \int d^4\xi \sin\theta \sum_k [L^2 \dot v_k^2 + v_k \nabla^2 v_k - v_k^2]. \label{eq: ypq ads action}
\end{equation}
The Laplacian on the squashed $S^3$ and its lens space counterpart is summarized in Appendix~\ref{app:lens}.  Comparison with (\ref{eq:Lp1met}) shows that the scalar harmonics satisfying $\nabla^2f=-\Lambda f$ can be read off from (\ref{eq:Lp1sys}) with the map
\begin{equation}
    p=\tilde c,\qquad \tilde\psi=\tilde\alpha,\qquad e^{2\tilde\lambda}=\fft{2\tilde y\ell}{\tilde c},
\end{equation}
along with an overall volume scaling.  In particular, the scalar harmonics take the form
\begin{equation}
    f_{lmn}(z,\phi,\alpha)=e^{im\phi}e^{2in\alpha/\ell}e^{-i\tilde cn\psi}z^{\fft12|m-\tilde cn|}(1-z)^{\fft12|m+\tilde cn|}{}_2F_1(a,b;c;z),
\label{eq:Lk1wf}
\end{equation}
with $a$, $b$ and $c$ given in the Appendix, and with the lens space conditions
\begin{equation}
    m+\tilde cn\in\mathbb Z,\qquad n\in\fft12\mathbb Z.
\label{eq:mnkcond}
\end{equation}
Note that these quantization conditions are compatible with the $Y^{p,q}$ periods under the action of the 3-torus given by (\ref{eq:e1e2e3}).  The corresponding eigenvalues are
\begin{equation}
    2\tilde y\tilde c\ell\Lambda_{lmn}=4l(l+1)-(2n)^2\left(\tilde c^2-\fft{\tilde c}{2\tilde y\ell}\right), \label{eq: scalar ypq spectrum raw}
\end{equation}
where
\begin{equation}
    l=\max(|m|,|\tilde cn|)+j,\qquad j=0,1,2,\ldots.
\label{eq:ldef}
\end{equation}
With a bit of manipulation, the eigenvalues can be re-expressed as
\begin{equation}
    \Lambda_{lmn}+1=\left(\left|\fft{n}{\tilde y\ell}\right|+1\right)^2+\fft1{2\tilde y\tilde c\ell}\left((2l+1)^2-(2|\tilde cn|+1)^2\right).
\label{eq:goodLlmn}
\end{equation}
We expand the AdS action, (\ref{eq: ypq ads action}), in terms of the lens space harmonics $f_{lmn}$ 
\begin{equation}
    S^{\mathrm{AdS}}_{2} = \dfrac{L^3 T_3}{6} |\tilde y| (1-\tilde y) \int dt \, \sum_{k,l,m,n}\Bigl[ \dot v_{(k)lmn}^2 - \dfrac{(\Lambda_{lmn}+1)}{L^2}\, v_{(k)lmn}^2\Bigr].
\end{equation}
To proceed with canonical quantization, we absorb the constant prefactor in the action by performing a rescaling $v_k \to \sqrt{3}\, v_k/\bigl(L^2\sqrt{T_3|\tilde y|(1-\tilde y)}\bigr)$; we then obtain the canonical momenta
\begin{equation}
    p_{(k)lmn} = \dfrac{1}{L}\dot v_{(k)lmn}.
\end{equation}
Then the Hamiltonian can be written as,
\begin{equation}
    H = \dfrac{1}{L} \sum_k\sum_{lmn} \dfrac{L^2}{2} p_{(k)lmn}^2 + \dfrac{(\Lambda_{lmn}+1)}{2L^2} v_{(k)lmn}^2.
\end{equation}
This is a Harmonic oscillator, and the quantization procedure closely follows our discussion for $T^{1,1}$ with appropriate modification for the different geometry. It is also convenient to transform to a helicity basis $a_{(\pm)}$ similar to \eqref{eq: helicity basis}. In terms of the corresponding number operators, $N_{(i\pm)lmn}$, the Hamiltonian takes the form,
\begin{equation}
    H  = \sum_{l,m,n} \sqrt{\Lambda_{lmn}+1}\, \sum_{i = 1}^2 (N_{(i+)lmn} + N_{(i-)lmn} ). \label{eq: ypq1 ads hamiltonian}
\end{equation}
In addition to the Hamiltonian, from the charges in (\ref{eq:charges}) and (\ref{eq:rstar}), and the wave function (\ref{eq:Lk1wf}), we can read off the charges for the AdS$_5$ scalar modes

\begin{align}
    J_i &= \sum_{l,m,n} (N_{(i+)lmn}-N_{(i-)lmn}),\nn\\
    F_A &= \sum_i\sum_{l,m,n} m(N_{(i+)lmn} + N_{(i-)lmn}),\nn\\
    F_B &= - \sum_i\sum_{l,m,n} pn (N_{(i+)lmn} + N_{(i-)lmn}),\nn\\
    r &= -\sum_i\sum_{l,m,n} (2\tilde c +q )n (N_{(i+)lmn} + N_{(i-)lmn}),\nn\\
    \fft32r_* &= -\sum_i\sum_{l,m,n} \fft{n}{\tilde y\ell} (N_{(i+)lmn} + N_{(i-)lmn}). \label{eq: ypq1 scalar charges}
\end{align}
This allows us to express the supersymmetric Hamiltonian \eqref{eq: SUSY H ypq} as
\begin{align}
    \mathcal{H} &=\sum_{l,m,n}\sum_i\biggl[\left(\sqrt{\Lambda_{lmn}+1}-1+\fft{n}{\tilde y\ell}\right)N_{(i+)lmn}\nn\\
    &\kern4em + \left(\sqrt{\Lambda_{lmn} + 1 } +1+\fft{n}{\tilde y\ell}\right)N_{(i-)lmn}\biggr],
\end{align}
where we have dropped the zero point energy of the oscillator modes.

In order to identify the BPS spectrum, we note that the BPS states saturate the bound $\mathcal H\ge0$.  From the explicit eigenvalues $\Lambda_{lmn}$ given in (\ref{eq:goodLlmn}), we see that the $v_{(i-)lmn}$ states are never BPS, while the $v_{(i+)lmn}$ states saturate the BPS bound when $n/\tilde y\le0$ and $l=|\tilde c n|$.  Recalling the definition of $l$ from (\ref{eq:ldef}), this requires $j=0$ and $|m|\le|\tilde cn|$.  Note that, in this case,  similar to what we argued for $T^{1,1}$, the hypergeometric function ${}_2F_1(a,b,c;z)=1$ is a constant, which is consistent with the notion that the BPS states are among the zero modes on the wrapped cycle. The zero mode ($j=0$) states and their quantum numbers are summarized in Table~\ref{tab: ypq1 AdS}.

\begin{table}[t]
    \centering
    \begin{tabular}{c c c c c c c c c c}
        \toprule
        {} & $H $ & $J_1$ & $J_2$ & $r_*$& $r$ &$F_A$& $F_B$ &BPS? \\
        \midrule
        $v_{(1+)\ell m n}$ & $|n/ \tilde y \ell| + 1$ & $1$ & $0$ & $-{2n}/{3 \tilde y \ell}$ & $-(2\tilde c + q)n$ & $m$ &$-pn$ & \checkmark\\
        $v_{(2+)\ell m n}$ & $|n/ \tilde y \ell| + 1$  & $0$ & $1$ & $-{2n}/{3 \tilde y \ell}$ & $-(2\tilde c + q)n$ & $m$ &$-pn$ & \checkmark\\
        $v_{(1-)\ell m n}$ & $|n/ \tilde y \ell| + 1$  & $-1$ & $0$ & $-{2n}/{3 \tilde y \ell}$ &$-(2\tilde c + q)n$ & $m$ &$-pn$ & {}\\
        $v_{(2-)\ell m n}$ & $|n/ \tilde y \ell| + 1$ & $0$ & $-1$ & $-{2n}/{3 \tilde y \ell}$ & $-(2\tilde c + q)n$ & $m$ &$-pn$ & {}\\
        \bottomrule
    \end{tabular}
    \caption{Quantum numbers of the zero mode ($j=0$) sector in the AdS fluctuations of the class one giants in $Y^{p,q}$.}
    \label{tab: ypq1 AdS}
\end{table}

\subsubsection*{BPS spectrum and the orbifold sum}
At this stage, it is perhaps worth returning to the individual description of $y_1$ and $y_2$ giants. For the $S_3$ giant located at $\tilde y=y_1<0$, the BPS condition requires $n\ge0$, $|m|\le(p+q)n$ and $l=(p+q)n$ along with the lens space conditions, (\ref{eq:mnkcond}).  For the $S_1$ giant located at $\tilde y=y_2>0$, the BPS condition instead requires $n\le0$, $|m|\le-(p-q)n$ and $l=-(p-q)n$ and the lens space conditions.  This is summarized as follows:
\begin{align}
    \mbox{$S_3$ giant at $y_1$}:&&n\ge0,&\qquad |m|\le(p+q)n,\qquad m-(p+q)n\in\mathbb Z,\qquad n\in\fft12\mathbb Z,\nn\\
    \mbox{$S_1$ giant at $y_2$}:&&n\le0,&\qquad |m|\le-(p-q)n,\qquad m+(p-q)n\in\mathbb Z,\qquad n\in\fft12\mathbb Z.
\label{eq:BPS12}
\end{align}
To understand these conditions better, we define the linear combinations of the $m$ and $n$ quantum numbers
\begin{equation}
    x=m-\tilde cn,\qquad y=-m-\tilde cn. \label{eq: ypq1 xy lattice}
\end{equation}
In particular, we have
\begin{align}
    \mbox{$S_3$ giant at $y_1$}:&\qquad x=(p+q)n+m\ge0,\qquad y=(p+q)n-m\ge0,\nn\\
    &\qquad x,y\in\mathbb Z,\qquad x+y\equiv 0\pmod{p+q},\nn\\
    \mbox{$S_1$ giant at $y_2$}:&\qquad x=-(p-q)n+m\ge0,\qquad y=-(p-q)n-m\ge0,\nn\\
    &\qquad x,y\in\mathbb Z,\qquad x+y\equiv0\pmod{p-q}. \label{eq: ypq1 xy lattice cndn}
\end{align}
The condition $x+y\equiv0\pmod{p\pm q}$ demonstrates that the sum over BPS states with $x,y\in\mathbb Z_{\ge0}$ takes the form of an orbifold sum, a crucial step in reconstructing the full giant graviton index.

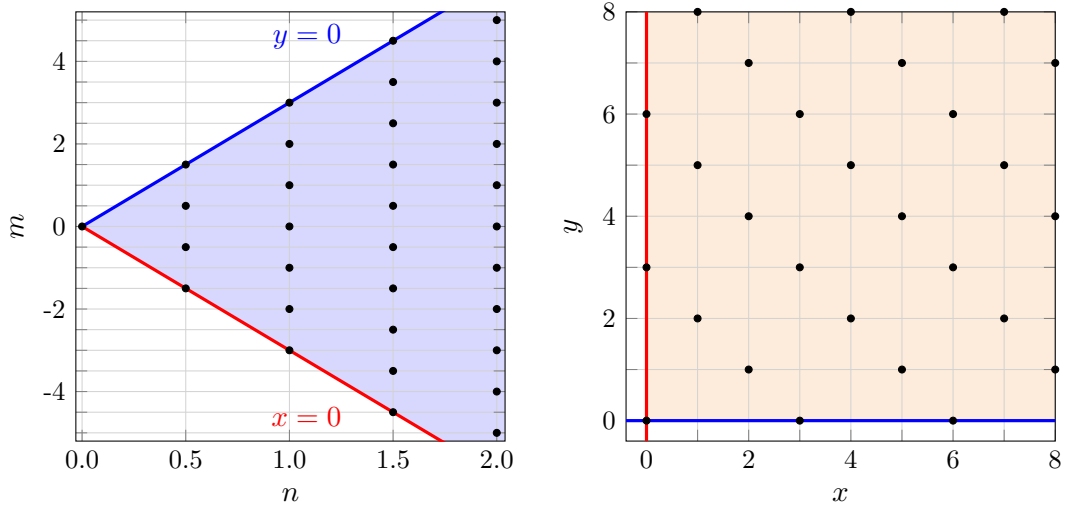
\begin{figure}[t]
    \begin{tikzpicture}
\begin{groupplot}[
    group style={group size=2 by 1, horizontal sep=1.6cm},
    width=0.48\linewidth, height=0.48\linewidth,
    grid=major,                                   
    major grid style={gray!35, line width=0.3pt}, 
    tick align=inside,
    tick label style={font=\small},
    clip mode=individual,
    set layers,
  ]
 
\nextgroupplot[
    xmin=-0.03, xmax=2.04,
    ymin=-5.2,  ymax=5.2,
    xtick={0,0.5,...,2},    xticklabels={0.0,0.5,1.0,1.5,2.0},
    ytick={-5,-4.5,...,5},  yticklabels={,,-4,,,,-2,,,,0,,,,2,,,,4,,},
    xlabel={$n$}, ylabel={$m$},
  ]
  \addplot[fill=wedgefill, draw=none, forget plot, on layer=axis background]
    coordinates {(0,0) (5.2/3,5.2) (2.04,5.2) (2.04,-5.2) (5.2/3,-5.2)} -- cycle;
  \addplot[blue, very thick, domain=0:2.1, samples=2] {3*x};
  \addplot[red,  very thick, domain=0:2.1, samples=2] {-3*x};
  \node[blue, anchor=east] at (axis cs:1.3,4.6) {$y=0$};
  \node[red,  anchor=east] at (axis cs:1.3,-4.6) {$x=0$};
  \addplot[only marks, mark=*, mark size=1.3pt, black] coordinates {
    (0,0)
    (0.5,-1.5) (0.5,-0.5) (0.5,0.5) (0.5,1.5)
    (1,-3) (1,-2) (1,-1) (1,0) (1,1) (1,2) (1,3)
    (1.5,-4.5) (1.5,-3.5) (1.5,-2.5) (1.5,-1.5) (1.5,-0.5)
    (1.5,0.5) (1.5,1.5) (1.5,2.5) (1.5,3.5) (1.5,4.5)
    (2,-5) (2,-4) (2,-3) (2,-2) (2,-1) (2,0)
    (2,1) (2,2) (2,3) (2,4) (2,5)
  };
 
\nextgroupplot[
    xmin=-0.4, xmax=8,
    ymin=-0.4, ymax=8,
    xtick={0,1,...,8}, xticklabels={0,,2,,4,,6,,8},
    ytick={0,1,...,8}, yticklabels={0,,2,,4,,6,,8},
    xlabel={$x$}, ylabel={$y$},
  ]
  \addplot[fill=quadfill, draw=none, forget plot, on layer=axis background]
    coordinates {(0,0) (8,0) (8,8) (0,8)} -- cycle;
  \addplot[red,  very thick] coordinates {(0,-0.4) (0,8)};  
  \addplot[blue, very thick] coordinates {(-0.4,0) (8,0)};  
  \addplot[only marks, mark=*, mark size=1.3pt, black] coordinates {
    (0,0) (0,3) (0,6)
    (1,2) (1,5) (1,8)
    (2,1) (2,4) (2,7)
    (3,0) (3,3) (3,6)
    (4,2) (4,5) (4,8)
    (5,1) (5,4) (5,7)
    (6,0) (6,3) (6,6)
    (7,2) (7,5) (7,8)
    (8,1) (8,4) (8,7)
  };
 
\end{groupplot}
\end{tikzpicture}
    \caption{BPS states among the AdS scalar modes of the $S_3$ giant for $Y^{2,1}$ represented in the $(m,n)$ and $(x,y)$ lattice. The allowed states have either $m,n\in\mathbb Z$ or $m,n\in\mathbb Z+\fft12$. The states populate a $\mathbb Z_3$ orbifold of the lattice. Note that the allowed states are either at integer values of $(m,n)$ or at the half integer values.}
    \label{fig: y21 lattice}
\end{figure}

To highlight the orbifold sum, we consider two simple cases, namely $Y^{2,1}$ (with $p \pm q$ odd) and $Y^{3,1}$ (with $p\pm q$ even), focusing on the $S_3$ giant localized at $y_1$. For $Y^{2,1}$, the zero mode states satisfy
\begin{equation}
    n\in\fft12\mathbb Z,\qquad m-3n\in\mathbb Z,
\end{equation}
which is equivalent to taking
\begin{equation}
    m,n\in\mathbb Z\qquad\mbox{or}\qquad m,n\in\mathbb Z+\fft12.
\end{equation}
On top of this quantization condition, the BPS states lie in the wedge defined by
\begin{equation}
    0\le|m|\le3n.
\end{equation}
This is shown in the left panel in Figure~\ref{fig: y21 lattice}.  The corresponding lattice in the $x$-$y$ plane with
\begin{equation}
    x=3n+m,\qquad y=3n-m,
\end{equation}
is shown in the right panel in the same figure.

For $Y^{3,1}$, on the other hand, the zero mode states are specified by
\begin{equation}
    n\in\fft12\mathbb Z,\qquad m-4n\in\mathbb Z,
\end{equation}
which can be stated simply as
\begin{equation}
    m\in\mathbb Z,\qquad n\in\fft12\mathbb Z.
\end{equation}
For $Y^{3,1}$ the BPS condition reads
\begin{equation}
    0\le|m|\le4n.
\end{equation}
%

Using (\ref{eq: ypq1 scalar charges}), we can read off the charges
\begin{align}
    F_A&=\fft12x-\fft12y, \quad
    F_B=\fft{p}{2\tilde c}x+\fft{p}{2\tilde c}y,\quad
    r=\fft{2\tilde c+q}{2\tilde c}x+\fft{2\tilde c+q}{2\tilde c}y,
\end{align}
where we recall that $r$ is the fake $R$-charge.  Furthermore, all these excitations are in the zero baryon number sector, $B=0$.  We can now compute the  contribution to the single letter index coming from AdS scalars (see Table \ref{tab: ypq1 AdS})
\begin{equation}
    f(\mathfrak p, \mathfrak q;v_i) = \sum_{\mathrm{BPS}}\left(\mathfrak p^{j_1} \mathfrak q^{j_2} \prod_iv_i^{R_i}\right)=\sum_{\mathrm{BPS}}\left(\mathfrak p^{j_1} \mathfrak q^{j_2} \rho^{\fft32r} u^{2F_A} v^{2F_B} \zeta^B\right),
\end{equation}
where we have used the fugacity map, (\ref{eq:fugmap}).  For the $S_3$ giant at $\tilde y=y_1$, we find
\begin{equation}
    S_3:\quad\prod v_I^{R_I}=(\rho^{\fft34(2p+q)}u^{(p+q)}v^{-p})^{\fft{x}{p+q}}(\rho^{\fft34(2p+q)}u^{-(p+q)}v^{-p})^{\fft{y}{p+q}}=(w_{3-\fft12})^{\fft{x}{p+q}}(w_{3+\fft12})^{\fft{y}{p+q}}.
\end{equation}
Hence
\begin{equation}
    f^{(3)}_{\mathrm{AdS}}(\mathfrak p,\mathfrak q;v_i)=\sum_{\genfrac{}{}{0pt}{1}{x,y\in Z_{\ge0}}{x+y\equiv0\pmod{p+q}}}(\mathfrak p+\mathfrak q)(w_{3-\fft12})^{\fft{x}{p+q}}(w_{3+\fft12})^{\fft{y}{p+q}}.
\end{equation}
This orbifold sum can be rewritten as a sum over all $x,y\in\mathbb Z_{\ge0}$ projected onto the sector with $x+y\equiv0\pmod{p+q}$.  In other words,

\begin{align}
    f^{(3)}_{\mathrm{AdS}} (\mathfrak p, \mathfrak q;v_i) &= \fft1{p+q}\sum_{n=0}^{p+q-1}(\mathfrak p + \mathfrak q)\sum_{x,y\ge0}\omega_3^{n(x+y)}(w_{3-\fft12})^{\fft{x}{p+q}}(w_{3+\fft12})^{\fft{y}{p+q}}\nn\\
    &=\fft1{p+q}\sum_{n=0}^{p+q-1}\fft{\mathfrak p + \mathfrak q}{\left(1-\omega_3^n\,(w_{3-\fft12})^{\fft1{p+q}}\right)\left(1-\omega_3^n\,(w_{3+\fft12})^{\fft1{p+q}}\right)},
\end{align}
where $\omega_3=e^{2\pi i/(p+q)}$ is a primitive $(p+q)$-th root of unity.

For the $S_1$ giant at $\tilde y=y_2$, we find on the other hand (see Table \ref{tab: ypq1 AdS})
\begin{equation}
    S_1:\quad\prod v_I^{R_I}=(\rho^{\fft34(2p-q)}u^{(p-q)}v^{p})^{\fft{x}{p-q}}(\rho^{\fft34(2p-q)}u^{-(p-q)}v^{p})^{\fft{y}{p-q}}=(w_{1+\fft12})^{\fft{x}{p-q}}(w_{1-\fft12})^{\fft{y}{p-q}}.
\end{equation}
The contribution to the single letter index is given by the $\mathbb Z_{p-q}$ orbifold sum
\begin{equation}
    f^{(1)}_{\mathrm{AdS}}(\mathfrak p, \mathfrak q;v_i)=\fft1{p-q}\sum_{n=0}^{p-q-1}\fft{\mathfrak p + \mathfrak q}{\left(1-\omega_1^n(w_{1+\fft12})^{\fft1{p-q}}\right)\left(1-\omega_1^n(w_{1-\fft12})^{\fft1{p-q}}\right)},
\end{equation}
where now we denoted $\omega_1=e^{2\pi i/(p-q)}$. 

Comparing the results for $S_3$ and $S_1$, the contribution of the AdS scalars to the single letter index of the class one giants can be written in the general form
\begin{equation}
    f^{(I)}_{\mathrm{AdS}} = \dfrac{1}{k_I}\sum_{n = 0}^{k_I - 1} \dfrac{\mathfrak p +\mathfrak q}{\left(1-w_{(n)}\right)\left(1-w_{(n)}'\right)},  \quad I =1,3.\label{eq: ypq1 index generic}
\end{equation}
where we have introduced the notation
\begin{equation}
    w_{(n)} = \omega_I^n (w_{I+\frac{1}{2}})^{1/k_I}, \qquad w_{(n)}' = \tilde\omega_I^n (w_{I-\frac{1}{2}})^{1/k_I}. \label{eq: ypq1 index generic1}
\end{equation}
Here $\omega_I,\tilde\omega_I$ are the primitive $k_I$-th roots of unity and for the class one giants we need $\tilde\omega_I = \omega_I$. This puts it in the toric notation that was discussed in the beginning of section~\ref{sec:Ypq-review}, suitable for comparison to field theory results.

\subsubsection*{Scalar fluctuations into \texorpdfstring{$Y^{p,q}$}{Y(p,q)}}
The remaining two scalars are internal fluctuation into $Y^{p,q}$, and are described by the quadratic action,
\begin{equation}
    S^{Y^{p,q}}_2 =  \dfrac{L^3 T_3}{6} |\tilde y| (1-\tilde y) \int d^4 \xi \sin\theta \biggl[L^2 |\dot W|^2 - |\mathcal{D}W|^2 - M_0^2 |W|^2 + 4L \, \mathrm{Im}(\bar W \dot W)\biggr]. \label{eq: ypq fluctuations}
\end{equation}
The spectrum of the twisted Laplacian associated with the covariant derivative $\mathcal{D}_i$ on this Lens space is studied in detail in Appendix \ref{app: internal modes}. We see that the field $W$ can be expanded in twisted harmonics,
\begin{equation}
    f_{l m n}(z,\phi,\tilde\alpha) = e^{im\phi} e^{2in\alpha/\ell} e^{-i\tilde c n \psi} z^{\frac{1}{2}|m- \tilde c n - \sigma|} (1-z)^{\frac{1}{2}|m +\tilde c n +\sigma|}{}_2F_1(a,b;c;z), \label{eq: twisted harmonics}
\end{equation}
with $\sigma = \pm1$, $a,b,c$ are given in the Appendix and we assume the familiar Lens space quantization conditions for $m,n$. The eigenvalues of these states are given by,
\begin{equation}
    2\tilde y \tilde c \ell \, \Lambda_{lmn} = 4l(l+1) - 4(\tilde c n + \sigma)^2 + 2 \tilde y \tilde c \ell \left(\dfrac{n}{\tilde y \ell} + 2\sigma\right)^2,
\end{equation}
where
\begin{equation}
    l = j + \max(|m|,|\tilde c n + \sigma|), \quad j = 0,1,2,\ldots.
\end{equation}
The internal modes can be expanded in these twisted harmonics,
\begin{align}
    W = \sum_{lmn}W_{lmn} f_{lmn}, \quad W_{lmn} = X^{(a)}_{lmn} + iX^{(b)}_{lmn},
\end{align}
in which case the action \eqref{eq: ypq fluctuations} reduces to
\begin{align}
    S_2^{\mathrm{Y^{p,q}}} = \dfrac{T_3L^3}{3} |\tilde y| (1-\tilde y ) \int dt \sum_{l m n} \biggl[&\dfrac{L^2}{2} (\dot X^{(a)}_{lmn}{}^2 + \dot X^{(b)}_{lmn}{}^2) - \dfrac{1}{2}(\Lambda_{lmn} + M_0^2)(X^{(a)}_{lmn}{}^2 + X^{(b)}_{lmn}{}^2) \nn \\
    &+ 2 L (X^{(a)}_{lmn}\dot X^{(b)}_{lmn} - X^{(b)}_{lmn}\dot X^{(a)}_{lmn})\biggr].
\end{align}
This is again a simple quantum mechanical system that can be canonically quantized using the same procedure as for $T^{1,1}$. Without elaborating further, in terms of the number operators in the helicity basis, we can write the Hamiltonian as
\begin{equation}
    H = \sum_{lmn} \sqrt{\Lambda_{lmn} + M_0^2 + 4} \,(N_{(+)lmn} + N_{(-)lmn})-2(N_{(+)lmn} - N_{(-)lmn}),\label{eq: ypq internal hamiltonian}
\end{equation}
where we have dropped the zero point energy of the oscillator modes. It is interesting to note that the factor in the square root reduces to,
\begin{equation}
    \Lambda_{lmn} + M_0^2 + 4 = \dfrac{2}{\tilde y \tilde c \ell} (l(l+1) - (\tilde c n)^2) + \left(\dfrac{n}{\tilde y \ell}\right)^2 + 1.
\end{equation}
From \eqref{eq: scalar ypq spectrum raw} and \eqref{eq:goodLlmn} it follows that this is exactly the $(\Lambda_{lmn} + 1)$ factor which appeared in the AdS modes above. For later convenience, with a little algebra we can rewrite this as,
\begin{equation}
    \Lambda_{lmn} + M_0^2 + 4 = \left(\left|\dfrac{n}{\tilde y \ell}\right| - 1 \right)^2 + \dfrac{1}{2\tilde y \tilde c \ell} \left((2l + 1)^2 - (2|\tilde c  n| - 1)^2\right). \label{eq: ypq spectrum convenient}
\end{equation}
Since these fluctuations are in $Y^{p,q}$ they do not carry any AdS angular momentum $(J_1 = J_2 = 0)$. They do, however, carry the $r$ charge and the mesonic charges. From \eqref{eq:charges} and \eqref{eq: twisted harmonics} we find 
\begin{align}
    F_A &= \sum_{lmn} m (N_{(+)lmn} + N_{(-)lmn}), & &r = -\sum_{\sigma = \pm}\sum_{lmn}[(2\tilde c +q)n + 2\sigma] \, N_{(\sigma)lmn}, \nn\\
    F_B &= -\sum_{lmn} pn (N_{(+)lmn} + N_{(-)lmn}), & &\dfrac{3}{2} r_* = -\sum_{\sigma = \pm}\sum_{lmn} \left(\dfrac{n}{\tilde y \ell} + 3\sigma \right) N_{(\sigma)lmn}.\label{eq: ypq1 internal charges}
\end{align}
From \eqref{eq: ypq internal hamiltonian}, \eqref{eq: ypq spectrum convenient} and the form of the true $R$-charge \eqref{eq: ypq1 internal charges}, it immediately follows that the `$-$' states $(\sigma = -1)$ can never saturate the BPS bound $\mathcal H \geq 0$. For the `$+$' modes $(\sigma = +1)$, \eqref{eq: ypq spectrum convenient} shows that the BPS bound is saturated precisely when $\tilde c n \le -1$ and $l = - \tilde c n -1$ with $|m| \le l$. These conditions imply $n/\tilde y \leq 0$. These are again the zero modes $(j = 0)$ and their quantum numbers are summarized in Table \ref{tab: ypq1 internal}.
\begin{table}[t]
    \centering
    \begin{tabular}{c c c c c c c}
        \toprule
        {} & $H$ & $r_*$& $r$ &$F_A$& $F_B$ &BPS? \\
        \midrule
        $X_{(+)lmn}$ & $|n/\tilde y\ell| - 3$  & $- (2n/3\tilde y\ell) - 2$ & $-(2\tilde c + q)n - 2$& $m$ & $-pn$ & \checkmark\\
        $X_{(-)lmn}$ & $\sqrt{\Lambda_{lmn}+M_0^2+4}+2$  & $- (2n/3\tilde y\ell) + 2$ & $-(2\tilde c + q)n + 2$&$m$ & $-pn$ & {} \\
        \bottomrule
    \end{tabular}
    \caption{Quantum numbers of the internal fluctuations of the class one giants in $Y^{p,q}$. The $X_{(+)}$ entries are evaluated on the BPS states, $\tilde cn\le-1$, $l=-\tilde cn-1$, $|m|\le l$. For $X_{(-)}$ the harmonics have $l = j+\max(|m|,|\tilde c n-1|)$, so these modes satisfy $\mathcal H>0$ and are not BPS.}
    \label{tab: ypq1 internal}
\end{table}

We now turn to constructing the single letter index for the internal modes. Compared to the discussion for the AdS scalars above, here the BPS modes satisfy $l = -\tilde c n - 1$ with $\tilde c  n \le -1$, which changes the bounds of the summation over $m$. For the $S_3$ and $S_1$ giants, we summarize this as follows:
\begin{align}
    \mbox{$S_3$ giant at $y_1$}:&&n\ge0,&\qquad |m|\le(p+q)n-1,\kern2.7em m-(p+q)n\in\mathbb Z,\qquad n\in\fft12\mathbb Z,\nn\\
    \mbox{$S_1$ giant at $y_2$}:&&n\le0,&\qquad |m|\le-(p-q)n - 1,\qquad m+(p-q)n\in\mathbb Z,\qquad n\in\fft12\mathbb Z.\nn
\end{align}
With the same definition of $x$, $y$ as in \eqref{eq: ypq1 xy lattice}, we see the $x$-$y$ lattice remains the same as \eqref{eq: ypq1 xy lattice cndn} with the modification $x,y \geq 1$.

From \eqref{eq: ypq1 internal charges}, in terms of $x$ and $y$ we can write the charges as,
\begin{align}
    F_A&=\fft12x-\fft12y, \quad
    F_B=\fft{p}{2\tilde c}x+\fft{p}{2\tilde c}y,\quad
    r=\fft{2\tilde c+q}{2\tilde c}x+\fft{2\tilde c+q}{2\tilde c}y-2.
\end{align}
Recall that these excitations do not carry any baryon charges. Following our discussion on the AdS scalars, the contribution to the single letter index from the internal scalar modes can be expressed as an orbifold sum. For the $S_3$ giant, this is given by
\begin{align}
    f^{(3)}_{Y^{p,q}} (\mathfrak p, \mathfrak q;v_i) &= \fft1{p+q}\sum_{n=0}^{p+q-1}\frac{1}{(\mathfrak p \mathfrak q)}\sum_{x,y\ge1}\omega_3^{n(x+y)}(w_{3-\fft12})^{\fft{x}{p+q}}(w_{3+\fft12})^{\fft{y}{p+q}}\nn\\
    & =\fft1{p+q}\sum_{n=0}^{p+q-1}\fft{\omega_3^{2n}\,(w_{3-\fft12}w_{3+\fft12})^{\fft1{p+q}}/(\mathfrak p \mathfrak q)}{\left(1-\omega_3^n\,(w_{3-\fft12})^{\fft1{p+q}}\right)\left(1-\omega_3^n\,(w_{3+\fft12})^{\fft1{p+q}}\right)}.
\end{align}
Similarly, for the $S_1$ giant one obtains a sum with appropriate edge fugacities, but over the $\mathbb Z_{p-q}$ orbifold. We conclude that, using \eqref{eq: ypq1 index generic1}, the generic form of the index from the internal modes takes the form
\begin{align}
    f^{(I)}_{Y^{p,q}} (\mathfrak p, \mathfrak q;v_i) =\fft1{k_I}\sum_{n=0}^{k_I-1}\fft{(w_{(n)} w'_{(n)}/\mathfrak p \mathfrak q)}{\left(1-w_{(n)}\right)\left(1-w_{(n)}'\right)},  \quad I =1,3. \label{eq: ypq1 index int generic}
\end{align}
%
\subsubsection{Vector fluctuations}
\label{sec:class1vectors}

We now turn to the worldvolume vector fluctuations.
Since our giants are D3 branes, these fluctuations are described by the standard Maxwell theory on this lens space background. Similar to the case of $T^{1,1}$, we choose a Coulomb gauge which leaves us with the transverse polarizations given by the action \eqref{eq: t11 vect action} but with the lens space metric. The two polarizations $(\pm)$ transform in the $l$ of $SU(2)_l$ with appropriate $U(1)_r$ quantum numbers. One can perform a mode expansion in these harmonics.
Taking into account the overall volume factor of $2\tilde y \tilde c \ell$, the spectrum can be written as
\begin{align}
    \Lambda^{(\pm)}_{lmn} &= \dfrac{1}{2\tilde y \tilde c \ell} \left((2l+1)^2 - (2|n\tilde c| - 1)^2\right) + \left(\left|\dfrac{n}{\tilde y \ell} \right|-1\right)^2 + 1\nn\\
    & \mp 2 \sqrt{ \dfrac{1}{2\tilde y \tilde c \ell} \left((2l+1)^2 - (2|n\tilde c| - 1)^2\right) + \left(\left|\dfrac{n}{\tilde y \ell} \right|-1\right)^2},\nn\\
    \text{where }&\quad |\tilde c n| \leq l\mp1, \quad |m| \leq l, \quad l = j + |m|, \quad j = 0,1,2,\ldots
    , \label{eq: ypq vecctor spectrum}
\end{align}
where $m,n$ follow the quantization conditions,
\begin{equation}
    m + \tilde c n \in \mathbb Z, \quad  n \in \frac{1}{2} \mathbb Z.
\end{equation}
We can expand the Maxwell action in the Coulomb gauge in these lens space vector harmonics and integrate out the spatial part. The resulting system can be canonically quantized to give
\begin{equation}
    H =  \sum_{l mn}\left(\sqrt{\Lambda^{(+)}_{l mn}} \, n^{(+)}_{l mn}+ \sqrt{\Lambda^{(-)}_{l mn}} \, n^{(-)}_{l mn}\right).
\end{equation}
The longitudinal modes do not carry any AdS angular momenta $(J_1 = J_2 = 0)$ but carry the mesonic and R-charges,
\begin{align}
    F_A &=  \sum_{lmn} m\, (n^{(+)}_{lmn} + n^{(-)}_{lmn}), &  &r = -\sum_{lmn} (2\tilde c + q)n (n^{(+)}_{lmn} + n^{(-)}_{lmn}),\nn\\
    F_B &= - \sum_{lmn} pn\, (n^{(+)}_{lmn} + n^{(-)}_{lmn}), & &\dfrac{3}{2}r_* = - \sum_{lmn} \dfrac{n}{\tilde y \ell} (n^{(+)}_{lmn} + n^{(-)}_{lmn}).
\end{align}
From \eqref{eq: ypq vecctor spectrum} we see that the modes in the `$+$' polarization do not saturate the BPS bound, but for the `$-$' polarization, the energy of the states with $|\tilde c n| = l + 1$ reduces to
\begin{equation}
    \sqrt{\Lambda^{(-)}_{lmn}} = \left|\dfrac{n}{\tilde y \ell}\right|.
\end{equation}
It follows that these states are BPS when $n/\tilde y  \leq 0$.

Since the BPS modes satisfy $l = |\tilde c n| - 1$, in terms of the $x$, $y$ quantum numbers, the single letter index is a sum over the orbifold lattice with $x,y\geq 1$. This is very similar to what we have seen for the internal modes. Without going into further details, the contribution to the single-letter index takes the general form
\begin{align}
    f_I^{\mathrm{vector}} (\mathfrak p, \mathfrak q;v_i) =\fft1{k_I}\sum_{n=0}^{k_I-1}\fft{w_{(n)}w_{(n)}'}{\left(1-w_{(n)}\right)\left(1-w_{(n)}'\right)},  \quad I =1,3,\label{eq: eq: ypq1 index vector generic}
\end{align}
where $w_{(n)}$ and $w'_{(n)}$ are defined in (\ref{eq: ypq1 index generic1}).

\subsubsection{Fermion fluctuations}
\label{sec:ypqfermions}

Finally, we consider the fluctuations of the worldvolume fermions.
The reduction of the D3-brane fermionic action was discussed in detail for the case of $T^{1,1}$ in Section~\ref{sec:fermions}. Following the conventions of \cite{Deddo:2025lfm}, the fermion action for class one giants reduces exactly to the form in \eqref{eq: fermion action}, but with the world volume given by the Lens space \eqref{eq:lsm} and $\fsl\nabla$ the Dirac operator on this background. 

The spectrum of the Dirac operator on a squashed three sphere is obtained in Appendix~\ref{app:spinor}. Comparing the class one world volume metric (\ref{eq:lsm}) with the squashed metric (\ref{eq:squashed}), and using the lens space metric (\ref{eq:Lp1met}) we identify $p = \tilde c$, $\tilde\psi \sim \psi_{\mathrm{eff}}$ and the squashing parameter 
\begin{equation}
    e^{2\lambda}\equiv2\tilde y\tilde c\ell,
\end{equation}
together with an overall volume factor which is also equal to $2\tilde y\tilde c\ell$. 

Here it is convenient to work with the unwrapped fiber coordinate $\psi_{\mathrm{eff}} =\tilde\alpha /\tilde c$, so that the orbital $U(1)$ weight is $\tilde cn$.  As discussed above for the internal $Y^{p,q}$ scalar fluctuations and in more detail in Appendix~\ref{app: internal modes}, the normal bundle of the giant is non-trivially fibered over the base (see \eqref{eq: ypq1 nontrivial fiber}). The connection is given by \eqref{eq: ypq cov der} which can be rewritten as
\begin{equation}
    \mathcal A=-2\tilde y \tilde c \ell \,\hat e^3+\cos\theta\,d\phi,\qquad \hat e^3\equiv d\psi_{\mathrm{eff}}+\cos\theta\,d\phi.
\label{eq:ypqAsplit}
\end{equation}
This makes it different from the case of $T^{1,1}$, where the normal bundle connection was proportional to the fiber one-form alone.  Here it splits into a piece along the fiber and a genuine monopole term $\cos\theta\,d\phi$ on the $(\theta,\phi)$ sphere. The two pieces act differently; the monopole shifts the \emph{angular} weight, which controls the range of the harmonics, while the fiber piece shifts the \emph{fiber} weight appearing in the squashing term. 

To make this clearer, we look back to the Dirac equation, \eqref{eq:2x2mat}. Let $\Sigma$ denote the total $U(1)$ fiber weight, that is the orbital weight $\tilde cn$ together with the frame spin $s_3$ carried by the spinor and $\sigma$ be the normal plane charge. This amounts to replacing $k_3 \to \Sigma$ in \eqref{eq:2x2mat}. The fiber piece of the connection enters the equation through a term proportional to $\sigma^3$ weighted by $\sigma (2\tilde y \tilde c \ell)$, which shifts only the diagonal of the $2 \times 2$ block. This is exactly what we saw in the case of $T^{1,1}$. The corresponding modification to the spectrum can be obtained from \eqref{eq:evD} by the replacement,
\begin{equation}
    \Sigma \to \check\Sigma=\Sigma+ \sigma\,(2\tilde y \tilde c \ell)\qquad\mbox{(fiber),} \label{eq: dirac mod1}
\end{equation}
on the squashing term alone. On the other hand, the monopole term enters the Dirac equation through $\sigma^\pm$ shifting the off diagonal terms. This can be accounted for by replacing $\Sigma$ in the non-squashing term by
\begin{equation}
    \Sigma \to \hat\Sigma=\Sigma+\sigma\qquad\mbox{(angular)}.\qquad\qquad \label{eq: dirac mod2}
\end{equation}
For the internal scalars, we have $s_3=0$ and $\sigma=\pm1$; this reproduces the twisted harmonics of Appendix~\ref{app: internal modes}.  For the spinors we have instead $\sigma = Q = \pm\ft12$.

With the modification to the spectrum due to \eqref{eq: dirac mod1} and \eqref{eq: dirac mod2} and restoring the overall volume factor, the fermionic Hamiltonian takes the form
\begin{equation}
    H=\fft12\pm2\sqrt{\fft{j(j+1) + \ft14-\hat\Sigma^2}{2\tilde y \tilde c \ell} + \fft{\check\Sigma^2}{(2\tilde y \tilde c \ell)^2}}\;-2Q,
\label{eq:ypqfermH}
\end{equation}
with the ranges
\begin{equation}
    \mbox{`$+$':}\quad|\hat\Sigma|\le j+\ft12,\qquad \mbox{`$-$':}\quad|\hat\Sigma|\le j-\ft12,\qquad |m|\le j.
\end{equation}
The charges follow from (\ref{eq:charges}) and (\ref{eq:rstar}) in the same way as for the scalars and vectors, with the frame spin included in $\Sigma$,
\begin{equation}
    \ft32r_*=-\left(\fft{\Sigma}{\tilde y \tilde c \ell}+3Q\right),\quad F_A=m,\quad F_B=-\fft{p\,\Sigma}{\tilde c},\quad r = -\dfrac{(2\tilde c + q)}{\tilde c} \Sigma - 2 Q,
\end{equation}
together with $J_1,J_2=\pm\ft12$ from the spinor representations in AdS${}_5$

We are now in a position to identify the BPS modes. For the `$+$' branch, note that for the highest state with $|\hat \Sigma| = j + \frac{1}{2}$, the Hamiltonian reduces to
\begin{equation}
    H  = \dfrac{1}{2} + \left| \dfrac{\check \Sigma}{\tilde y \tilde c \ell} \right| - 2 Q.
\end{equation}
This gives us
\begin{equation}
    \mathcal{H} = \left| \dfrac{\check \Sigma}{\tilde y \tilde c \ell} \right| + \dfrac{\check \Sigma}{\tilde y \tilde c \ell} + \dfrac{1}{2} - Q - J_1 - J_2.
\end{equation}
The BPS condition $\mathcal H=0$ therefore requires $J_1+J_2=\ft12-Q$ and $\check \Sigma/(\tilde y \tilde c \ell) \le 0$, giving one tower for $Q=-\ft12$ and two for $Q=+\ft12$. 

On the `$-$' branch, consider the states with $|\hat \Sigma| = j - \frac{1}{2}$. For the quantity under the square root of \eqref{eq:ypqfermH} to be a perfect square, we require the correlated choice $\hat\Sigma=\pm(j-\ft12)$ with $Q=\mp\ft12$, in which case we have,
\begin{equation}
    H = - \left|\dfrac{j}{\tilde y \tilde c \ell} + 1\right| + \dfrac{1}{2} - 2Q.
\end{equation}
These states have negative energy; they are the filled Dirac-sea modes, and the physical excitations are the corresponding antiparticles, with all charges reversed.  Taking $\hat\Sigma=j-\ft12$ and $Q=-\ft12$, the antiparticle has $Q=+\ft12$ and 
\begin{equation}
    H  = \left|\dfrac{j}{\tilde y \tilde c \ell} + 1\right| - \dfrac{3}{2} = \dfrac{j}{\tilde y\tilde c\ell}-\dfrac12 = \left|\dfrac{\check\Sigma}{\tilde y \tilde c \ell}\right| + \dfrac{1}{2},
\end{equation}
These states saturate the BPS bound for $J_1=J_2=+\ft12$ and $\check \Sigma/(\tilde y \tilde c \ell) \le 0$. Since the `$-$' branch requires $j\ge\ft12$, this tower is missing its lowest state. When we calculate the single letter index below, we will see that subtracting out this contribution is necessary to avoid overcounting the BPS states. The four BPS towers are collected in Table~\ref{tab: D3 fermions ypq}.

\begin{table}[t]
    \centering
    \begin{tabular}{c c c c c c c c}
        \toprule
        {} & branch & $Q$ & $H$ & $J_1$ & $J_2$ & $\ft32r_*$ & range\\
        \midrule
        $\psi_{(1)}$ & $+$ & $-\ft12$ & $|\check \Sigma/(\tilde y \tilde c \ell)|+\ft32$ & $\ft12$ & $\ft12$ & $-\check \Sigma/(\tilde y \tilde c \ell)+\ft12$ & $x,y\ge0$\\
        $\psi_{(2)}$ & $+$ & $+\ft12$ & $|\check \Sigma/(\tilde y \tilde c \ell)|-\ft12$ & $\ft12$ & $-\ft12$ & $-\check \Sigma/(\tilde y \tilde c \ell)-\ft12$ & $x,y\ge1$\\
        $\psi_{(3)}$ & $+$ & $+\ft12$ & $|\check \Sigma/(\tilde y \tilde c \ell)|-\ft12$ & $-\ft12$ & $\ft12$ & $-\check \Sigma/(\tilde y \tilde c \ell)-\ft12$ & $x,y\ge1$\\
            $\psi_{(4)}$ & $-$ & $+\ft12$ & $|\check \Sigma/(\tilde y \tilde c \ell)|+\ft12$ & $\ft12$ & $\ft12$ & $-\check \Sigma/(\tilde y \tilde c \ell)- \ft12$ & $x,y\ge0$, $x+y>0$\\
        \bottomrule
    \end{tabular}
    \caption{The four BPS towers of fermionic fluctuations for the class one giant. The last entry is the antiparticle of a `$-$' branch mode; the restriction $j\ge\ft12$ on that branch removes its lowest state.}
    \label{tab: D3 fermions ypq}
\end{table}

It remains to assemble the contribution to the single-letter index.  Exactly as for the scalars, we introduce
\begin{equation}
    x=m-\Sigma,\qquad y=-m-\Sigma,\qquad x+y=-2\Sigma.
\end{equation}
Since the $\mathbb Z_{|\tilde c|}$ quotient acts along the fiber, it acts on a spinor through the \emph{total} weight $\Sigma$ rather than the orbital weight alone; this is also what keeps both states of a given $2\times2$ block in the spectrum.  The lens space projection is therefore the same as in the bosonic sector, $x+y\equiv0\pmod{|\tilde c|}$. Using the bound on the quantum numbers $m$, $\hat \Sigma$ and using $\Sigma=\hat\Sigma-Q$, we find the range of allowed lattice points in $x$, $y$. This is summarized in Table~\ref{tab: D3 fermions ypq}.

Finally, using the fugacity map (\ref{eq:fugmap}) together with $\rho^3 = \mathfrak p \mathfrak q$ we can compute the index. Note that a fermionic letter of normal charge $Q$ carries the extra factor $(\mathfrak p \mathfrak q)^{-Q}$ relative to a bosonic one. Then for the $S_3$ giant
\begin{equation}
    \prod_Iv_I^{R_I }\Big|_{\mathrm{fermion}}=\mathfrak p^{J_1}\mathfrak q^{J_2}(\mathfrak p \mathfrak q)^{-Q}(w_{3-\fft12})^{\fft x{p+q}} (w_{3+\fft12})^{\fft y{p+q}}.
\end{equation}
The case of $S_1$ giant also follows similarly with the replacement $p + q \to p - q$ and $w_{3\pm\fft12} \to w_{1\pm\fft12}$.

In general, in terms of $w_{(n)},w'_{(n)}$ defined in \eqref{eq: ypq1 index generic1} and performing the sums over the ranges in Table~\ref{tab: D3 fermions ypq}, the four towers contribute $\mathfrak p \mathfrak q$, $w_{(n)}w'_{(n)}/\mathfrak q$, $w_{(n)}w'_{(n)}/\mathfrak p$ and $1/(1-w_{(n)})(1-w'_{(n)})-1$ respectively, so that
\begin{equation}
    f_{\mathrm{fermion}}= \dfrac{1}{k_I}\sum_{n = 0}^{k_I - 1}\fft{w_{(n)}+w'_{(n)}-w_{(n)} w'_{(n)} +\mathfrak p \mathfrak q+\fft{w_{(n)} w'_{(n)}}{\mathfrak p \mathfrak q}(\mathfrak p+ \mathfrak q)}{(1-w_{(n)})(1-w'_{(n)})},\quad I = 1,3.
\label{eq:ypqffermion}
\end{equation}
As in the $T^{1,1}$ case, the term $-w_{(n)}w'_{(n)}$ in the numerator is the removal of the lowest state of the fourth tower, and prevents an over-counting of BPS states.  

\subsubsection{The single-letter index for class one giants}

We are now ready to assemble the single-letter index for class one giants by adding together the AdS scalar \eqref{eq: ypq1 index generic}, internal scalar \eqref{eq: ypq1 index int generic}, vector \eqref{eq: eq: ypq1 index vector generic} and fermion \eqref{eq:ypqffermion} mode contributions. We have,
\begin{align}
    f^{(I)} &= f^{(I)}_{\mathrm{AdS}} + f^{(I)}_{Y^{p,q}} + f^{(I)}_{\mathrm{vector}} - f^{(I)}_{\mathrm{fermion}}\nn\\
     &= 1 - \dfrac{1}{k_I} \sum_{n = 0}^{k_I - 1}\dfrac{(1 - w_{(n)}w'_{(n)}/(\mathfrak p \mathfrak q))(1-\mathfrak p)(1 - \mathfrak q)}{(1-w_{(n)})(1-w'_{(n)})}, \quad I =1,3.
\label{eq:class1f}
\end{align}
We now see that this agrees with the dual field theory computation \eqref{eq:orbindex}. Here $I = 1,3$ corresponds to the single letters $f_{(1,0,0,0)}$ and $f_{(0,0,1,0)}$ in the standard notation. However, unlike the simpler spherical giants discussed in \cite{Deddo:2025lfm} or the giants in $T^{1,1}$ from the previous section, they are not related to each other by a simple permutation of the fugacities. 

We can recover the full index of these single giants by plethystic exponentiation
\begin{equation}
    \hat I^{\mathrm{GG}}_{(I)}=k_I\mathrm{PE}(f^{(I)}),
\label{eq:IGGPE}
\end{equation}
where the multiplicity $k_I$ is the vacuum degeneracy associated to the holonomy of flat connections of the D3-brane worldvolume gauge field.  We can similarly construct the index of multi giant configurations like $\hat I^{GG}_{(m_1,0,m_3,0)}$. However, as noted in the case for $T^{1,1}$, our approach so far is not sufficient to recover the index of intersecting configurations.

\subsection{Class two giants}

We now turn to the class two giants, $S_4$ and $S_2$, which are localized at $\theta=0$ or $\pi$ respectively and wrap the three cycle defined by $(y,\tilde\psi,\alpha)$ with metric
\begin{equation}
    ds^2 = \dfrac{dy^2}{w(y)q(y)} + \dfrac{q(y)}{9} d\tilde\psi^2 + w(y) (d\alpha + f(y) d\tilde\psi)^2.
\end{equation}
Here $\tilde\psi = \psi \mp \phi$, where the top sign corresponds to $\theta = 0$ and the bottom sign to $\theta=\pi$. The $(\tilde\psi,\alpha)$ is a  2-torus  in the interval $[y_1,y_2]$. Since $q(y_i) = 0$, one of the cycles degenerates at the end points. Hence this is topologically a lens space. We can show this more explicitly by viewing the metric as a $U(1)$ bundled over a topological $S^2$ given by $(y,\tilde\psi)$,
\begin{equation}
    ds^2 \sim ds^2_{S^2} + w(y) (d\alpha + A)^2,
\end{equation}
where $A = f(y) d\tilde \psi$.
Following a discussion in \cite{Gauntlett:2004yd}, we have
\begin{equation}
    \dfrac{1}{4\pi}\int_{S^2} dA = \dfrac{p\ell}{2}.
\end{equation}
So one could rewrite the metric in a more suggestive form,
\begin{equation}
    ds^2 \sim ds^2_{S^2} + \dfrac{w(y) \ell^2}{4} (d\hat{\alpha} + p\hat{A})^2
\end{equation}
where $\hat{\alpha} = 2\alpha/\ell$ and $\hat{A} = 2A/(p\ell)$. Written in this manner, the fiber coordinate has the period $\hat{\alpha} \sim \hat{\alpha} + 4\pi$. It is topologically a lens space and admits an inhomogeneously squashed metric, but it is not the homogeneous lens-space metric studied earlier for the class one giants because $w(y)$ is not constant.

\subsubsection{Scalar fluctuations}
Let us consider small transverse fluctuations of this giant. There are four AdS scalar modes and two internal $Y^{p,q}$ modes. As before, we introduce rectangular coordinates $(A,B)$ centered at the north (or south) pole of the $(\theta,\phi)$ two sphere and define the complex combination $Z = A+iB$. Let the AdS fluctuations be $\delta v_k = \epsilon v_k$ and the internal fluctuations be $\delta A = \epsilon A$ and $\delta B = \epsilon B$.  Then, to leading order, the components of the DBI action take the form,
\begin{align}
    \sqrt{-\det P[g]} = \dfrac{L^3}{3} &+ \dfrac{L^3}{3} \Biggl[\dfrac{L^2}{12} (y - 1) |\dot Z|^2 - \dfrac{1}{12} \mathrm{Re}(\bar Z \tilde\nabla^2 Z) + \dfrac{1}{2} \mathrm{Im} (\bar Z \partial_{\tilde\psi}Z)\Biggr]\nn\\
    &- \dfrac{L}{6}\sum_k\Bigl[ L^2 \dot v_k^2 + v_k \nabla^2 v_k -v_k^2\Bigr],
\label{eq:class2DBI}
\end{align}
whereas the Wess-Zumino term in an appropriate gauge choice is
\begin{equation}
    P[C_4] = \dfrac{L^4}{9} (y - 1) \mathrm{Im} (\bar Z \dot Z).
\end{equation}
Note that we have introduced two scalar Laplacians, $\tilde\nabla$ for the two $Y^{p,q}$ modes and $\nabla$ for the four AdS scalars; their explicit forms are: 
\begin{align}
\label{eq:case2Ylap}
    \tilde\nabla^2 f &= 2\partial_y[(a-y^2)q(y)\, \partial_y]f\nn\\
    &\quad+ \dfrac{(1-y)}{4 (a-y^2)q(y)} \Biggl((a-y^2) (6 \partial_{\tilde\psi} - \partial_\alpha)^2 + 2 (y-1)\partial_\alpha\Bigl((2y-1) \partial_\alpha - 12 y \partial_{\tilde\psi}\Bigr)\Biggr)f,\\
    \nabla^2 f &= 2\partial_y\Biggl[\dfrac{(a-y^2)q(y)}{(1-y)}\, \partial_y\Biggr]f,\nn\\
    &\quad+ \dfrac{1}{4 (a-y^2)q(y)} \Biggl((a-y^2) (6 \partial_{\tilde\psi} - \partial_\alpha)^2 + 2 (y-1)\partial_\alpha\Bigl((2y-1) \partial_\alpha - 12 y \partial_{\tilde\psi}\Bigr)\Biggr)f.
\label{eq:case2Alap}
\end{align}
Similar to what we saw for the class one, these differences arise because the giants are wrapped non trivially on the base. This is apparent for the internal fluctuations which probe this geometry, while the AdS modes takes a relatively simpler form. Hence we study them separately.

\subsubsection*{Scalar fluctuations into AdS}

The fluctuations into AdS, up to quadratic order are  described by the action
\begin{equation}
    S_2^{\mathrm{AdS}} = \dfrac{L T_3}{6} \int d^4\xi  \sum_k (L^2 \dot v_k^2 + v_k \nabla^2 v_k - v_k^2). \label{eq: ypq2 ads action}
\end{equation}
This is very similar in structure to what we had for $T^{1,1}$ and the class one giants but with a different Laplacian (\ref{eq:case2Alap}), requiring a more sophisticated harmonic analysis. In terms of the zeroes $y_i$ of $q(y)$, this Laplacian can be expressed as 
\begin{align}
    \nabla^2&=4\partial_y\fft{(y-y_1)(y-y_2)(y-y_3)}{1-y}\partial_y
   +\fft{1-y}{2(a-y^2)}\partial_\alpha^2\nn\\
   & + \fft{9(a-y^2)}{2(y-y_1)(y-y_2)(y-y_3)}\left(\partial_{\tilde\psi}-\fft{a-2y+y^2}{6(a-y^2)}\partial_\alpha\right)^2,
\end{align}
where we have factored the cubic polynomial $(a-y^2)q(y)$.  We can obtain the scalar spectrum by solving for the eigenfunctions of the Laplacian, $\nabla^2f = -\Lambda f$.  After separation of variables, we are left with an ODE of Fuchsian type with five regular singular points at $y=y_1,y_2,y_3,1$ and $\infty$. One can compute the indices, and it turns out that the point at $y=1$ is an apparent singularity as its indices are $0$ and $2$. Solving the indicial equation at the other three singular points suggests that we write
\begin{equation}
    f_{lmn}(y,\tilde\psi,\alpha)=\left(\prod_{i = 1}^3(y-y_i)^{r_i}\right)e^{im\tilde\psi}e^{in\alpha/\ell}g(y), \quad r_i = \pm \dfrac{1}{2} \left(m + \dfrac{1-y_i}{6\ell y_i} n\right),
\label{eq:Phiy}
\end{equation}
where, in terms of $p,q$ we have
\begin{equation}
    r_1=\pm\fft12\left(m-\fft{p+q}2n\right),\quad r_2=\pm\fft12\left(m+\fft{p-q}2n\right),\quad r_3=\pm\fft12\left(m+n\left(q-\fft1{2\ell}\right)\right).
\label{eq:indices}
\end{equation}
As the resulting ODE for $f(y)$ is of Heun class, we do not aim to obtain the entire spectrum.  Instead, taking motivation from the analysis of $T^{1,1}$ and the class one giants in $Y^{p,q}$, we look for solutions of the special form $g(y)=1$, corresponding to zero mode states. These are expected to exhaust the BPS spectrum. In what follows we restrict our discussion to this sub-sector.  

In this case, we find that (\ref{eq:Phiy}) with $g(y)=1$ is indeed a solution, provided the $\pm$ signs in (\ref{eq:indices}) are all positive or all negative.  The corresponding eigenvalues are
\begin{equation}
    \Lambda_{mn}=\left(3m-\fft{n}{2\ell}\pm1\right)^2-1,
\label{eq:BPSeval}
\end{equation}
where all $\pm$ signs in (\ref{eq:Phiy}) and (\ref{eq:indices}) are correlated.

Restricting to the zero mode sector and integrating out the spatial parts, the quadratic action for the AdS fluctuations (\ref{eq: ypq2 ads action}) reduces to
\begin{equation}
    S^{AdS}_2 \supset \dfrac{L^3T_3}{6} \int dt \sum_{k,mn} \left[ \dot v_{(k)mn}^2 - \dfrac{(\Lambda_{mn}+1)}{L^2} v_{(k)mn}^2 \right],
\end{equation}
where we have only summed over the zero modes.  With an appropriate rescaling, $v_k \to \sqrt{3}\, v_k/L^2$, this system reduces to a simple harmonic oscillator. Following the same steps as before, we perform the standard canonical quantization to obtain,
\begin{equation}
    H = \sum_{mn}\sqrt{\Lambda_{mn}+1}\sum_i (N_{(i+)mn} + N_{(i-)mn}),
\label{eq:class2H}
\end{equation}
where the number operators $N_{(i\pm)mn}$ are in the helicity basis (similar to \eqref{eq: helicity basis}). From (\ref{eq:charges}) and (\ref{eq:rstar}), we can read off the charges carried by these modes
\begin{align}
    J_i &= \sum_{mn}( N_{(i+)mn} - N_{(i-)mn}),\nn \\
    F_A &= \mp\sum_i\sum_{mn} m(N_{(i+)mn} + N_{(i-)mn}),\nn\\
    F_B &= -\fft12\sum_i \sum_{mn}pn (N_{(i+)mn} + N_{(i-)mn}),\nn\\
    r &= \sum_i \sum_{mn}\left(2m-\fft12qn\right) (N_{(i+)mn} + N_{(i-)mn}),\nn\\
    \fft32r_* &= \sum_i\sum_{mn} \left(3m - \dfrac{n}{2\ell}\right) (N_{(i+)mn} + N_{(i-)mn}).
\label{eq:class2q}
\end{align}
Note that the sign on $F_A$ corresponds to the $\theta=0$ ($S_4$) versus $\theta=\pi$ ($S_2$) giant. From (\ref{eq:BPSeval}), along with the last equation in (\ref{eq:class2q}), we make the observation that in this restricted sector the spectrum is related to the true R-charge by
\begin{equation}
    \sqrt{\Lambda_{mn}+1} = \left|\fft32r_*\pm1\right|,
\label{eq:deltamn}
\end{equation}
where the sign is correlated with the sign of the exponents in the wavefunction, (\ref{eq:Phiy}) (this should not be confused with the sign of $F_A$ which is correlated with whether we consider the $S_2$ or $S_4$ giant), and where
\begin{equation}
    \fft32r_*=3m - \dfrac{n}{2\ell}.
\end{equation}
The supersymmetric Hamiltonian is given by \eqref{eq: SUSY H ypq}. Using \eqref{eq:class2H}, \eqref{eq:class2q} and \eqref{eq:deltamn}, $\mathcal{H}$ can be written as
\begin{align}
    \mathcal H=\sum_i&\sum_{mn}\Biggl[\left(\left|\fft32r_*\pm1\right|-\left(\fft32r_*+1\right)\right)N_{(i+)mn}\nn\\
    &\kern1em+\left(\left|\fft32r_*\pm1\right|-\left(\fft32r_*-1\right)\right)N_{(i-)mn}\Biggr].
\label{eq:susyHc2}
\end{align}
The regularity of the wavefunction, (\ref{eq:Phiy}), at the $y=y_1$ and $y=y_2$ poles demands $r_1\ge0$ and $r_2\ge0$.  For the top sign in (\ref{eq:indices}), this corresponds to a wedge
\begin{equation}
    m\ge\fft{p+q}2n\quad\mbox{and}\quad m\ge-\fft{p-q}2n,
\label{eq:mnwed}
\end{equation}
while for the bottom sign the inequalities are reversed.  The allowed range of quantum numbers is perhaps more easily stated in terms of the linear combinations
\begin{equation}
    x=m-\fft{p+q}2n,\qquad y=m+\fft{p-q}2n.
\label{eq:xydef}
\end{equation}
This defines an $x$-$y$ lattice similar to the one that we constructed for the class one giants.
Regularity of the wave function \eqref{eq:Phiy} with the top sign for $r_i$ in (\ref{eq:indices}) corresponds to the first quadrant, $x\ge0$ and $y\ge0$, while regularity with the bottom sign corresponds to the third quadrant, $x\le0$ and $y\le0$.  The zero mode spectrum in the first quadrant is summarized in Table~\ref{tab: D3 fluctuations ypq2}.

Notice that in the class one giants, we were able to solve for the full spectrum, which imposed constraints on the corresponding quantum numbers $l,m,n$. These constraints along with the BPS bound gave a natural choice for the quadrants in the $x$-$y$ lattice \eqref{eq: ypq1 xy lattice cndn}. Since we do not have this level of control over the full spectrum here, we rely on the sign of the true $R$-charge to get the correct choice. Since our BPS states saturate $\mathcal{H} \geq 0$ and for all the states we have $H \geq 0$, we see that $r_* \geq 0$ becomes a necessary condition.

\begin{table}[t]
    \centering
    \begin{tabular}{c c c c c c c c c c}
        \toprule
        {}  & $H $ & $J_1$ & $J_2$ & $r_*$& $r$& $F_A$& $F_B$ & BPS? \\
        \midrule
        $v_{(1+) m n}$ & $3m - \frac{n}{2\ell} + 1$ & $1$ & $0$ & $2m - \frac{n}{3\ell}$ & $2m - \frac{qn}{2}$ & $\mp m$ &$-\frac{pn}2$ & \checkmark\\
        $v_{(2+) m n}$ & $3m - \frac{n}{2\ell} + 1$  & $0$ & $1$ & $2m - \frac{n}{3\ell}$ &$2m - \frac{qn}{2}$ &  $\mp m$ &$-\frac{pn}2$ & \checkmark\\
        $v_{(1-) m n}$ & $3m - \frac{n}{2\ell} + 1$  & $-1$ & $0$ & $2m - \frac{n}{3\ell}$ & $2m - \frac{qn}{2}$ & $\mp m$ &$-\frac{pn}2$ & {}\\
        $v_{(2-) m n}$ & $3m - \frac{n}{2\ell} + 1$ & $0$ & $-1$ & $2m - \frac{n}{3\ell}$ &$2m - \frac{qn}{2}$ &  $\mp m$ &$-\frac{pn}2$ & {}\\
        \bottomrule
    \end{tabular}
    \caption{Quantum numbers of the zero mode ($j = 0$) sector in the AdS fluctuation of the $S_2$ and $S_4$ giants in $Y^{p,q}$. Here we implicitly assume the restriction \eqref{eq:mnwed} on $m$ and $n$.}
\label{tab: D3 fluctuations ypq2}
\end{table}

Note that the true $R$-charge, in terms of the $x$, $y$ quantum numbers and $(p,q)$ using \eqref{eq: ell}, can be expressed as
\begin{equation}
    \fft32r_*=\fft1{2q}\left(\left(3q-2p+\sqrt{4p^2-3q^2}\right)x+\left(3q+2p-\sqrt{4p^2-3q^2}\right)y\right).
\end{equation}
It is easy to show that
\begin{equation}
    2p-3q<\sqrt{4p^2-3q^2}<2p+3q.
\end{equation}
Hence regular states in the first quadrant have positive $r_*$ charge while regular states in the third quadrant have negative $r_*$ charge.  For the first quadrant states, we take positive $r_*$ and the top sign in (\ref{eq:susyHc2}) so the supersymmetric Hamiltonian takes the form (dropping the zero point energy of the oscillator modes)
\begin{equation}
    \mathcal H=\sum_i\sum_{mn}(0N_{(i+)mn} + 2N_{(i-)mn}).
\end{equation}
This demonstrates that the $v_{(i+)mn}$ states in the first quadrant are BPS while the $v_{(i-)mn}$ are not.  For regularity in the third quadrant, we take negative $r_*$ and the bottom sign in (\ref{eq:susyHc2}).  This gives instead
\begin{equation}
    \mathcal H=\sum_i\sum_{mn}\left(3|r_*|N_{(i+)mn}+(3|r_*|+2)N_{(i-)mn}\right).
\end{equation}
Clearly none of these states are BPS as one would expect since the $R$-charge is negative.

\subsubsection*{BPS spectrum and the orbifold sum}
We are now in a position to compute the contribution of the AdS scalars to the single-letter index of the $S_2$ and $S_4$ giants.  To summarize, the BPS states populate the lattice defined by $x \geq 0$, $y \geq 0$.

To complete the picture, we also need to understand the quantization condition on $x$ and $y$.  From Ref.~\cite{Martelli:2004wu}, we see that the effectively acting 3-torus (with closed orbits) is, up to $SL(3;\mathbb Z)$ transformations, given by
\begin{equation}
    e_1=\partial_\phi+\partial_\psi,\qquad e_2=\partial_\phi-\fft{p-q}2\ell\partial_\alpha,\qquad e_3=\ell\partial_\alpha.
\label{eq:e1e2e3}
\end{equation}
The action of this torus on the wavefunction (\ref{eq:Phiy}) is
\begin{equation}
    e_1:m\mp m,\qquad e_2:\mp m-\fft{p-q}2n,\qquad e_3:n.
\end{equation}
Demanding integer quantization (so the wavefunction is well defined) gives the equivalent conditions
\begin{equation}
    n\in\mathbb Z,\qquad m+\fft{p+q}2n\in\mathbb Z,
\end{equation}
independent of whether we are looking at the $S_2$ or $S_4$ giant.  From (\ref{eq:xydef}), the second condition ensures that both $x$ and $y$ are integer quantized and that $x-y=-pn$.  Since $n$ is an integer, this demonstrates that $x\equiv y\mod p$.  As a result, the BPS spectrum corresponds to a $\mathbb Z_p$ orbifold with $x$ and $y$ on the integer lattice subject to $y-x\equiv 0\mod p$.  The single letter index then takes the form of an orbifold projection, (\ref{eq:orbindex}), in this case with $\tilde\omega=\omega^{-1}$.

In terms of $x$, $y$ and using \eqref{eq:class2q}, we can write the mesonic and $r$-charge as
\begin{align}
    F_A&=\mp\fft{p-q}{2p}x\kern.4em\mp\fft{p+q}{2p}y,\quad 
    F_B=\fft12 x -\fft12 y,\quad
    r=\fft{2p-q}{2p}x+\fft{2p+q}{2p}y,
\end{align}
where  the $\pm$ sign corresponds to the $S_4$ and  $S_2$ giants, respectively. Using the fugacity map \eqref{eq:fugmap}, we find
\begin{equation}
    \prod v_I^{R_I}=\left(\rho^{\fft34(2p-q)}u^{\mp(p-q)}v^p\right)^{\fft{x}p}\left(\rho^{\fft34(2p+q)}u^{\mp(p+q)}v^{-p}\right)^{\fft{y}p}.
\end{equation}
In terms of the edge fugacities (\ref{eq:wtoquv}) this can be rewritten in the compact form
\begin{equation}
    \prod v_I^{R_I}=\begin{cases}(w_{\fft12})^{\fft{x}p}(w_{4-\fft12})^{\fft{y}p},&S_4\mbox{ giant at }\theta=0;\\
    (w_{2-\fft12})^{\fft{x}p}(w_{2+\fft12})^{\fft{y}p},&S_2\mbox{ giant at }\theta=\pi.\end{cases} \label{eq: ypq2 edge fugacities}
\end{equation}
Hence, the single-letter index can be expressed as,
\begin{align}
    f^{(I)}_{\mathrm{AdS}} (\mathfrak p, \mathfrak q, v_i) = \sum_{\genfrac{}{}{0pt}{1}{x,y\in Z_{\ge0}}{x - y\equiv0\pmod{p}}}(\mathfrak p+\mathfrak q)(w_{I+\fft12})^{\fft{x}{p}}(w_{I-\fft12})^{\fft{y}{p}}, \quad I = 2,4.
\end{align}
Rewriting this orbifold sum as a sum over the full lattice with a projection to $y = x \pmod p$, we have,
\begin{align}
    f^{(I)}_{\mathrm{AdS}} &= \dfrac{1}{p} \sum_{n=0}^{p - 1} (\mathfrak p + \mathfrak q) \sum_{x,y \geq 0} \omega_I^{nx} \tilde\omega_I^{ny} (w_{I+\frac{1}{2}})^{\frac{x}{p}} (w_{I-\frac{1}{2}})^{\frac{y}{p}}\nn\\
    &= \dfrac{1}{k_I} \sum_{n = 0}^{k_I - 1} \dfrac{\mathfrak p + \mathfrak q}{(1 - w_{(n)}) (1 - w'_{(n)})}, \quad I = 2,4, \label{eq: ypq2 index generic}
\end{align}
where $\omega_I=\tilde \omega_I^{-1}$ is a primitive $k_I$-th root of unity ($k_I = p$ for these giants). In the second line we have used \eqref{eq: ypq1 index generic1} to present this in the toric notation and to make the comparison with the index of the first class and field theory more obvious.

\subsubsection*{Scalar fluctuations into \texorpdfstring{$Y^{p,q}$}{Y(p,q)}}

The $Y^{p,q}$ fluctuations of this class of giants are described by the quadratic action,
\begin{align}
    S_2^{Y^{p,q}} = -\dfrac{L^3T_3}{3} \int d^4\xi \Biggl[&\dfrac{L^2}{12} (y-1)|\dot Z|^2 - \dfrac{1}{12}\mathrm{Re}(\bar Z\tilde\nabla^2Z) \nn\\
    &+ \dfrac{1}{2}\mathrm{Im}(\bar Z \partial_{\tilde \psi}Z) + \dfrac{L}{3} (1-y) \mathrm{Im} (\bar Z \dot Z)\Biggr]. \label{eq: ypq2 int action}
\end{align}
For these fluctuations two features have to be handled with care, and it is worth being explicit about both since they determine the operator whose spectrum we would require.

First, the giant is non trivially wrapped, similar to what we had for class one. So one has to be cautious about treating $Z$ simply as a complex scalar.  Near the pole $\theta=0$ we may take $Z=\theta e^{i\phi}$, so that the normal-plane metric is $\ft{1-y}6|dZ|^2$, while $\cos\theta\,d\phi=d\phi-\ft12\theta^2d\phi+\mathcal O(\theta^4)$ gives
\begin{equation}
    d\psi-\cos\theta\,d\phi=d\tilde\psi+\ft12\mathrm{Im}(\bar ZdZ)+\mathcal O(Z^4),\qquad\tilde\psi=\psi-\phi.
\end{equation}
The fluctuation therefore enters the $Y^{p,q}$ metric (\ref{eq:Ypq}) only through the replacement
\begin{equation}
    d\tilde\psi\ \longrightarrow\ d\tilde\psi+\ft12\mathrm{Im}(\bar ZdZ)
\end{equation}
in the fiber one-form, together with the normal-plane term $\ft{1-y}6|dZ|^2$.  Since the phase of $Z$ is $\phi$, and $\tilde\psi=\psi-\phi$, a change of the reference $\phi$ rotates $Z$ and shifts $\tilde\psi$ simultaneously. So $Z$ is a section of the normal bundle carrying charge $\sigma=\pm1$, and its connection lies along $d\tilde\psi$.  The cross term generated by the replacement above is precisely the coupling $\ft12\mathrm{Im}(\bar Z\partial_{\tilde\psi}Z)$ appearing in (\ref{eq: ypq2 int action}).  Consequently the operator governing $Z$ is not the scalar Laplacian but the twisted one,
\begin{equation}
    \Delta_\sigma=\tilde\nabla^2+6i\sigma\,\partial_{\tilde\psi},
\label{eq:twistedlap2}
\end{equation}
the first order term being the connection contribution $-2i\sigma\mathcal A^i\partial_i$ with
\begin{equation}
    \mathcal{A} = - \dfrac{3}{1-y} \left(\left(\dfrac{q(y)}{9}  +w(y)f(y)^2\right) d\tilde \psi  + w(y) f(y) d\alpha\right). \label{eq: ypq2 connection}
\end{equation}
This is the class two counterpart of the twisted Laplacian used for the class one giants in Appendix~\ref{app: internal modes}, and of the $\Lambda\to\Lambda+6n$ shift found for $T^{1,1}$ in Section~\ref{sec:scalars}.  Note that $\Delta_\sigma$ is odd in $\sigma$, so that $Z$ and $\bar Z$ see opposite twists and the spectrum is \emph{not} symmetric under $(m,n)\to(-m,-n)$.

The second feature is the factor $\frac16  (1-y)$ multiplying the normal-plane metric, which carries over to the kinetic term in (\ref{eq: ypq2 int action}).  The harmonic analysis is therefore a Sturm-Liouville problem with weight $(1-y)$,
\begin{equation}
    \Delta_\sigma\Phi=-(1-y)\Lambda\Phi,
\end{equation}
rather than an ordinary eigenvalue problem. With the Ansatz $\Phi=e^{im\tilde\psi}e^{in\alpha/\ell}f(y)$, the weighted equation reduces to
\begin{align}
    &2 \partial_y [(a - y^2) q(y)\partial_y]f + [-6\sigma m + \Lambda (1-y)]f\nn\\
    &- \dfrac{1-y}{4(a-y^2) q(y)} \left((a-y^2) \left(6m - \dfrac{n}{\ell}\right)^2 + 2(y-1)\frac{n}{\ell} \left((2y-1)\frac{n}{\ell} - 12 y m\right)\right)f=0.
\label{eq:c2intode}
\end{align}
This is of Heun type, with regular singular points at $y_1$, $y_2$, $y_3$ and $\infty$.  The indicial equation at these singular points $y_i$ suggests
\begin{equation}
    \Phi_{lmn}(y)=\left(\prod_{i = 1}^3(y-y_i)^{r_i}\right) e^{im\tilde \psi} e^{in\alpha/\ell}g(y),\qquad
    r_i=\sigma\left(\dfrac{n}{\ell}\dfrac{1-y_i}{12y_i}+\dfrac m2\right).
\label{eq:c2intexp}
\end{equation}
Similar to what we saw in the case of the AdS modes, it is important that the three signs are correlated, and moreover fixed by the bundle charge. For a given $\sigma$ the regular constant solution exists only for the choice in (\ref{eq:c2intexp}), and not for the opposite one.  This is a consequence of the  connection term, $6\sigma m$, in (\ref{eq:c2intode}) being odd, and is the point at which the twisting genuinely matters.  Restricting, as for the AdS$_5$ modes, to the zero mode ($g(y) = 1$) sub-sector, one can find solutions with the spectrum given by
\begin{equation}
    \Lambda_{mn}+4=\left(3m+2\sigma-\dfrac{n}{2\ell}\right)^2.
\label{eq:c2intspec}
\end{equation}
Expanding in these twisted harmonics $f_{mn}$ in the zero mode sector 
\begin{equation}
    Z = \sum_{mn} Z_{mn} f_{mn}, \quad Z_{mn} = Y^{(a)}_{mn} + i Y^{(b)}_{mn},
\end{equation}
and integrating out the spatial factors, the action \eqref{eq: ypq2 int action} reduces to,
\begin{align}
    S_2^{Y^{p,q}} \supset \dfrac{L^3 T_3}{18} \int dt \sum_{mn}& \Biggl[\dfrac{L^2}{2} ( \dot Y^{(a)}_{mn}{}^2 + \dot Y^{(b)}_{mn}{}^2) - \dfrac{\Lambda_{mn}}{2} (Y^{(a)}_{mn}{}^2 + Y^{(b)}_{mn}{}^2) \\
    &- 2L (Y^{(a)}_{mn}\dot Y^{(b)}_{mn} - Y^{(b)}_{mn}\dot Y^{(a)}_{mn})\biggr].
\end{align}
After canonical quantization, in the helicity basis, the Hamiltonian reads
\begin{equation}
    H = \sum_{mn} \sqrt{\Lambda_{mn} + 4} \, (N_{(+)mn} + N_{(-)mn} + 1) - 2(N_{(+)mn} - N_{(-)mn}).
\end{equation}
These internal modes do not carry AdS angular momenta. However,
from (\ref{eq:charges}) and (\ref{eq:rstar}) we find
\begin{align}
    F_A &= \mp \sum_{\sigma = \pm}  \sum_{mn}(m+\sigma) \, N_{(\sigma)mn} , & r& = \sum_{mn} \left(2m-\dfrac{qn}{2}\right)(N_{(+)mn} + N_{(-)mn}), \nn \\
    F_B &=  - \dfrac{1}{2} \sum_{mn} pn (N_{(+)mn} + N_{(-)mn}),&  \dfrac{3}{2}r_* &= \sum_{mn} \left(3m - \dfrac{n}{2\ell}\right) (N_{(+)mn} + N_{(-)mn}),
\end{align}
where, as noted before, the sign of $F_A$ corresponds to the $S_4$ and $S_2$ giants. We also observe that the spectrum and the true R-charge are related by
\begin{equation}
    \sqrt{\Lambda_{mn}+4}=\left|\ft32r_*+2\sigma\right|,
\label{eq:c2intdelta}
\end{equation}
which is the direct counterpart of (\ref{eq:deltamn}) for fluctuations into AdS. Regularity of the wavefunction requires $r_1,r_2\ge0$. In terms of $x$, $y$ in \eqref{eq:xydef}, this selects the $x \ge 0$, $y\ge 0$ quadrant for $\sigma = + 1$ and $x \le 0$, $y \le 0$ for $\sigma = -1$. Above, for AdS fluctuations, we showed that the true $R$-charge is positive in the first quadrant and negative in the third quadrant.

In the first quadrant only the $\sigma=+1$ modes are regular. For these, (\ref{eq:c2intdelta}) gives $\sqrt{\Lambda_{mn}+4}=\ft32r_*+2$, so that their energy is $H=\ft32r_*$ and they saturate the BPS bound, $\mathcal H=0$. In the third quadrant only the $\sigma=-1$ modes are regular. For these, $\sqrt{\Lambda_{mn}+4}=\left|\ft32r_*-2\right|=2+\ft32|r_*|$, so that $H=4+\ft32|r_*|$ and the supersymmetric Hamiltonian is
\begin{equation}
    \mathcal H = \sum_{mn} \left(4+3|r_*|\right) N_{(-)mn},
\end{equation}
which is strictly positive. Hence none of the third-quadrant modes are BPS, as expected from the sign of the R-charge. This is summarized in Table~\ref{tab: D3 fluctuations int ypq2}.

\begin{table}[t]
    \centering
    \begin{tabular}{c c c c c c c}
        \toprule
        {}  & $H$ & $\mathcal H$ & $J_1,J_2$ & $r_*$ & & BPS? \\
        \midrule
        \multicolumn{7}{l}{First quadrant: $x \geq 0,\ y \geq 0$, $r_* \geq 0$}\\
        \midrule
        $Y_{(+) m n}$ & $3m - \frac{n}{2\ell}$ & $0$ & $0$ & $2m - \frac{n}{3\ell}$ & regular & \checkmark\\
        $Y_{(-) m n}$ & $\left|3m - \frac{n}{2\ell}-2\right|+2$ & $\geq 0$ & $0$ & $2m - \frac{n}{3\ell}$ & irregular & {} \\
        \midrule
        \multicolumn{7}{l}{Third quadrant: $x \leq 0,\ y \leq 0$, $r_* \leq 0$}\\
        \midrule
        $Y_{(+) m n}$ & $\left|3m - \frac{n}{2\ell}+2\right|-2$ & -- & $0$ & $2m - \frac{n}{3\ell}$ & irregular & {}\\
        $Y_{(-) m n}$ & $4-\left(3m - \frac{n}{2\ell}\right)$ & $4+3|r_*|$ & $0$ & $2m - \frac{n}{3\ell}$ & regular & {} \\
        \bottomrule
    \end{tabular}
    \caption{Energies and true R-charges of the zero-mode ($g(y)=1$) sector of the internal fluctuations of the $S_2$ and $S_4$ giants in $Y^{p,q}$. Only the regular $Y_{(+)}$ modes in the first quadrant are BPS. The energies of the irregular modes are listed for completeness only. The remaining charges are given in the text.}
    \label{tab: D3 fluctuations int ypq2}
\end{table}

We now have all the ingredients to find the single-letter index for these fluctuations into $Y^{p,q}$. These states populate the $\mathbb Z_p$ orbifold of the lattice $x \ge 0$, $y \ge 0$. In terms of $x$, $y$ using \eqref{eq:class2q}, the mesonic and $R$-charge takes the form,
\begin{equation}
    F_A=\mp\fft{p-q}{2p}x\kern.4em\mp\fft{p+q}{2p}y \mp 1,\quad 
    F_B=\fft12 x -\fft12 y,\quad
    r=\fft{2p-q}{2p}x+\fft{2p+q}{2p}y,
\end{equation}
where the $\pm$ signs are respectively for the $S_4$ and $S_2$ giants. We also find
\begin{equation}
    \prod v_I^{R_I} = u^{\mp2} \left(\rho^{\fft34(2p-q)}u^{\mp(p-q)}v^p\right)^{\fft{x}p}\left(\rho^{\fft34(2p+q)}u^{\mp(p+q)}v^{-p}\right)^{\fft{y}p}.
\end{equation}
Following our discussion on the AdS modes, from \eqref{eq: ypq2 edge fugacities} we can write the $Y^{p,q}$ scalar contribution to the single letter index as
\begin{align}
    f^{(I)}_{Y^{p,q}} = \dfrac{1}{k_I} \sum_{n = 0}^{k_I - 1} \dfrac{u^{\mp 2}}{(1 - w_{(n)}) (1 - w'_{(n)})}, \quad I = 2,4,
\end{align}
where $w_{(n)},w'_{(n)}$ are defined in \eqref{eq: ypq1 index generic1}. One can show that
\begin{equation}
    u^{\mp 2} = \dfrac{w_{(n)}w'_{(n)}}{\mathfrak p \mathfrak q},
\end{equation}
for both the $S_2$ and $S_4$ giants. This allows us to write the index in the generalized form,
\begin{align}
    f^{(I)}_{Y^{p,q}} = \dfrac{1}{k_I} \sum_{n = 0}^{k_I - 1} \dfrac{(w_{(n)}w'_{(n)}/\mathfrak p \mathfrak q)}{(1 - w_{(n)}) (1 - w'_{(n)})}, \quad I = 2,4.\label{eq: ypq2 index int}
\end{align}

\subsubsection{Vector fluctuations}
\label{sec:class2vectors}

As in the previous cases, the linearized world-volume vector fluctuations reduce to Maxwell theory \eqref{eq: t11 vect action}. In Coulomb gauge we need the transverse one-form harmonics on the wrapped cycle.  As discussed in Appendix~\ref{app:lens}, these are most efficiently obtained from the curl operator, since $\delta d=\mathrm{curl}^2$ on transverse forms and the construction makes no assumption about the geometry. Since this class of giants wraps a topological lens space with an inhomogeneous metric, as was the case for the scalars, the harmonic analysis in the Appendix does not immediately follow.

It is convenient to work in coordinate components, $X=X_ydy+X_{\tilde\psi}d\tilde\psi+X_\alpha d\alpha$, since the volume form is simply $\ft13dy\wedge d\tilde\psi\wedge d\alpha$, so that $\sqrt g=1/3$ is constant. We solve the eigenvalue problem for the $curl$ operator with the Ansatz
\begin{equation}
    X_\mu = \hat X_\mu(y) e^{im\tilde\psi} e^{in\alpha/\ell}.
\end{equation}
This gives an algebraic equation
\begin{equation}
    \hat X_y=\fft{3i}{\kappa\,wq}\left(m\hat X_\alpha-\fft n\ell\hat X_{\tilde\psi}\right),
\label{eq:class2Xy}
\end{equation}
which can be used to eliminate $\hat X_y$ leaving us with first order system with \emph{rational} coefficients,
\begin{equation}
    \fft{d}{dy}\begin{pmatrix}\hat X_{\tilde\psi}\\ \hat X_\alpha\end{pmatrix}=\fft{M(y)}{(a - y^2)q(y)}\begin{pmatrix}\hat X_{\tilde\psi}\\ \hat X_\alpha\end{pmatrix},
\label{eq:class2rat}
\end{equation}
where $M(y)=M_0+M_1y+M_2y^2$ is quadratic with
\begin{align}
    M_0 &= \dfrac{1}{12\ell^2 \kappa}
    \begin{pmatrix}
        6\ell(3mn - a\ell\kappa^2) & \ell^2((2 + a)\kappa^2 - 18m^2)\\
        6(3n^2 -6a\ell^2\kappa^2) & 6\ell(a\ell\kappa^2-3mn)
    \end{pmatrix},\nn\\
    M_1 &= \dfrac{1}{12\ell^2\kappa}
    \begin{pmatrix}
        6\ell(2\ell\kappa^2 - 3mn) & \ell^2(18m^2 - 6\kappa^2)\\
        - 18 n^2 & 6\ell (3mn - 2\ell\kappa^2)
    \end{pmatrix},\nn\\
    M_2 &= \dfrac{\kappa}{4}
    \begin{pmatrix}
        -2 & 1\\
        12 & 2
    \end{pmatrix}.
\end{align}
This Fuchsian system has regular singular points only at $y_1$, $y_2$, $y_3$ and infinity, and one can check that the indicial exponents at $y_i$ are exactly the scalar exponents $r_i$ in (\ref{eq:indices}). So the solutions have to be of the form,
\begin{equation}
    \begin{pmatrix}\hat X_{\tilde\psi}\\ \hat X_\alpha\end{pmatrix}=\prod_i(y-y_i)^{r_i}\,\mathbf v (y),
\end{equation}
where $\mathbf v$ is a vector field regular at the poles. Similar to the scalars, we can look for solutions that restrict $\mathbf v(y)$ to be a constant vector, $\mathbf c$. We expect all the BPS modes to be contained in this sub-sector. For this sub-sector, Eq. (\ref{eq:class2rat}) requires 
\begin{align}
    M(y) \mathbf c = 2\sum_ir_i\prod_{j\ne i}(y-y_j)\; \mathbf c.
\end{align}
Since $M(y)$ is polynomial in $y$, $\mathbf c$ must be a simultaneous eigenvector of $M_0$, $M_1$ and $M_2$.  The quadratic coefficient $M_2$ is traceless with determinant $-\kappa^2$, so its eigenvalues are $\pm\kappa$. It immediately follows that
\begin{equation}
    \kappa=\pm2\left(r_1+r_2+r_3\right)=\pm\left(3m-\fft n{2\ell}\right)=\pm\ft32r_*, \label{eq: ypq2 polarization}
\end{equation}
where we have used (\ref{eq:indices}).  Imposing in addition that $\mathbf c$ be an eigenvector of $M_1$ and $M_0$ selects the upper sign, and gives $\mathbf c\propto(1,6)$.  The vector harmonics of the zero mode sector are therefore
\begin{equation}
    X=\prod_i(y-y_i)^{r_i}\,e^{im\tilde\psi}e^{in\alpha/\ell}\left[d\tilde\psi+6\,d\alpha
    +\fft{3i}{\kappa\,wq}\left(6m-\fft n\ell\right)dy\right],\qquad\kappa=\ft32r_*.
\label{eq:class2bpsvec}
\end{equation}

Following our discussion in Appendix~\ref{app:lens}, since $\Lambda=\kappa^2$, these modes have $H=\sqrt\Lambda=\ft32|r_*|$. These modes do not carry AdS angular momentum, $J_1 = J_2 = 0$, so the supersymmetric Hamiltonian reduces to (dropping the zero point energy),
\begin{align}
    \mathcal{H} = \sum_{mn} \fft32 (|r_*| - r_*)(n^{(+)}_{mn} + n^{(-)}_{mn}).
\end{align}
The results we concluded for the scalars apply in this case as well. In terms of the $x$, $y$ quantum numbers \eqref{eq:xydef}, the true R-charge, $r_*$, is positive in the first quadrant and negative on the third. So the BPS modes populate the first quadrant. Out of the two polarizations only the `$+$' (correlated with the `$+$' sign in \eqref{eq: ypq2 polarization}) is regular here, so that only the $n^{(+)}_{mn}$ in the first quadrant saturates the BPS bound.

It remains to give the range of $x$, $y$.  Although the exponents $r_i$ coincide with those of the scalars, the regularity condition is more delicate. This is because the frame components of a one-form differ from the coordinate components by inverse powers of the degenerating vielbein.  Near $y=y_1$, let $\rho\propto\sqrt{y-y_1}$ be the local radial coordinate. Then, $e^2=\frac{1}{3}\sqrt{q}\,d\tilde\psi\sim\rho\,d\tilde\psi$. So a scalar behaving as $(y-y_1)^{r_1}=\rho^{x}$ corresponds to frame components $X_1,X_2\sim\rho^{x-1}$.  
Hence regularity requires $x\ge1$ rather than $x\ge0$, and likewise $y\ge1$ at the other pole.  

We can now compute the contribution of these vector modes to the single letter index. Note that in terms of $x$, $y$ the mesonic and $R$-charge take the form
\begin{equation}
    F_A=\mp\fft{p-q}{2p}x\kern.4em\mp\fft{p+q}{2p}y ,\quad 
    F_B=\fft12 x -\fft12 y,\quad
    r=\fft{2p-q}{2p}x+\fft{2p+q}{2p}y,
\end{equation}
where the $\pm$ signs are respectively for the $S_4$ and $S_2$ giants. It can now be shown that the index takes the general form
\begin{equation}
    f_{\mathrm{vector}}^{(I)}=\dfrac{1}{k_I}\sum_{n = 0}^{k_I - 1}\fft{w_{(n)}w'_{(n)}}{(1-w_{(n)})(1-w'_{(n)})}, \quad I = 2,4,
\label{eq:class2fvector}
\end{equation}
where $w_{(n)},w'_{(n)}$ are defined in \eqref{eq: ypq1 index generic1}.

\subsubsection{Fermion fluctuations}

We now complete the analysis of the class two giants by considering the fermion fluctuations.
The reduction of the fermionic action is insensitive to the internal geometry. As in Sections~\ref{sec:fermions} and \ref{sec:ypqfermions}, the background carries only the metric and the self-dual five-form, so $\kappa$-symmetry and the ten-dimensional Weyl condition leave the fermionic action in the same form as \eqref{eq:D3fermionact}.
As for $T^{1,1}$ and the class one giants, $\fsl\nabla$ must include the connection on the normal bundle,
\begin{equation}
    \fsl\nabla\ \to\ \fsl\nabla-iQ\gamma^a\mathcal A_a .
\end{equation}
The wrapped cycle, however, is no longer a simple lens space.  It is a $U(1)$ bundle over the \emph{topological} two-sphere spanned by $(y,\tilde\psi)$, whose metric is not round and whose fiber $\sqrt{w(y)}$ varies over the base.  The harmonic analysis of Appendix~\ref{app:lens} therefore does not apply, and the Dirac operator leads to a radial problem of the same Heun class encountered for the scalars, with regular singular points at $y=y_1,y_2,y_3,1$ and $\infty$.  As in the scalar case we do not attempt the full spectrum.  Instead we adopt the same working hypothesis as above, namely that the BPS states are the zero modes, for which the radial function is trivial and the entire $y$ dependence resides in the exponents at the singular points.

Granting this, the structure of the fermionic BPS spectrum is fixed by two inputs, neither of which requires solving the radial equation. The first is the assignment of quantum numbers to the towers.  This follows from the decomposition of the ten-dimensional Majorana-Weyl spinor together with the $\kappa$-symmetry projection, and is therefore common to every D3-brane giant in AdS$_5\times SE_5$; it knows nothing about the internal geometry.  In the case of $T^{1,1}$ and the first class where the geometry was tractable enough to check it directly, the supersymmetric Hamiltonian evaluated on the extremal modes reduced to
\begin{equation}
    \mathcal H=\ft12-Q-J_1-J_2,
\label{eq:univBPS}
\end{equation}
independently of the squashing, and hence of $p$ and $q$.  We therefore expect the same four BPS towers here: one with $Q=-\ft12$ and $J_1+J_2=1$, and three with $Q=+\ft12$, two of which have $J_1+J_2=0$ while the fourth arises as the antiparticle of a negative-energy mode.

The second input is the range of the quantum numbers, which is where the geometry does enter. This is fixed by regularity of the wave function at the two poles $y=y_1$ and $y=y_2$.  Recall from (\ref{eq:Phiy}) and (\ref{eq:xydef}) that a scalar zero mode behaves as $(y-y_1)^{x/2}(y-y_2)^{y/2}$.  Near $y=y_1$ the base is smooth, and in a local polar coordinate $\rho\propto\sqrt{y-y_1}$ this is $\rho^{x}$, so that $x$ is the angular momentum of the mode about that pole; regularity gives the wedge $x$, $y\ge0$ of (\ref{eq:mnwed}).  A spinor at such a pole behaves differently. In a local frame its two components carry angular momenta shifted by $\mp\ft12$ relative to the orbital value, so that they go as $\rho^{x-1}$ and $\rho^{x}$.  Regularity therefore imposes $x\ge1$ on one component and $x\ge0$ on the other, and likewise at $y=y_2$ for $y$.  These are exactly the two lattice offsets found in the $T^{1,1}$ and class one analyses, where they could be derived rather than inferred.

It is worth recording the Dirac operator itself, both because it fixes the second input independently and because it shows why the construction of Section~\ref{sec:class2vectors} does not carry over. Consider the Ansatz $e^{im\tilde\psi} e^{in\alpha/\ell} \chi(y)$ for the spinor modes.
In the frame adapted to the fibration, 
\begin{equation}
    e^1 = \dfrac{dy}{\sqrt{w(y)q(y)}}, \quad e^2 = \dfrac{\sqrt{q(y)}}{3} d\tilde\psi, \quad e^3 = \sqrt{w(y)} (d\alpha + f(y) d\tilde \psi),
\end{equation}
Cartan's equation gives
\begin{equation}
    \omega_{12}=-\mathcal P e^2-\ft{\mathcal Q}2e^3,\qquad
    \omega_{13}=-\ft{\mathcal Q}2e^2-\mathcal R e^3,\qquad
    \omega_{23}=\ft{\mathcal Q}2e^1,
\end{equation}
with $\mathcal P=\sqrt w(\sqrt q)'$, $\mathcal R=\sqrt q(\sqrt w)'$ and $\mathcal Q=3wf'$, while the normal bundle connection \eqref{eq: ypq2 connection} in the frame components is given by,
\begin{equation}
    \mathcal{A} = - \dfrac{\sqrt{q(y)}}{1 - y} e^2 - \dfrac{3\sqrt{w(y)}\, f(y)}{1 - y} e^3.
\end{equation}
They contribute $-iQ \fsl{\mathcal{A}} + \ft12[(\mathcal P+\mathcal R)\gamma^1+\ft{i\mathcal Q}2]$ to $\fsl\nabla$. Since $\mathcal P+\mathcal R=(qw)'/2\sqrt{qw}$, the substitution $\chi=(qw)^{-1/4}\eta$ removes the first order friction term, leaving
\begin{align}
    \sqrt{wq}\,\eta' = \left[\hat\mu\gamma^3-\hat\nu\gamma^2+i(\Lambda-\ft{\mathcal Q}4)\gamma^1\right]\eta,
\label{eq:class2dirac}
\end{align}
where $\Lambda$ is the eigenvalue of $-i\fsl\nabla$ and $\hat\mu$, $\hat\nu$ are the class two counterparts of \eqref{eq: munu class one}
\begin{equation}
    \hat\mu = \dfrac{3 (m - nf(y)/\ell)}{\sqrt{q(y)}} + \dfrac{Q \sqrt{q(y)}}{1 - y}, \quad \hat{\nu} = \dfrac{n}{\ell \sqrt{w(y)}} + \dfrac{3Q \sqrt{w(y)}\, f(y)}{1 - y}.
\end{equation}
The indicial exponents of (\ref{eq:class2dirac}) are $\eta_1\sim(y-y_1)^{\pm x/2}$ at $y=y_1$ and $\eta_1\sim(y_2-y)^{\pm y/2}$ at $y=y_2$, with $\eta_2$ taking the opposite sign in each case.  The spinor modes, therefore, live on the same $x$-$y$ lattice as the scalars, the two ends entering asymmetrically as one expects for a spinor, which confirms our earlier discussion about regularity.

Unlike the vector problem, however, (\ref{eq:class2dirac}) cannot be brought to a system with rational coefficients, and the obstruction is intrinsic.  For the vectors the reduction was possible because $\sqrt g=1/3$ is constant, so that the coordinate components of a one-form see the metric functions only through the combination $wq$ which is rational.  The Dirac operator instead requires the frame itself, and $\sqrt w$ and $\sqrt q$ enter separately making the analysis nontrivial. We present the relevant details in  Appendix \ref{App:FermionsClassII}.

Following the discussion in the Appendix we find that the BPS spectrum is given by
\begin{align}
    \Lambda = \dfrac{3}{2} r_* + \dfrac{1}{2} + Q.
\end{align}
It remains to fix the range.  The frame $(e^1,e^2)$ is a polar frame about each pole, rotating with $K_1=\partial_{\tilde\psi}-\ft{p+q}2\ell\partial_\alpha$ at $y_1$ and $K_2=\partial_{\tilde\psi}+\ft{p-q}2\ell\partial_\alpha$ at $y_2$.  The spinor's angular momenta about the poles are therefore
\begin{equation}
    j_1=m-\fft{p+q}2n,\qquad j_2=m+\fft{p-q}2n\ \in\ \mathbb Z+\ft12 .
\end{equation}
The BPS spinor behaves as $\rho^{j_1\mp\frac12}$ near $y_1$, and likewise at $y_2$, so regularity requires $j_1,j_2\ge\ft12$.  In terms of the integer labels $x=j_1+Q$, $y=j_2+Q$, this is $x,y\ge0$ for $Q=-\ft12$ and $x,y\ge1$ for $Q=+\ft12$, which are the two offsets mentioned above.  Moreover $y-x=pn$, as for the bosons. We summarize the BPS towers in  Table~\ref{tab: D3 fermions ypq2}.

\begin{table}[t]
    \centering
    \begin{tabular}{c c c c c c}
        \toprule
        {} & $Q$ & $J_1$ & $J_2$ & lattice & letter\\
        \midrule
        $\psi_{(1)mn}$ & $-\ft12$ & $\ft12$ & $\ft12$ & $x,y\ge0$ & $\mathfrak p\mathfrak q$\\
        $\psi_{(2)mn}$ & $+\ft12$ & $\ft12$ & $-\ft12$ & $x,y\ge1$ & $1/\mathfrak q$\\
        $\psi_{(3)mn}$ & $+\ft12$ & $-\ft12$ & $\ft12$ & $x,y\ge1$ & $1/\mathfrak p$\\
        $\psi_{(4)mn}$ & $+\ft12$ & $\ft12$ & $\ft12$ & $x,y\ge0$, $(x,y)\ne(0,0)$ & $1$\\
        \bottomrule
    \end{tabular}
    \caption{The four BPS towers of fermionic fluctuations for the class two giant, and the fugacity carried by each in addition to $w^{x/p}w'^{y/p}$.  The first three towers satisfy $\mathcal H=0$ by (\ref{eq:univBPS}); the fourth is the antiparticle of a negative-energy mode, for which $\mathcal H=H-J_1-J_2-\ft32r_*=0$ directly.}
    \label{tab: D3 fermions ypq2}
\end{table}

The extra fugacity carried by each letter follows as before.  Using the map (\ref{eq:fugmap}) together with $\rho^3=\mathfrak p \mathfrak q$ in the $B=0$ sector, a fermionic letter of normal charge $Q$ carries the factor $\mathfrak p^{J_1}\mathfrak q^{J_2}(\mathfrak p \mathfrak q)^{-Q}$ relative to a bosonic one (the normal charge contributes $-Q$ to $F_A$, and the shift $x=j_1+Q$ supplies $(w_{\frac12}w_{3+\frac12})^{-Q/p}=(\mathfrak p \mathfrak q)^{-Q}u^{2Q}$, so that the $u$ dependence cancels), which gives the last column of Table~\ref{tab: D3 fermions ypq2}.  Writing in terms of $w_{(n)}$ and $w'_{(n)}$ defined in \eqref{eq: ypq1 index generic1} and summing each tower over its lattice, we obtain
\begin{equation}
    f_{\mathrm{fermion}}^{(I)}= \fft1{k_I}\sum_{n = 0}^{k_I - 1}\fft{w_{(n)}+w'_{(n)}-w_{(n)}w'_{(n)}+ \mathfrak p \mathfrak q+\fft{w_{(n)}w'_{(n)}}{\mathfrak p \mathfrak q}(\mathfrak p + \mathfrak q)}{(1-w_{(n)})(1-w'_{(n)})}, \quad I = 2,4.
\label{eq:class2ffermion}
\end{equation}
As in the class one case, the $-w_{(n)}w'_{(n)}$ term in the numerator arises from the removal of the lowest state of the fourth tower.  Note that (\ref{eq:class2ffermion}) is identical in form to the class one result (\ref{eq:ypqffermion}), as it must be.

\subsubsection{The single letter index for class two giants}
We can now write down the single letter index for the class two giants by combining the AdS scalar \eqref{eq: ypq2 index generic}, internal scalar \eqref{eq: ypq2 index int}, vector \eqref{eq:class2fvector} and fermion \eqref{eq:class2ffermion} contributions to obtain
\begin{equation}
    f^{(I)}=1-\fft1{k_I}\sum_{n = 0}^{k_I - 1}\fft{(1-w_{(n)}w'_{(n)}/\mathfrak p \mathfrak q)(1- \mathfrak p)(1-\mathfrak q)}{(1-w_{(n)})(1-w'_{(n)})}, \qquad I = 2,4,
\label{eq:class2f}
\end{equation}
which is again the single-letter index entering the orbifold formula (\ref{eq:orbindex}), here with a $\mathbb Z_p$ projection corresponding to the lens space $L(p,1)$ partition function.

\subsection{The giant graviton indices on \texorpdfstring{$Y^{p,q}$}{Y(p,q)}}
While the fluctuation spectra of the class one and class two giants are distinct, the corresponding single letter indices, (\ref{eq:class1f}) and (\ref{eq:class2f}), have the identical form
\begin{equation}
    f^{(I)}=1-\fft1{k_I}\sum_{n = 0}^{k_I - 1}\fft{(1-w_{(n)}w'_{(n)}/\mathfrak p \mathfrak q)(1- \mathfrak p)(1-\mathfrak q)}{(1-w_{(n)})(1-w'_{(n)})}, \qquad I = 1,\ldots,4,
\end{equation}
where
\begin{equation}
    k_1=p-q,\qquad k_2=p,\qquad k_3=p+q,\qquad k_4=p,
\end{equation}
and
\begin{equation}
    w_{(n)} = \omega_I^n (w_{I+\frac{1}{2}})^{1/k_I}, \qquad w_{(n)}' = \tilde\omega_I^n (w_{I-\frac{1}{2}})^{1/k_I},
\end{equation}
with the $w_{I\pm\fft12}$ defined in (\ref{eq:wfug}).  Each giant graviton corresponds to a D3-brane wrapped on a lens space $L(k_I,1)\simeq S^3/\mathbb Z_{k_I}$, and the orbifold projection is enforced by taking $\omega_I$ to be a primitive $k_I$-th root of unity.  One slight distinction between the two classes of giants is that $\tilde\omega_I=\omega_I$ for the first class of giants ($I=1,3$) while $\tilde\omega_I=\omega_I^{-1}$ for the second class ($I=2,4$). The single-giant contribution, $\hat I^{\mathrm{GG}}_{(I)}$, then follows from \eqref{eq:IGGPE}.

\section{Conclusions}
\label{sec:conclusions}

In this work, we have derived single-giant contributions to
superconformal indices directly from the quantization of probe
D3-branes in $\mathrm{AdS}_{5}\times T^{1,1}$ and
$\mathrm{AdS}_{5}\times Y^{p,q}$. These backgrounds provide a
nontrivial extension of previous analyses in sphere compactifications:
they preserve only $\mathcal N=1$ supersymmetry, and the giant
gravitons wrap three-cycles with nontrivial topology, squashing, and
normal-bundle structure. Nevertheless, their quantized worldvolume
degrees of freedom reproduce the single-letter indices predicted by
the giant-graviton expansions of the dual toric quiver gauge theories \cite{Arai:2019aou}.

For the four maximal giants in $T^{1,1}$, we determined the scalar,
vector, and fermionic fluctuation spectra and isolated the modes
saturating the superconformal BPS bound. Their contributions combine
into the expected single-letter index, including its dependence on the
mesonic and baryonic fugacities. This provides a direct bulk derivation
of all four single-giant sectors of the 
giant-graviton expansion for the Klebanov-Witten theory.

The $Y^{p,q}$ family exhibits a richer structure. The cycles associated
with the external vertices of the toric diagram are lens spaces, and
the four corresponding giant gravitons divided into two geometrically
distinct classes. For class one giants, the relevant modes can be
obtained by harmonic analysis on squashed lens spaces. The global
identifications of the worldvolume produce the
$\mathbb Z_{p+q}$ and $\mathbb Z_{p-q}$ projections expected from the
field-theory expansion. For class two giants, the fluctuation equations
involve the nontrivial $y$ direction and reduce to Heun-type problems.
Although their complete spectra are not generally available in closed
form, the BPS sector remains exactly accessible. In the bosonic
sectors, the superconformal bound and Sturm-Liouville theory show that
the regular nodeless zero modes are the unique BPS states at fixed
charges. The corresponding fermionic modes are selected by their
regularity and saturation of the BPS bound. The resulting charge
lattice implements the required $\mathbb Z_p$ projection.

A central outcome of our analysis is the separation between the local
worldvolume spectrum and the global data of the wrapped cycle. Before
projection, the scalar, vector, and fermionic contributions combine
into the same function,
\begin{equation}
  f(\mathfrak{p},\mathfrak{q};w,w')
  =
  1-
  \frac{(1-ww'/\mathfrak{pq})(1-\mathfrak{p})(1-\mathfrak{q})}
       {(1-w)(1-w')},
  \label{eq:conclusion-universal-letter}
\end{equation}
for every giant considered here. The geometry enters globally through
the allowed lattice of charges and the corresponding orbifold
projection. Thus, the universal part of the single-letter index is
encoded in the local supersymmetric D3-brane fluctuation theory, while
the topology and embedding of the cycle determine how those local
degrees of freedom are assembled into globally well-defined states.

This structure suggests several directions for further investigation.
First, it would be useful to obtain a more conceptual characterization
of the BPS zero modes. Their relation to the toric charge lattice
suggests that they should admit a description in terms of holomorphic
functions or sections on the Calabi-Yau cone following the original formulation in \cite{Mikhailov:2000ya, Beasley:2002xv} (see also \cite{Butti:2006au}) for the half-BPS sector and as recently applied in \cite{GonzalezLezcano:2026wje} for the analogue of the half-BPS index. Such a formulation could
explain the universality of
\eqref{eq:conclusion-universal-letter} without requiring a
case-by-case solution of the worldvolume fluctuation equations, possibly appealing to a sort of cohomological toric analysis.

Second, the analysis should be extended beyond $T^{1,1}$ and
$Y^{p,q}$ to generic  toric Sasaki-Einstein manifolds; we expect the same unprojected
single-letter function persists and the complete effect of the
global geometry is always captured by a projection of the charge
lattice. Given the overall description of microstates in terms of giant gravitons, it would also be interesting to understand how much of the
construction survives for backgrounds without a toric description or even in non-supersymmetric backgrounds; a giant graviton expansion for thermal partition functions of $\mathcal{N} = 1$ theories was recently constructed in \cite{Budzik:2026vqe}.

Finally, the present calculation concerns a single probe D3-brane.
Recovering sectors with multiple coincident giant gravitons requires an
interacting non-Abelian worldvolume description (see \cite{Budzik:2025vzr} for a complementary perspective in terms of brane/negative-brane systems). Understanding this
sector is necessary for a complete gravitational derivation of the
finite-$N$ superconformal index and may clarify the relation between
open-string degrees of freedom on the branes, closed-string states in
the bulk, and the trace relations of the boundary theory. The agreement
established here at the single-giant level provides a concrete
foundation for pursuing this broader gravitational formulation of
finite-$N$ supersymmetric state counting.

\section*{Acknowledgments}
We are particularly grateful to Alfredo Gonz\'alez Lezcano for discussions.  This work is partially supported by the
U.S. Department of Energy under grants DE-SC0007859 and DE-SC0026542. ED and SJ were supported by Leinweber summer fellowships.

\subsection*{AI statement:} The manuscript was written by the authors with AI assistance in drafting the fermion fluctuation sections; the authors take full responsibility for any errors. AI helped with normal bundle issues, simplification of vector harmonics, the fermion fluctuations, and with editing of the manuscript.

\appendix

\section{Lens spaces}
\label{app:lens}

Before considering the lens space $L(p,1)$, we introduce the squashed $S^3$, which plays an important role in constructing the lens space harmonics. We can define the unit three-sphere by taking $|z_1|^2+|z_2|^2=1$ in $\mathbb C^2$.  If we let
\begin{equation}
    z_1=\cos\left(\fft\theta2\right)e^{i(\psi+\phi)/2},\qquad z_2=\sin\left(\fft\theta2\right)e^{i(\psi-\phi)/2},
\label{eq:z1z2}
\end{equation}
then the round three-sphere metric inherited from the Euclidean metric on $\mathbb C^2$ is
\begin{equation}
    ds^2=\fft14\left(d\theta^2+\sin^2\theta\, d\phi^2+(d\psi+\cos\theta d\phi)^2\right).
\end{equation}
This is a Hopf fibration written as a $U(1)$ bundle over $\mathbb P_1$.  From (\ref{eq:z1z2}), we see that the angles $\phi$ and $\psi$ are periodic with the identifications
\begin{equation}
    (\phi,\psi)=(\phi+2\pi,\psi+2\pi)=(\phi+2\pi,\psi-2\pi).
\end{equation}
This is equivalent to taking
\begin{equation}
    (\phi,\psi)=(\phi+2\pi,\psi+2\pi)=(\phi,\psi+4\pi).
\label{eq:periods}
\end{equation}
The first identification allows us to restrict $\phi\in[0,2\pi)$, and the second allows us to take $\psi\in[0,4\pi)$.  Note that, for the round sphere, $\phi$ and $\psi$ are interchangeable (corresponding to $z_2\to\bar z_2$), but this discrete isometry will be broken by squashing  along the $U(1)$ fiber.  In particular, we parametrize the squashing according to
\begin{equation}
    ds^2=\fft14\left(d\theta^2+\sin^2\theta\, d\phi^2+e^{2\lambda}(d\psi+\cos\theta d\phi)^2\right),
\label{eq:squashed}
\end{equation}
where $\lambda$ is a constant.

We can highlight the residual $SU(2)_l\times U(1)_r\subset SU(2)_l\times SU(2)_r$ isometry of the squashed sphere by introducing the left-invariant one-forms
\begin{equation}
    \sigma_L^1=\sin\psi\,d\theta-\cos\psi\sin\theta\,d\phi,\qquad\sigma_L^2=\cos\psi\,d\theta+\sin\psi\sin\theta\,d\phi,\qquad\sigma_L^3=d\psi+\cos\theta\,d\phi,
\end{equation}
that satisfy $d\sigma_L^a=-\fft12\epsilon_{abc}\sigma_L^b\wedge\sigma_L^c$.  The metric is then
\begin{equation}
    ds^2=\fft14\left((\sigma_L^1)^2+(\sigma_L^2)^2+e^{2\lambda}(\sigma_L^3)^2\right)=(e^1)^2+(e^2)^2+(e^3)^2,
\label{eq:sLframe}
\end{equation}
where we have introduced the natural frame basis.  The corresponding spin connection can be read off from the $SU(2)$ algebra
\begin{equation}
    \omega=-e^\lambda\begin{pmatrix}0&(2e^{-2\lambda}-1)e^3&- e^2\\-(2e^{-2\lambda}-1)e^3&0& e^1\\ e^2&- e^1&0\end{pmatrix}.
\label{eq:spinconn}
\end{equation}

For later use, we introduce the left-invariant vector fields $\xi^a$ dual to $\sigma_L^a$ satisfying $[\xi^a,\xi^b]=\epsilon_{abc}\xi^c$.  Explicitly, we have
\begin{equation}
    \xi^1=\sin\psi\partial_\theta-\cos\psi(\csc\theta\partial_\phi-\cot\theta\partial_\psi),\quad \xi^2=\cos\psi\partial_\theta+\sin\psi(\csc\theta\partial_\phi-\cot\theta\partial_\psi),\quad\xi^3=\partial_\psi.
\label{eq:xivec}
\end{equation}
These generate the $SU(2)_r$ under which the frame rotates.   Note that, in the following, we will choose a conventional $SU(2)$ presentation by defining the generators
\begin{equation}
    J^1=-i\xi^1,\qquad J^2=i\xi^2,\qquad J^3=-i\xi^3,
\label{eq:Jvec}
\end{equation}
so that we have $[J^a,J^b]=i\epsilon^{abc}J^c$.

We now introduce the lens space $L(p,1)$ which shows up when considering the class one giants on $Y^{p,q}$.  The general lens space $L(p,q)\simeq S^3/\mathbb Z_p$ is obtained by the identification
\begin{equation}
    (z_1,z_2)=(\omega z_1,\omega^q z_2),
\end{equation}
where $\omega=e^{2\pi i/p}$ is a primitive $p$-th root of unity.  For $q=1$, examination of (\ref{eq:z1z2}) then shows that $L(p,1)$ is given by making the identifications
\begin{equation}
    (\phi,\psi)=(\phi+2\pi,\psi+2\pi)=(\phi,\psi+4\pi/p),
\label{eq:Lp1per}
\end{equation}
which generalizes (\ref{eq:periods}).

\subsection{Scalar harmonics}

The scalar Laplacian on this squashed $S^3$ is
\begin{equation}
    \nabla^2=4\left(\fft1{\sin\theta}\partial_\theta\sin\theta\partial_\theta+\fft1{\sin^2\theta}(\partial_\phi-\cos\theta\partial_\psi)^2+e^{-2\lambda}\partial_\psi^2\right).
\end{equation}
Now define $z=\sin^2\left(\fft\theta2\right)$ and take a wavefunction of the form
\begin{equation}
    \Phi(z,\phi,\psi)=e^{im\phi}e^{in\psi}f(z).
\end{equation}
The Laplacian acting on $f(z)$ then takes the form
\begin{equation}
    \nabla^2=4\left(\partial_zz(1-z)\partial_z-\fft1{4z(1-z)}(m-n+2zn)^2-e^{-2\lambda}n^2\right).
\end{equation}
This is a second order ODE with three regular singular points.  Solving the indicial equation at $z=0$ and $1$ suggests that we take
\begin{equation}
    f(z)=z^{r_1}(1-z)^{r_2}g(z),
\end{equation}
with
\begin{equation}
    r_1=\pm\fft12(m-n),\qquad r_2=\pm\fft12(m+n).
\end{equation}
The eigenvalue equation $(\nabla^2+\Lambda)f=0$ then turns into the hypergeometric equation
\begin{equation}
    z(1-z)g''+(c-(a+b+1)z)g'-abg=0 \label{eq: hypergeomeq},
\end{equation}
with
\begin{align}
    a,b&=\fft12\left(1+2r_1+2r_2\pm\sqrt{\Lambda+4(1-e^{-2\lambda})n^2+1}\,\right),\nn\\
    c&=2r_1+1. \label{eq: hypergeom params}
\end{align}
To be explicit, we take $a$ to have the positive sign of the square root.

For solutions that are regular at the ($z=0$ and $1$) poles, we restrict to
\begin{equation}
        r_1=\fft12|m-n|,\qquad r_2=\fft12|m+n|,
\end{equation}
in which case we end up with
\begin{align}
    a,b&=\fft12\left(1+2\max(|m|,|n|)\pm\sqrt{\Lambda+4(1-e^{-2\lambda})n^2+1}\,\right),\nn\\
    c&=|m-n|+1,
\label{eq:abc}
\end{align}
and we find the wavefunction
\begin{equation}
    \Phi(z,\phi,\psi)=e^{im\phi}e^{in\psi}z^{\fft12|m-n|}(1-z)^{\fft12|m+n|}{}_2F_1(a,b;c;z).
\label{eq:Phiwf}
\end{equation}
The hypergeometric function is regular at $z=0$.  Regularity at $z=1$ can be obtained by making sure $b$ is a non-positive integer, so that we have a polynomial solution to the hypergeometric equation.  (Here $b$ is chosen since it is the smaller of the two $a$ and $b$.)  As a result, we find the discrete spectrum
\begin{equation}
    \Lambda_{m,n,k}=(2\max(|m|,|n|)+2k+1)^2-1-4(1-e^{-2\lambda})n^2,\qquad k=0,1,2,\ldots.
\label{eq:Lmnk}
\end{equation}
This can also be written as
\begin{equation}
    \Lambda_{l,m,n}=l(l+2)-4(1-e^{-2\lambda})n^2,\qquad l=2\max(|m|,|n|)+2k.
\label{eq:Llmn}
\end{equation}
Note that the $n$ dependence drops out for the round $S^3$ when $\lambda=0$.

So far, we have not been explicit about the quantization conditions on $m$ and $n$.  Returning to the periods, (\ref{eq:periods}), we see that single-valuedness of $\Phi$ demands
\begin{equation}
    m+n\in\mathbb Z,\qquad 2n\in\mathbb Z.
\end{equation}
This is equivalent to taking
\begin{equation}
    (m,n)\in\mathbb Z\quad\mbox{or}\quad (m,n)\in\mathbb Z+\ft12. \label{eq: mn quantization}
\end{equation}
The low lying spectrum is given in Table~\ref{tbl:round}.

\begin{table}[t]
\centering
\begin{tabular}{l|l|l|l}
$l$&$(l/2,l/2)$&$(m,n)_k$&$\Lambda_{l,m,n}$\\
\hline
$0$&$(0,0)$&$(0,0)_0$&$0$\\
\hline
$1$&$(\fft12,\fft12)$&$(\pm\fft12,\pm\fft12)_0$&$3-(1-e^{-2\lambda})$\\
\hline
$2$&$(1,1)$&$(0,0)_1$&$8$\\
&&$(\pm1,0)_0$&$8$\\
&&$(0,\pm1)_0$&$8-4(1-e^{-2\lambda})$\\
&&$(\pm1,\pm1)_0$&$8-4(1-e^{-2\lambda})$
\end{tabular}
\caption{Low lying spectrum on the squashed $S^3$.  The angular momentum $l$ representations on the round sphere correspond to the $(l/2,l/2)$ representation of $SO(4)\simeq SU(2)_l\times SU(2)_r$.  Here, squashing breaks $SU(2)_r$ with corresponding weight $n$.  Note that the $\pm$ signs are uncorrelated.}
\label{tbl:round}
\end{table}

Alternatively, we can obtain the spectrum of scalar harmonics algebraically by using (\ref{eq:xivec}) and (\ref{eq:Jvec}) to write the scalar Laplacian as
\begin{equation}
    -\nabla^2=4\left(|\vec J|^2+(e^{-2\lambda}-1)(J^3)^2\right).
\end{equation}
This evaluates the quadratic Casimir of $SU(2)_r$ plus a squashing induced shift.  The spectrum is then given by  $-\nabla^2\Phi=\Lambda_{j,j_3}\Phi$ where
\begin{equation}
    \Lambda_{j,j_3}=4\left(j(j+1)+(e^{-2\lambda}-1)(j_3)^2\right)\qquad\mbox{for $SU(2)_r$ state}\quad|j,j_3\rangle.
\label{eq:jj3spec}
\end{equation}
The relation between this spectrum and (\ref{eq:Llmn}) determined by solving the Sturm-Liouville problem is easily derived.  In particular, we identify $j=l/2$ and $j_3=n$.  The latter follows directly from the generator $J^3=-i\partial_\psi$.  As in Table~\ref{tbl:round}, the spectrum falls in $SU(2)_l\times SU(2)_r$ representations $(l/2,l/2)=(j,j)$ with third components $(m,n)$.  Because $SU(2)_l$ is unbroken, each state enumerated in (\ref{eq:jj3spec}) transforms in the spin-$j$ representation of $SU(2)_l$ and is $(2j+1)$-fold degenerate corresponding to the degeneracy in $m$ of (\ref{eq:Llmn}).

Finally, we turn to the spectrum of scalar harmonics on the lens space $L(p,1)$, which is a $\mathbb Z_p$ orbifold of the three-sphere given by (\ref{eq:Lp1per}).  Since the orbifold acts freely, there are no twisted sectors, and the only effect of the orbifolding is to project onto singlets under the $\mathbb Z_p$ action.  This means the scalar spectrum is still given by (\ref{eq:Llmn}) for the squashed sphere.  However, the quantum numbers $m$ and $n$ must now satisfy the lens space periods, (\ref{eq:Lp1per}).  In particular, this gives
\begin{equation}
    m+n\in\mathbb Z,\qquad n\in\fft{p}2\mathbb Z.
\label{eq:lsper}
\end{equation}
Note that $n$ is no longer quantized in units of $1/2$, but is instead restricted to be multiples of $p/2$.  This highlights the $\mathbb Z_p$ orbifold nature of the lens space.  In particular, instead of taking $2n$ to be an integer, we restrict to $2n\equiv0\mod p$.

Alternatively, for $L(p,1)$, we can rewrite the metric (\ref{eq:squashed}) as
\begin{equation}
    ds^2=\fft14\left(d\theta^2+\sin^2\theta\, d\phi^2+e^{2\tilde\lambda}(d\tilde\psi+p\cos\theta d\phi)^2\right),
\label{eq: lens space general}
\end{equation}
where
\begin{equation}
    e^{2\tilde\lambda}=\fft1{p^2}e^{2\lambda},\qquad\tilde\psi=p\psi.
\label{eq:Lp1met}
\end{equation}
Note that $\tilde\psi$ has period $4\pi$.  Since this is just a field redefined version of (\ref{eq:squashed}), the spectrum is identical to what we obtained above.  In particular, from (\ref{eq:Phiwf}), (\ref{eq:Llmn}) and (\ref{eq:lsper}) we find
\begin{align}
    &\Phi(z,\phi,\tilde\psi)=e^{im\phi}e^{in\tilde\psi}z^{\fft12|m-pn|}(1-z)^{\fft12|m+pn|}{}_2F_1(a,b;c;z),\nn\\
    & \Lambda_{l,m,n}=l(l+2)-4(p^2-e^{-2\tilde\lambda})n^2,\qquad l=2\max(|m|,p|n|)+2k,\nn\\
    &m+pn\in\mathbb Z,\qquad n\in\fft12\mathbb Z,
\label{eq:Lp1sys}
\end{align}
where we made the replacement $n\to pn$.

\subsection{Vector harmonics}
In a similar manner, one could consider the vector harmonic problem in the squashed sphere \eqref{eq:squashed},
\begin{align}
    (g^{ij}\nabla^2 - R^{ij})X_j \equiv \Delta^{ij}X_j= -\Lambda X^i,
\end{align}
where $R^{ij}$ is the Ricci curvature associated with the squashed sphere metric $g$, $X^i$ is a vector field on the squashed sphere and $\Delta^{ij}$ is our notation for the \textit{vector} Laplace-Beltrami operator. More abstractly one could rewrite this as,
\begin{equation}
    \Delta X_{[1]} = -(d \delta + \delta d)X_{[1]} = - \Lambda X_{[1]}.
\end{equation}
The problem of finding the spectrum simplifies further if we restrict to the space of transverse vector fields,
\begin{align}
    \nabla^iX_i = 0,\, \qquad (\delta X_{[1]} = 0).
\end{align}
In this space, the vector Laplacian reduces to,
\begin{equation}
    \Delta X_{[1]} = -\delta d X_{[1]}.
\end{equation}
We use the following method to find the spectrum. Since this is a three-dimensional manifold, consider the \textit{curl} operator defined as,
\begin{equation}
    \text{curl} (X_{[1]}) = \star d X_{[1]}.
\end{equation}
Since $\delta=\star d\star$ on two-forms in three dimensions, we have
\begin{equation}
    \delta d X_{[1]} = \text{curl}(\text{curl}(X_{[1]})),
\end{equation}
so \textit{curl} commutes with $\delta d$, and hence with the Laplacian on the space of transverse one-forms. Moreover, \textit{curl} is self-adjoint on this space, so it can be diagonalized within each eigenspace of $\Delta$: if $\text{curl}(X)=\kappa X$, then $\Delta X=-\kappa^2X$, i.e.\ $\Lambda=\kappa^2$. Further, the manifold admits no harmonic one-forms, so $\kappa\neq0$ and every eigenspace of $\Delta$ is reached by eigenforms of \textit{curl}. It is therefore enough to solve the first-order differential equation
\begin{align}
    \text{curl} (X) = \kappa X,
\end{align}
to study the spectrum of transverse vector harmonics on this squashed $S^3$.

We choose the left invariant orthonormal frame,
\begin{align}
    e^1 = \dfrac{d\theta}{2}, \qquad e^2 = \dfrac{\sin\theta \, d\phi}{2}, \qquad e^3 = \dfrac{e^\lambda}{2} (d\psi + \cos\theta \,d\phi),
\end{align}
and torsion free connections adapted to this frame.
A general one form can be written as $X = X_a e^a$ where we use the ansatz 
\begin{equation}
    X_a(z,\phi,\psi) = e^{im\phi} e^{in\psi} f(z).
\end{equation}
For convenience we introduce,
\begin{align}
    \mu(\theta) := 2i \dfrac{(m-n\cos\theta)}{\sin\theta}, \qquad \nu := 2in\, e^{-\lambda}, \label{eq: munu class one}
\end{align}
which are just the $e^2,e^3$ frame components of $d$ acting on our ansatz. With this, $\text{curl}(X) = \kappa X$ gives a system of equations
\begin{align}
     \kappa f_1 &= \mu f_3 - \nu f_2 \nn, \\ 
    \kappa f_2 &= \nu f_1 - 2f_3', \nn\\
    \kappa f_3 &= 2f_2'-\mu f_1 + 2 \cot\theta f_2 - 2e^\lambda f_3.
\end{align}
Eliminating $f_1, f_2$ and changing the variable to $z = \sin^2\frac{\theta}{2}$ we have,
\begin{align}
    z(1-z)f_3(z)'' &+ (1-2z)f_3'(z) \\
    &+ \left(\dfrac{\kappa^2+2\kappa e^\lambda+4n^2 (1-e^{-2\lambda})}{4} +\dfrac{(m+n)^2}{4(z-1)} - \dfrac{(m-n)^2}{4z}\right)f_3(z) = 0. \label{eq: vector harmonic eq}
\end{align}
Similar to the scalar case, this is a Fuchsian equation with regular singular points at $z = 0,1,\infty$ and solving the indicial equation suggests,
\begin{equation}
    f_3(z)= z^{r_1} (1-z)^{r_2}g(z), \quad r_1 = \pm \dfrac{1}{2}(m-n), \quad r_2 = \pm \dfrac{1}{2} (m+n).
\end{equation}
This reduces \eqref{eq: vector harmonic eq} to the standard hypergeometric equation \eqref{eq: hypergeomeq} for $g$ with,
\begin{align}
    a,b &= \dfrac{1}{2} (1 + 2 \max(|m|,|n|) \pm \sqrt{1 + \kappa^2+2\kappa e^\lambda+4n^2 (1-e^{-2\lambda})}),\nn\\
    c &= |m-n| + 1,
\end{align}
and
\begin{equation}
    X_3(z,\phi,\psi) = e^{im\phi} e^{in\psi} z^{\frac{1}{2}|m-n|}(1-z)^{\frac{1}{2}|m+n|}{}_2F_{1}(a,b;c;z). \label{eq: vect sol}
\end{equation}
Requiring regularity at $z=1$ demands $b$ is a non positive integer. This results in a discrete spectrum,
\begin{equation}
    \kappa_\pm = - e^{\lambda} \pm \sqrt{e^{2\lambda} + 4l(l+1) - 4n^2(1-e^{-2\lambda})}, \label{eq: kappa}
\end{equation}
and
\begin{equation}
    \Lambda^{(\pm)}_{l,m,n} = 4l (l + 1) + 2 e^{2\lambda} - 4n^2 (1-e^{-2\lambda} ) \mp 2e^{\lambda} \sqrt{e^{2\lambda} + 4l (l + 1) - 4n^2(1-e^{-2\lambda})},  \label{eq: vect spectrum}
\end{equation}
where
\begin{equation}
    l = s + \max(|m|,|n|), \qquad s = 0,1,2,\ldots, \label{eq: lmn quantization}
\end{equation}
with the quantization conditions for $m,n$ are dictated by the periodicity of $\phi,\psi$ and is given by \eqref{eq: mn quantization}.
From \eqref{eq: vector harmonic eq}, we can restore $f_1$ and $f_2$ to be,
\begin{align}
    f_2^\pm (\theta) &= \dfrac{\mu \nu f_3(\theta) - 2\kappa_\pm f_3'(\theta)}{\kappa_\pm^2 - 4n^2 e^{-2\lambda}},\nn\\
    f_1^\pm(\theta) &= \dfrac{\mu f_3(\theta) - \nu f_2^\pm(\theta)}{\kappa_\pm} ,\label{eq: f1 f2 eq}
\end{align}
so that the full solution comes in two polarizations,
\begin{equation}
    X^{(\pm)}_{\ell,m,n}(\theta,\phi,\psi) = X_1^\pm (\theta,\phi,\psi) e^1 + X_2^\pm (\theta,\phi,\psi) e^2 + X_3 (\theta,\phi,\psi) e^3.
\end{equation}
It is worth noting that in the limit where we get the round sphere $(\lambda \to 0)$ the spectrum reduces to $\Lambda^{(+)}_l = 4l^2$ and $\Lambda^{(-)}_l = 4(l + 1)^2$. But in the round sphere case one would expect the transverse vector harmonics to be degenerate for a given $l$. The inconsistency can be fixed by observing that $\Lambda^{(+)}_l = \Lambda^{(-)}_{l - 1}$. So when one studies the effect of squashing as lifting of the degeneracy in one of the polarization directions one should be comparing the $\Lambda^{(+)}$ with $l$ quantum number and $\Lambda^{(-)}$ with $l - 1$ quantum number.

In the round sphere limit, each component $X_a$ are themselves scalar harmonics that transforms in the $(l,l)$ of $SU(2)_l \times SU(2)_r$ (there is a factor of 2 difference here in comparison to the scalar harmonics discussion in the previous section; that is because here $l$ can be $\mathbb Z/2$) and the co-frame is a vector that transforms under the $(0,1)$, since we chose it to be left invariant. Hence the vector harmonics themselves transform in $(l,l \pm 1), (l, l)$, where the latter are the longitudinal modes which we are not interested in. This suggests that when the $SU(2)_r$ is broken to $U(1)_r$, the corresponding quantum number $n$ should be bounded by $|n| \leq l - 1$ and $|n| \leq l +1$. This appears to be in contradiction with the quantization conditions derived in \eqref{eq: lmn quantization}.

This is resolved by looking at edge cases of \eqref{eq: vect sol}. Note that when $|n| = l$, \eqref{eq: kappa} reduces to $\kappa_+ = 2l e^{-\lambda}$ and $\kappa_- = -2( e^{\lambda}+l e^{-\lambda})$. The `$-$' polarization remains safe but from \eqref{eq: f1 f2 eq}, $f^+_2$ and hence $f^+_1$ becomes singular, making the `$+$' polarization ill defined. This suggests that $|n| = l$ is not allowed, and $|n| \leq l - 1$ for the `$+$' polarization.

We are missing the remaining $|n| = l + 1$ modes in the `$-$' polarization. This is because we can have an additional solution to the system of equations \eqref{eq: vector harmonic eq} with $f_3 = 0$. For $f_3 = 0$ the system reduces to a first order equation 
\begin{equation}
    -\nu f_2 = \kappa f_1, \quad \nu f_1 = \kappa f_2, \quad 2f_2' = \mu f_1 - 2\cot\theta\, f_2.
\end{equation}
This has to satisfy $\kappa = \pm 2ne^{-\lambda}$ and gives two solutions,
\begin{align}
    f_1(\theta) &\sim \sin(\theta)^{n-1}\tan \left(\dfrac{\theta}{2}\right)^{-m},\nn\\
    \tilde f_1(\theta) &\sim  \sin(\theta)^{-(n+1)}\tan \left(\dfrac{\theta}{2}\right)^{m}.
\end{align}
Regularity at both the poles demands $|m| \leq n-1$  for the first solution and $|m| \leq -(n+1)$. Define $|n| = l + 1$, then both solutions are valid for $|m| \leq l$. These states have $\Lambda 
= \kappa^2 = 4(l+1)^2 e^{-2\lambda}$. Although \eqref{eq: vect spectrum} was derived for $f_3\neq0$, evaluating its `$-$' branch at $|n|=l+1$ gives $\Lambda^{(-)}=(e^\lambda+|e^\lambda-2(l+1)e^{-\lambda}|)^2$, which agrees with $4(l+1)^2e^{-2\lambda}$ provided $e^{2\lambda}\le2(l+1)$. This holds for all $l$ whenever $e^{2\lambda}\le2$, which is the case for every squashing considered in this paper: $e^{2\lambda}=\ft23$ for $T^{1,1}$ and $e^{2\lambda}=2\tilde y\tilde c\ell=\ft23(1-\tilde y)<1$ for the class one giants. We summarize these results below.

\paragraph{The `$+$' polarization:} The harmonics are given by
\begin{equation}
    X^{(+)}_{l m n} = X_1^+ (\theta,\phi,\psi) e^1 + X_2^+ (\theta,\phi,\psi) e^2 + X_3 (\theta,\phi,\psi) e^3,
\end{equation}
where $X_3(z,\phi,\psi) = e^{im\phi}e^{in\psi} z^{\frac{1}{2}|m-n|}(1-z)^{\frac{1}{2}|m+n|}{}_2F_1(a,b;c;z)$ and $X_1, X_2$ follows from $X_3$ \eqref{eq: f1 f2 eq}. The spectrum is given by
\begin{align}
    \Lambda^{(+)}_{lmn} &= 4l (l + 1) + 2 e^{2\lambda} - 4n^2 (1-e^{-2\lambda} ) - 2e^{\lambda} \sqrt{e^{2\lambda} + 4l (l + 1) - 4n^2(1-e^{-2\lambda})},\nn\\
    l &= s +|m|, \quad s = 0,1,2,\cdots, \quad |n| \leq l - 1, \quad m + n \in \mathbb Z, \quad n \in \frac{1}{2}\mathbb Z.
\end{align}

\paragraph{The `$-$' polarization:} For $|n| \leq l$, the harmonics are given by
\begin{equation}
    X^{(-)}_{l m n}(\theta,\phi,\psi) = X_1^- (\theta,\phi,\psi) e^1 + X_2^- (\theta,\phi,\psi) e^2 + X_3 (\theta,\phi,\psi) e^3,
\end{equation}
where $X_3(z,\phi,\psi) = e^{im\phi}e^{in\psi} z^{\frac{1}{2}|m-n|}(1-z)^{\frac{1}{2}|m+n|}{}_2F_1(a,b;c;z)$ and $X_1, X_2$ follows from \eqref{eq: f1 f2 eq}. 

For $|n| = l + 1$, the harmonics are given by,
\begin{align}
    X^{(-)}_{l m n}(\theta,\phi,\psi) = X_1(\theta,\phi,\psi) e^1 + X_2(\theta,\phi,\psi) e^2,
\end{align}
where
\begin{align}
    X_1(\theta,\phi,\psi) &\sim \sin(\theta)^{n-1}\tan(\frac{\theta}{2})^{-m} e^{im \phi} e^{in\psi}, \quad n = l + 1,\nn\\
    X_1(\theta,\phi,\psi) &\sim \sin(\theta)^{-(n+1)}\tan(\frac{\theta}{2})^{m} e^{im \phi} e^{in\psi}, \quad n = - (l + 1).
\end{align}
$X_2$ is related to $X_1$ by $X_2 = - \kappa X_1/\nu$. For $e^{2\lambda}\le2$, the spectrum for all the $X^{(-)}_{l m n}$ takes the form
\begin{align}
    \Lambda^{(-)}_{lmn} &= 4l (l + 1) + 2 e^{2\lambda} - 4n^2 (1-e^{-2\lambda} ) + 2e^{\lambda} \sqrt{e^{2\lambda} + 4l (l + 1) - 4n^2(1-e^{-2\lambda})},\nn\\
    l &= s +|m|, \quad s = 0,1,2,\cdots, \quad |n| \leq l + 1, \quad m + n \in \mathbb Z, \quad n \in \frac{1}{2}\mathbb Z.
\end{align}

Following the discussion of the scalar spectrum on the corresponding lens space $L(p,1)$, (\ref{eq: lens space general}) and (\ref{eq:Lp1met}), we find that replacing $n\to pn$ and $e^{2\lambda}\to p^2e^{2\tilde\lambda}$ in the harmonics and eigenvalues of the squashed sphere gives their lens space analogues.  In particular, for the spectrum we find
\begin{equation}
    \Lambda^{(\pm)}_{lmn}=4l(l+1)+2p^2e^{2\tilde\lambda}-4n^2\left(p^2-e^{-2\tilde\lambda}\right)\mp2pe^{\tilde\lambda}\sqrt{p^2e^{2\tilde\lambda}+4l(l+1)-4n^2\left(p^2-e^{-2\tilde\lambda}\right)}.
\label{eq:lensvect}
\end{equation}
Note that one has to be consistent with the quantization conditions and the bounds on $m$ and $n$, which are in general different for the two polarizations.

\subsection{Spinor harmonics}
\label{app:spinor}

The Dirac operator on the squashed sphere (\ref{eq:squashed}) can be analyzed algebraically using the frame (\ref{eq:sLframe}), which makes $SU(2)_l\times U(1)_r$ explicit.  Taking the Dirac matrices on $S^3$ to be
\begin{equation}
    \gamma^1=\sigma^1,\qquad\gamma^2=-\sigma^2,\qquad\gamma^3=\sigma^3,
\end{equation}
the Dirac operator takes the form
\begin{equation}
    -i\slashed\nabla=2\vec\sigma\cdot\vec J+2(e^{-\lambda}-1)\sigma^3J^3+\fft12(e^\lambda+2e^{-\lambda}),
\label{eq:DiracO}
\end{equation}
where the middle term encodes the squashing and the final term comes from the spin connection, (\ref{eq:spinconn}).

The key observation for the spinor harmonics is the addition of angular momentum, $\vec K=\vec J+\vec S$ where $\vec J$ is the $SU(2)_r$ angular momentum and $\vec S=\fft12\vec\sigma$ acts on the spinor.  For the round sphere ($\lambda=0$), we then have
\begin{equation}
    -i\slashed\nabla=4\vec S\cdot\vec J+\fft32=2\left(|\vec K|^2-|\vec J|^2-|\vec S|^2+\fft34\right)=2\left(|\vec K|^2-|\vec J|^2\right)=2(k(k+1)-j(j+1)).
\label{eq:spinorbit}
\end{equation}
Since $k=j\pm\fft12$, we find the spectrum $-i\slashed\nabla\Psi=\Lambda_j^{(\pm)}\Psi$ with $\Lambda_j^{(+)}=2j+\fft32$ with $j\ge0$ and $\Lambda_j^{(-)}=-(2j+\fft12)$ with $j\ge\fft12$.  Taking into account the $SU(2)_l$ degeneracy, the spinor harmonics on the round $S^3$ fall in the $(j,j+\fft12)$ and $(j,j-\fft12)$ representations of $SU(2)_l\times SU(2)_r$ with eigenvalues $\Lambda_j^{(+)}$ and $\Lambda_j^{(-)}$, respectively.  This just reproduces the familiar spectrum of spinor harmonics on the round $S^3$.

We are, of course, mainly interested in the squashed sphere.  In this case, the mixed $\sigma^3J^3$ term in (\ref{eq:DiracO}) does not commute with the Dirac operator itself.  Hence we cannot use the spin-orbit coupling inspired procedure of (\ref{eq:spinorbit}) to directly obtain the spectrum.  Nevertheless, with $\vec K=\vec J+\vec S$, we can work in a basis of $\{|\vec J|^2,|\vec S|^2,|\vec K|^2,K^3\}$.  Since $s=\fft12$, we work in a two-dimensional subspace corresponding to spin-up and spin-down.

To highlight the form of the Dirac operator in this two-dimensional subspace, we introduce the raising and lowering components
\begin{equation}
    \sigma_\pm=\fft12(\sigma^1\pm i\sigma^2),\qquad J_\pm=J^1\pm iJ^2,
\end{equation}
so that
\begin{align}
    -i\slashed\nabla&=2(\sigma_+J_-+\sigma_-J_++e^{-\lambda}\sigma^3J^3)+\fft12(e^\lambda+2e^{-\lambda})\nn\\
    &=2\begin{pmatrix}e^{-\lambda}J^3&J_-\\J_+&-e^{-\lambda}J^3\end{pmatrix}+\fft12(e^\lambda+2e^{-\lambda})I_{2\times2}.
\end{align}
Acting on the two-component state
\begin{equation}
    \Psi=\begin{pmatrix}|k,k_3|j,s,j_3=k_3-\fft12,s_3=\fft12\rangle\\[1mm]
    |k,k_3|j,s,j_3=k_3+\fft12,s=-\fft12\rangle\end{pmatrix},
\end{equation}
we find the matrix
\begin{equation}
    -i\slashed\nabla=\begin{pmatrix}e^{-\lambda}(2k_3-1)&2\sqrt{j(j+1)+\fft14-(k_3)^2}\\2\sqrt{j(j+1)+\fft14-(k_3)^2}&-e^{-\lambda}(2k_3+1)\end{pmatrix}+\fft12(e^\lambda+2e^{-\lambda})I_{2\times2}.
\label{eq:2x2mat}
\end{equation}
Diagonalizing this matrix gives the eigenvalues of the Dirac operator
\begin{equation}
    \Lambda_{j,k_3}^{(\pm)}=\fft12e^\lambda\pm2\sqrt{j(j+1)+\ft14-(k_3)^2(1-e^{-2\lambda})}.
\label{eq:evD}
\end{equation}
Considering the addition of spin-$j$ with spin-$1/2$, the eigenstates of the Dirac operator transform under $SU(2)_l\times SU(2)_r$ as the $(j,k)$ representation with $k=j\pm\fft12$ and corresponding eigenvalues $\Lambda_{j,k_3}^{(\pm)}$.  The squashing explicitly breaks $SU(2)_r\supset U(1)_r$, and the spectrum is summarized in Table~\ref{tbl:spinorharm}.

\begin{table}[t]
\centering
\begin{tabular}{l|c|l}
$SU(2)_l\times SU(2)_r$&$(m,k_3)$&$\Lambda$\\
\hline
$(l/2,(l+1)/2)$&$|m|\le l/2,\quad |k_3|\le (l+1)/2$&$\Lambda_{l/2,k_3}^{(+)}$\\
$(l/2,(l-1)/2)$&$|m|\le l/2,\quad |k_3|\le (l-1)/2$&$\Lambda_{l/2,k_3}^{(-)}$
\end{tabular}
\caption{Spectrum of spinor harmonics on the squashed $S^3$.  We have set $j=l/2$ with integer $l\ge0$ for the $(+)$ tower and $l\ge1$ for the $(-)$ tower.  The quantum numbers $m$ and $k_3$ refer to the weights of $SU(2)_l$ and $SU(2)_r$, respectively, and the eigenvalues are given explicitly in (\ref{eq:evD}).}
\label{tbl:spinorharm}
\end{table}

Finally, we will need the case in which the spinor is in addition charged under a $U(1)$ bundle whose connection is proportional to the fiber one-form $e^3$; this is precisely what happens for the normal bundle of the giant, as discussed in Section~\ref{sec:fermions}.  Such a connection enters the Dirac operator through a term proportional to $\sigma^3$, which only shifts the diagonal of the $2\times2$ block in (\ref{eq:2x2mat}).  The resulting spectrum is obtained from (\ref{eq:evD}) by the replacement $k_3\to k_3+q$ in the squashing term alone, so that
\begin{equation}
    \Lambda_{j,k_3}^{(\pm)}=\ft12e^\lambda\pm2\sqrt{j(j+1)+\ft14-(k_3)^2+e^{-2\lambda}(k_3+q)^2},
\label{eq:diractwist}
\end{equation}
where $q$ is the charge of the spinor under the bundle in units set by the normalization of the connection.

\section{Internal modes of the class one giants and twisted harmonics} \label{app: internal modes}
The geometry of the base of $Y^{p,q}$ is more complicated than that of $T^{1,1}$ even though both are topologically $S^2 \times S^2$. The $Y^{p,q}$ metric is given by \eqref{eq:Ypq}. The $(y,\psi)$ two sphere is non trivially fibered over the $(\theta,\phi)$ two sphere. Because of this, one has to pay special attention while studying the  modes that fluctuate into $(y,\psi)$ (internal modes). It turns out that one has to treat these fluctuations as true sections of a complex line bundle, not just as scalar functions. This complication does not arise for the fluctuations into AdS.

Here we present a careful treatment of the internal modes of the class one giants localized at $y = \tilde y = y_1 \text{ or } y_2$. The class one giants wrap the Lens space $L(|\tilde c|,1)$ described by \eqref{eq:lsm}.  Near $y = \tilde y$, the $(y,\psi)$ metric can be written in the plane polar coordinates 
\begin{equation}
    dR^2 + \dfrac{q'{}^2(\tilde y) w(\tilde y) R^2}{36} (d\psi - \cos\theta d\phi)^2, \qquad R = \sqrt{\dfrac{4(y - \tilde y)}{w(\tilde y)q'(\tilde y)}}. \label{eq: y to R}
\end{equation}
For every point on the brane (fixed $(\theta,\phi)$), $\psi$ is a genuine polar angle with period $2\pi$. But because of the wrapping factor $\cos\theta d\phi$, the local frames at two points are related to each other by a position dependent rotation. Let us introduce rectangular coordinates $A, B$ such that,
\begin{equation}
    W = A + iB = R e^{i\psi} \label{eq: R to W}
\end{equation}
Under rotations $\psi \to \psi + \chi(\xi)$, $W \to e^{i\chi(\xi)}\,W$. So $W$ transforms like a complex scalar that is charged under a $U(1)$ gauge field ($\bar W$ transforms with the opposite charge, so in general we write $W \to e^{i\sigma \chi(\xi)}W$ with $\sigma = \pm1$ to keep track of both signs). Define the gauge field to be $\mathcal{A}_0$ with the transformation $\mathcal{A}_0 \to \mathcal{A}_0 +  d\chi$ and the covariant derivative $\mathcal{D}_i = \partial_i - i \sigma \mathcal{A}_0$. 

From the wrapping factor of the $(y,\psi)$ sphere,  the natural choice of the gauge connection is $\mathcal{A}_0 =  \cos\theta \, d\phi$\footnote{Notice that this resembles the gauge field of a Dirac monopole on $S^2$}. After a few lines of algebra, near $y = \tilde y$, one can rewrite the $Y^{p,q}$ metric as
\begin{align}
    ds^2 &= \dfrac{1-y}{6} (d\theta^2 + \sin^2 \theta\, d\phi^2) + |\mathcal{D}W|^2 + \left(\dfrac{q(y)}{9|W|^2} - 1\right) \dfrac{(\Im(\bar W \mathcal{D}W))^2}{|W|^2}\nn\\
    &\quad + w(y) \left(\dfrac{\ell}{2}(d\tilde\alpha + \tilde c  \mathcal{A}_0) + \dfrac{2f(y) + \ell \tilde c}{2|W|^2} \Im(\bar W \mathcal{D}W)\right)^2,
\end{align}
where $y$ is implicitly a function of $W$ though \eqref{eq: y to R} and \eqref{eq: R to W}. Since we are interested in constructing a quadratic action down the line, this can be written in a more suggestive form keeping only terms that are up to quadratic order in fluctuations away from $y = \tilde y$,
\begin{align}
    ds^2 &= \dfrac{1-\tilde y + 3\tilde y |W|^2}{6} (d\theta^2 + \sin^2\theta \, d\phi^2) + |\mathcal{D} W|^2 \nn\\
    \quad&+  (\tilde y \ell)^2 (1 + 3|W|^2) \left(d\tilde\alpha + \tilde c\mathcal{A}_0  + \dfrac{2}{\tilde y \ell} \Im (\bar W \mathcal{D} W)\right)^2. \label{eq: ypq1 nontrivial fiber}
\end{align}
The four form flux takes the form,
\begin{equation}
    C_{[4]} = \dfrac{L^4}{9} [(2-y)y - (2 - \tilde y)\tilde y] \sin\theta\, d\theta \wedge d\phi \wedge d\psi \wedge d\alpha ,
\end{equation}
where we have made an appropriate gauge choice.

The world-volume component of the DBI action up to quadratic order in these fluctuations takes the form,
\begin{align}
    \sqrt{-\det P[g]} = \dfrac{L^3}{3} |\tilde y| (1-\tilde y) \sin\theta \biggl[1 - \frac{1}{2}\left(L^2 |\dot W|^2 - |\mathcal{D} W|^2 - M_0^2 |W|^2\right) \biggr],
\end{align}
where we have modified $\mathcal{D}_i$ by redefining the gauge connection,
\begin{equation}
    \mathcal{A} = \mathcal{A}_0- 2 \tilde y \ell (d\tilde\alpha + \tilde c \mathcal{A}_0),
\end{equation}
and introduced an effective mass,
\begin{equation}
    M_0^2 = \dfrac{7 \tilde y -1}{1 - \tilde y}.
\end{equation}
Here one would be interested in the spectrum of the twisted Laplacian defined as,
\begin{align}
    \Delta_{\mathcal{A}}\Phi &= \dfrac{1}{\sqrt{\gamma}}\mathcal{D}_i (\sqrt{\gamma}\, \gamma^{ij} \mathcal{D}_j \Phi).
\end{align}
Define $z = \sin^2(\frac{\theta}{2})$ and consider the ansatz,
\begin{equation}
    \Phi(z,\tilde\alpha,\phi) = e^{im\phi} e^{in\tilde \alpha} f(z).
\end{equation}
Then, the eigenvalue problem $\Delta_{\mathcal{A}}\Phi = -\Lambda \Phi$ reduces to,
\begin{equation}
    \partial_zz(1-z)\partial_zf(z) - \left[\dfrac{(m - \hat n(1-2z))^2}{4z(1-z)} +\dfrac{\tilde c(n+2\sigma \tilde y \ell)^2}{2\ell\tilde y}- \dfrac{\tilde y \tilde c \ell}{2} \Lambda\right]f(z) = 0 ,\label{eq: twisted harmonic eqn}
\end{equation}
where
\begin{equation}
    \hat n = \tilde c n + \sigma, \quad \sigma = \pm 1.
\end{equation}
This is very similar to the scalar Laplacian, except for the shifts in $n$ captured by the charge under the gauge field. The indicial equation evaluated at the regular singular points suggest $f(z)$ takes the form,
\begin{equation}
    f(z) = z^{r_1} (1-z)^{r_2} g(z), \quad r_1 = \pm \frac{1}{2}(m-\hat n), \quad r_2 = \pm \frac{1}{2} (m+\hat n).
\end{equation}
This reduces \eqref{eq: twisted harmonic eqn} to the standard hypergeometric equation for $g$ with,
\begin{align}
    a,b &= \dfrac{1}{2} \left(1 + 2r_1 + 2r_2 \pm \sqrt{2\tilde c \tilde y \ell \left(\Lambda-\left(\dfrac{n}{\tilde y \ell} + 2\sigma\right)^2\right)+ 4\hat n^2 + 1}\,\right),\nn\\
    c &= |m-\hat n| + 1,
\end{align}
and
\begin{equation}
    \Phi(z,\phi,\tilde \alpha) = e^{im\phi} e^{in\tilde \alpha} z^{\frac{1}{2}|m-\hat n|} (1-z)^{\frac{1}{2}|m+\hat n|} {}_2F_1(a,b;c;z). 
\end{equation}
Regularity of the hypergeometric function requires that $b$ is a non positive integer. This will give,
\begin{equation}
    \Lambda_{l m n} = \dfrac{1}{2\tilde c \tilde y \ell} [4l (l + 1) - 4\hat n^2] + \left(\dfrac{n}{\tilde y \ell} + 2\sigma\right)^2, \quad l = s + \max (|m|,|\hat n|), \quad s = 0,1,2,\ldots.
\end{equation}
The quantization conditions for $m,n$ are dictated by the periodicity in the Lens space and remains the same as that of the scalar harmonics discussed in the previous section,
\begin{equation}
    m + \tilde c n \in \mathbb Z, \quad n \in \frac{1}{2} \mathbb Z.
\end{equation}

\section{Fermions of the class two giants} \label{App:FermionsClassII}

Here we present an analysis of the spectrum of the Dirac equation for the class two giants. Starting from \eqref{eq:class2dirac}, with some algebra, we can eliminate $\eta_2$ in favor of $\eta_1$ and gauge fix $\eta_1 = \phi u$ with $\phi'/\phi = A =\hat\mu/\sqrt{w(y)q(y)}$. This reduce (\ref{eq:class2dirac}) to the single equation
\begin{equation}
    u''-\left(\fft{B'}B-2A\right)u'-BE\,u=0,\qquad
    B=\fft{i(\hat\nu+\Lambda-\ft{\mathcal Q}4)}{\sqrt{wq}},\quad E=\fft{i(\Lambda-\ft{\mathcal Q}4-\hat\nu)}{\sqrt{wq}}.
\end{equation}
Note that, here $BE=-[(\Lambda-\ft{\mathcal Q}4)^2-\hat\nu^2]/wq$ is rational, and one finds that $\sqrt q$ cancels from $B'/B-2A$ as well.  The single surviving irrationality is therefore $\sqrt w$, and the coefficients of the equation lie in the field
\begin{equation}
    K=\mathbb Q\left(y,\sqrt{(a-y^2)(1-y)}\right),
\end{equation}
since $\sqrt w=\sqrt{2(a-y^2)(1-y)}/(1-y)$.  The cubic $(a-y^2)(1-y)$ has roots $\pm\sqrt a$ and $1$, and discriminant $4(a-1)^2a$.  Now $a=3y_1^2-2y_1^3$ with $y_1<0$, so $a>0$ always. We cannot have $a=1$ since it forces $y_1=-\ft12$ and hence $q=p$, which lies outside the allowed range $0<q<p$.  The three roots are therefore distinct for every $Y^{p,q}$, and $K$ is the function field of a smooth elliptic curve,
\begin{equation}
    \varsigma^2=2(a-y^2)(1-y),
\label{eq:class2elliptic}
\end{equation}
of genus one.  A curve of genus one admits no rational parametrization, so no change of variable can rationalize (\ref{eq:class2dirac}).  This is precisely why the Ansatz that produced (\ref{eq:class2bpsvec}) for vectors has no analogue here.

What the argument does say, constructively, is that the Dirac problem is rational on the elliptic curve (\ref{eq:class2elliptic}).  This is enough to obtain the BPS spinors in closed form, as we now do.

The irrational exponent at $r_3$ at $y_3$ multiplies both $\eta_1$ and $\eta_2$, so it cancels in their ratio. Setting $\chi_R=\eta_2/(\sqrt{q}\,\eta_1)$, which removes $\sqrt q$ from the system, the pair (\ref{eq:class2dirac}) collapses to the Riccati equation
\begin{equation}
    \chi'_R=-\fft{i(\hat\nu+\Lambda-\ft{\mathcal Q}4)}{\sqrt w}\chi^2_R-\left(2A+\fft{q'}{2q}\right)\chi_R
    +\fft{i(\Lambda-\ft{\mathcal Q}4-\hat\nu)}{\sqrt w\,q},
\label{eq:class2ricc}
\end{equation}
whose coefficients are rational on (\ref{eq:class2elliptic}).  Seeking a solution that is likewise rational there, we consider the ansatz $\chi=(P(y)+Q(y)\varsigma)/(q(y)(a - y^2))$ with $P$, $Q$ cubic. One finds exactly two solution,
\begin{equation}
    P=\pm i(a-y^2),\qquad Q=iy,
\end{equation}
related by charge conjugation, $\chi\to\sigma^2\chi^*$.  Only the upper sign is regular in the first quadrant, and we keep it. Using $\varsigma=(1-y)\sqrt w$, it takes the remarkably simple form
\begin{equation}
    \fft{\eta_2}{\eta_1}=\fft{i}{\sqrt q}\left(1+\fft{2y}{\sqrt w}\right).
\label{eq:class2bpsspinor}
\end{equation}
We note that (\ref{eq:class2bpsspinor}) is \emph{universal}: it contains no reference to $m$, $n$ or to $p$ and $q$, entering only through the metric functions $w(y)$ and $q(y)$.  More importantly, it solves (\ref{eq:class2ricc}) if and only if
\begin{equation}
    \Lambda =\ft32r_*+\ft12+Q,\qquad \ft32r_*=3m-\fft n{2\ell};
\label{eq:class2diraceig}
\end{equation}
for any other $\Lambda$ the residual fails to vanish.  The ratio is unchanged by the normal-bundle connection, which shifts only the eigenvalue.  The world-volume energy is then
\begin{equation}
    H=\Lambda-2Q=\ft32r_*+\ft12-Q,
\end{equation}
so that $\mathcal H=H -J_1-J_2-\ft32r_*=\ft12-Q-J_1-J_2$, which is precisely (\ref{eq:univBPS}).  Without the normal-bundle connection one would instead find $\mathcal H=\ft12-2Q-J_1-J_2$, which is half-odd and never vanishes, the twist is essential for the fermions to be BPS.

\bibliographystyle{JHEP}
\bibliography{refs}

\end{document}